\documentclass[3p,times,twocolumn]{elsarticle}

 \pdfoutput=1
\usepackage{graphicx}
\makeatletter
\def\ps@pprintTitle{%
 \let\@oddhead\@empty
 \let\@evenhead\@empty
 \def\@oddfoot{}%
 \let\@evenfoot\@oddfoot}
\makeatother

\usepackage[binary-units=true]{siunitx}
\usepackage{siunitx}
\usepackage{color}
\usepackage{verbatim}
\usepackage {longtable}
\usepackage {caption}
\usepackage{subcaption}
\usepackage {hhline}
\usepackage{soul}
\usepackage{amssymb}
\usepackage{lineno}
\usepackage{graphicx}
\usepackage{tikz}
\newcommand*\circled[1]{\tikz[baseline=(char.base)]{
            \node[shape=circle,draw,inner sep=2pt] (char) {#1};}}
\usepackage{amsmath}
\usepackage[misc]{ifsym}

\begin{document}
\begin{frontmatter}

\title{The evolution of the temperature field during cavity collapse in liquid nitromethane. Part II: Reactive case}

\author[cav]{L.˜Michael}
\ead{lm355@cam.ac.uk}

\author[cav]{N.~Nikiforakis}
\ead{nn10005@cam.ac.uk}

\address[cav]{Laboratory for Scientific Computing, Cavendish Laboratory, Department of Physics, University of Cambridge, UK}

\begin{abstract}
This work is concerned with the effect of
cavity collapse in non-ideal explosives as a means of controlling their sensitivity. The main objective is to understand the origin of localised temperature peaks (hot spots) which play a leading order role at the early stages of ignition. To this end we perform two- and three-dimensional numerical simulations of shock induced single gas-cavity collapse in liquid nitromethane. Ignition is the result of a complex interplay between fluid dynamics and exothermic chemical reaction. In the first part of this work we focused on the hydrodynamic effects in the collapse process by switching off the reaction terms in the mathematical formulation. In this part, we reinstate the reactive terms and study the collapse of the cavity in the presence of chemical reactions. By using a multi-phase formulation which overcomes current challenges of cavity collapse modelling in reactive media we account for the large density difference across the material interface without generating spurious temperature peaks thus allowing the use of a temperature-based reaction rate law. 
The mathematical and physical models are validated against experimental and analytic data. In Part I, we demonstrated that, compared to experiments, the generated hot spots have a more complex topological structure and additional hot spots arise in regions away from the cavity centreline. Here, we extend this by identifying which of the previously-determined high-temperature regions in fact lead to ignition and comment on the reactive strength and reaction growth rate in the distinct hot spots. We demonstrate and quantify the sensitisation of nitromethane by the collapse of the isolated cavity by comparing the ignition times of nitrometane due to cavity collapse and the ignition time of the neat material. The ignition in both the centreline hot spots and the hot spots generated by Mach stems occurs in less than half the ignition time of the neat material. We compare two- and three-dimensional simulations to examine the change in topology, temperatures and reactive strength of the hot spots by the third dimension. It is apparent that belated ignition times can be avoided by the use of three-dimensional simulations. The effect of the chemical reactions on the topology and strength of the hot spots in the timescales considered is also studied, in a comparison between inert and reactive simulations where maximum temperature fields and their growth rates are examined. 
\end{abstract}

\begin{keyword}condensed phase explosives \sep cavity collapse \sep
temperature \sep nitromethane \sep hot spots \sep ignition\end{keyword}
\end{frontmatter}

\section{Introduction}
\label{intro}

This work is motivated by the necessity for optimising the performance of non-ideal, inhomogeneous explosives such as those 
used in mining. In order to increase their sensitivity and to 
control their performance, cavities are introduced in the bulk of the explosive, often by means of glass micro-balloons. 
When a precursor shock wave passes through the explosive, these cavities 
collapse, generating regions of locally high pressure and temperature, which are commonly 
referred to as hot spots. These lead to multiple local ignition sites, which cumulatively result in a shorter time to ignition than that of the neat material. Parameters such as the number, size, shape and 
distribution of the cavities affect the degree of sensitisation of the explosive. Understanding the correlation between these 
parameters and the reduced ignition time will allow better control of the 
behaviour of the explosive.

To this end, the collapse of cavities has been extensively studied in the past by means of experiment and numerical
simulation. An extensive literature review on previous studies in inert materials is given in Part I of this work \cite{michael2017parti}. Hence here we indicatively only mention some of the studies for cavities collapsing in reactive liquid, gelatinous and solid explosives.
To determine the effect of the collapse on the hot spot generation and explosive ignition, cavity collapse experiments in reactive materials were performed by Bourne and Field \cite{bourne1991bubble,bourne1999shock}. The principal ignition mechanism was determined to be the hydrodynamic heating due to jet impact. Conduction from the highly-compressed, heated gas inside the cavity does not occur in the short time-scales governing the particular shock wave-cavity configurations studied.

Limitation of computational power restricted early numerical work on cavity collapse in reactive materials. Amongst the first studies on cavity collapse in a reactive medium was the work by Bourne and Milne \cite{bourne2002cavity,bourne2003temperature} who observed numerically the same loci of hot spots as in experiments. More recently the authors presented limited results on simulating cavity collapse in reacting nitromethane \cite{michael2015DetSymp}. Kapila et al.\ \cite{kapila2015numerical} presented the collapse process and the detonation generation in an exemplary explosive and studied the effect of cavity shape in the detonation generation using a pressure-dependent reaction rate. In an elastoplastic framework, Tran and Udaykumar \cite{tran2006simulation2} studied the response of reactive HMX with micron-sized cavities, while Rai et al.\ \cite{rai2017high,rai2017collapse} considered the resolution required for reactive cavity collapse simulations and the sensitivity behaviour of elongated cavities in HMX.

Despite technological advances, performing complete numerical simulations of ignition due to shock-induced cavity collapse still poses many challenges. These include the use of complex equations of state to describe the explosive materials, maintaining oscillation-free pressure, velocity and temperature fields across the cavities material boundaries upon their interaction with shock waves, sustaining (at least) 1000:1 density difference across these boundaries,  retrieving physically accurate temperature fields in the explosive matrix and numerically modelling the ignition of the material as a temperature-driven phenomenon. Moreover, the computational power needed for accurately resolving the complete phenomenon in three-dimensions is still large. As the explosive initiation is a temperature-driven phenomenon, the challenges regarding the temperature field are of critical importance. These challenges are described in detail in art I of this work.

A complete physical simulation of the initiation of a condensed phase explosive due to cavity collapse has several requirements; a three-dimensional framework, realistic material models (equations of state), oscillation-free material interfaces, the ability to recover accurate and oscillation-free temperature fields. A temperature-dependent reaction rate law is also desirable, to mathematically describe ignition to be driven by the heating of the material. Moreover, each component used should be validated alone and in combination with all the components composing the numerical framework. 

Simulations presented in the literature satisfy some but not all of these requirements. In this work we exercise the mathematical model (MiNi16) proposed by the authors in previous work \cite{michael2016hybrid} to overcome the difficulties in numerically simulating the cavity collapse and move towards a complete simulation of explosive initiation due to cavity collapse. In Part I of this work \cite{michael2017parti} we simulated the three-dimensional collapse of isolated air cavities in nitromethane using a validated equation of state (Cochran-Chan). We looked in depth how the hydrodynamical effects in absence of reaction (e.g.\ generation and propagation of waves) lead to local temperature elevation. Such regions were identified as candidates for critical hot spots in a  reactive simulation and the necessity for three-dimensional (as opposed to two-dimensional) simulation was justified. 
In this second part of the work, we take advantage of oscillation-free and reliable temperature fields that can be recovered by using the MiNi16 model and extend the work of Part I to reactive scenarios. We perform three-dimensional simulations of the collapse of isolated air cavities in reacting, liquid nitromethane, using equations of state in Mie-Gr\"uneisen form and a temperature-dependent reaction-rate law. We study in detail the ignition process and we link the evolution of the reaction-progress variable to the temperature elevations and the wave-pattern generated during the collapse process. We identify the reacting hot spots and study their relative reactive strength and reaction growth rates. The effect of the cavity collapse on shortening the time to ignition is illustrated explicitly by comparing the ignition due to cavity collapse against the ignition of the neat material. We also demonstrate the necessity for three-dimensional simulations (compared to 2D) by looking at the percentage of burnt material over time in the two scenarios but also looking at the evolution of waves and temperature fields. Moreover, we compare inert and reactive simulations to examine the added effect of the reactions on the temperature fields and the topology of the hot spots at the timescales considered.

The rest of the paper is outlined as follows: the next section presents the underlying mathematical formulation in terms of the governing partial differential equations, the equations of state that close the system and the form and calibration of the reaction rate law for nitromethane combustion. A section on validation follows, where we compare numerical results against theoretical and experimental temperatures in shocked nitromethane and the CJ and von Neumann values for steady state detonation. The ignition regime is validated in a similar way, by comparing numerical and experimental times-to-ignition for various input pressures with the use of an ignition Pop-plot. 
In the results section, we consider the collapse of a gas cavity in liquid nitromethane, follow the events leading to the generation of local temperatures which are more than three times the post-shock temperature of the neat material and compare and analyse the difference between the 2D and 3D simulations as well as inert and reactive simulations in the context mentioned earlier.

\section{Mathematical and physical model}
A formulation which rectifies the issues commonly presented in shock-bubble simulations was proposed by Michael and Nikiforakis \cite{michael2016hybrid}. This formulation considers the cavity as an inert phase (\emph{phase 1}) and the surrounding material as a reacting phase (\emph{phase 2}) composed of two materials; the reactant nitromethane as \emph{material $\alpha$} and the gaseous products of reaction as \emph{material $\beta$}. Mixing rules are employed to determine the properties of phase 2 from the properties of the two materials. Mixing rules are also in effect between phase 1 and phase 2 across material interfaces where the diffusion zones lie. In this work, we neglect the effect of reaction products and thus use a reduced form of the formulation. 

Consider the gas inside the cavity to be \emph{phase 1} and the liquid nitromethane around the cavity  to be  \emph{phase 2}. Then,  
the governing equations for this system take the form:
\begin{eqnarray}
\frac{\partial z_1 \rho_1}{\partial t}+\nabla \cdot(z_1 \rho_1 \mathbf{u})& = & 0, \nonumber \\
\frac{\partial z_2 \rho_2}{\partial t}+\nabla \cdot(z_2 \rho_2 \mathbf{u})& = & 0, \nonumber\\ \label{Hyb:system}
\frac{\partial}{\partial t}(\rho \mathbf{u_k})+\nabla \cdot(\rho \mathbf{u_ku})+\frac{\partial p}{\partial{\mathbf{x_k}}} & = & 0, \\ 
\frac{\partial}{\partial t}(\rho E)+\nabla \cdot [ (\rho E+p)\mathbf{u}] & = & 0, \nonumber \\ 
\frac{\partial z_1 }{\partial t} + \mathbf{u} \cdot \nabla z_1  & = & 0, \nonumber \\ 
\frac{\partial z_2 \rho_2 \lambda}{\partial t}+\nabla \cdot(z_2 \rho_2 \mathbf{u}\lambda) & = & z_2 \rho_2 K, \nonumber 
\end{eqnarray}
where for $i=1,2$, $\rho_i$  are the densities for the air and nitromethane, $z_i$ are their corresponding volume fractions 
($z_1+z_2=1$),
$\rho$ is the total density given by $\rho=z_1\rho_1+z_2\rho_2$, $\mathbf{u}$
is the velocity vector and $p$ is the total pressure. The total specific energy is given by $E=\frac{1}{2}u^2+e$, where $e$ is the total specific internal
energy. Also, $\lambda$ is a reaction progress variable and $K$ represents the reaction source terms, to be defined later. 
 The mixture rule for the total internal energy is given 
by $\rho e=\rho_1z_1e_1+\rho_2z_2e_2$, where $e_i$ for $i=1,2$ are the specific internal energies for the air and nitromethane, 
given by their corresponding equations of state. A mixture rule for $\xi=\frac{1}{\gamma-1}$, where $\gamma$ is the total 
adiabatic index, is also required and in this case is given by $\xi=z_1\xi_1+z_2\xi_2$. 
The sound speed for the total mixture is given by 
\begin{equation}
\xi c^2 = y_1\xi_1c_1^2+y_2\xi_2c_2^2,
\end{equation}
 where $y_i$ is the mass fraction of \emph{phase i}, given by $y_i=\frac{\rho_iz_i}{\rho}$, for $i=1,2$.

This formulation can be considered an augmented two-phase model for condensed-phase explosives in the same way that the Euler equations have been augmented to 
study gaseous combustion problems (Nikiforakis and Clarke \cite{nikiforakis1996numerical}). 

The nitromethane is modelled by the Cochran Chan equation of state as presented in Part I of this work. In reactive simulations, a term $Q$ is included in the reference energy function of the reactant, representing the heat of detonation released upon reaction, such that $e_{\text{ref}} = e_{\text{ref}} + Q$. Alternatively, this term can be in incorporated as a source term to the energy equation.

\subsection{Reaction rates for nitromethane}
In order to model the reactions in nitromethane, a single-step, temperature-dependent Arrhenius reaction rate law is used, of the form
\begin{equation}
K=\frac{d\lambda}{dt}=-\lambda Ce^{-T_A/T_{NM}},
\end{equation}
where $C$ is a constant pre-exponential factor and $T_A$ is the activation
temperature of the material.

Many sets of the reaction rate parameters  are available in the literature for liquid nitromethane. For example, $(C,T_A)=(\SI{2.6e9}{\per \second}$, $\SI{11500}{\kelvin}$) is used by Menikoff and Shaw \cite{menikoff2011modeling}, $(\SI{6.9e10}{\per \second}, \SI{14400}{\kelvin}$)  by Tarver and Urtiew \cite{tarver2010theory}  and  $(\SI{1.27e12}{\per
\second}$, $\SI{20110}{\kelvin})$ by Ripley et al.\ \cite{ripley2006detonation}.
In this work we adopt the pre-exponential factor proposed by Menikoff and 
 Shaw \cite{menikoff2011modeling} and adjust the activation temperature to $T_A=$\SI{11350}{\kelvin} to match 
the experimentally calculated overtake time and the shape of the velocity versus distance graph of the shock-induced ignition 
experiment in Sheffield et al.\ \cite{sheffield2006homogeneous} (neat nitromethane, shocked at \SI{9.1}{\giga \pascal}).

In order to calculate the value for the heat of detonation, $Q$, we follow the approach described by Arienti \cite{arienti2004shock}. 
This involves varying the parameter $Q$ to match as closely as possible the experimental value of the pressure at the CJ 
point of \SI{12.5}{\giga \pascal} \cite{dattelbaum2009influence} on the $p\text{-}v$ plane. In general, varying the value 
of $Q$ shifts the reactive Hugoniots upwards. When the Rayleigh line becomes tangent to the reactive (pseudo-product) 
Hugoniot, the appropriate value for $Q$ is determined. Here, $Q=\SI{4.48e6}{\joule \per \kilogram}$. 

We note that the set of calibrated parameters given above is valid for steady-state detonation propagation of neat nitromethane  
and no further adjustment is necessary for ignition case studies, as the results in the validation section indicate. 

\section{Validation}
System  (\ref{Hyb:system}) is integrated numerically using a high resolution shock-capturing numerical scheme, namely the 
MUSCL-Hancock finite volume method with an underlying HLLC Riemann solver. Hierarchical, structured, adaptive mesh refinement 
(AMR) is used to dynamically increase the resolution locally \cite{bates2007richtmyer}.  The source terms that describe chemical reactions are integrated 
with a high-order Runge-Kutta scheme. 

The aim of this section is to validate the resulting code and also to assess whether the combination of parameters described in the 
previous section can match the experimentally determined von Neumann spike, the CJ values and the ignition 
pop-plot, without any further user adjustment. The inert formulation and physical model were validated in Part I of this work, showing non-oscillatory hydrodynamic fields and recovery of realistic temperature fields in shocked nitromethane.

\subsection{Steady ZND detonation}
Having established a reasonable post-shock temperature, we now assess whether the code predicts the correct CJ and 
von Neumann peak values. To this end, a one-dimensional slab of liquid nitromethane is shocked to \SI{23}{\giga \pascal}, 
a value close to the von Neumann pressure.  

Following an initial unsteady evolution, the simulation settles to a steady detonation wave, yielding a constant value of 
the von Neumann spike pressure of \SI{22.7}{\giga \pascal} and a pressure at the CJ point equal to \SI{12.6}{\giga \pascal}. 
These values fall within the range of values reported in the literature \cite{sheffield2002particle}.

\subsection{Comparison with experimental  ignition times }
For the final part of this assessment, it is worth recalling that we are mainly interested in the early ignition stages of 
the shock-to-detonation process. It is not unusual to find that, depending on the mathematical formulation of the model, 
the reaction rate parameters used for detonation modelling are not suitable for ignition studies. This is attributed to the fact that the parameters are adjusted 
to fit post-shock temperatures and steady detonation values. To assess whether the current set-up can be employed for arbitrary 
studies without any further adjustment, we compare our numerical results against experimental Pop-plot data that show induction 
time (i.e.\  time to ignition) versus input pressure.
\begin{figure}[!th]
\centering
\includegraphics[width=0.5\textwidth]{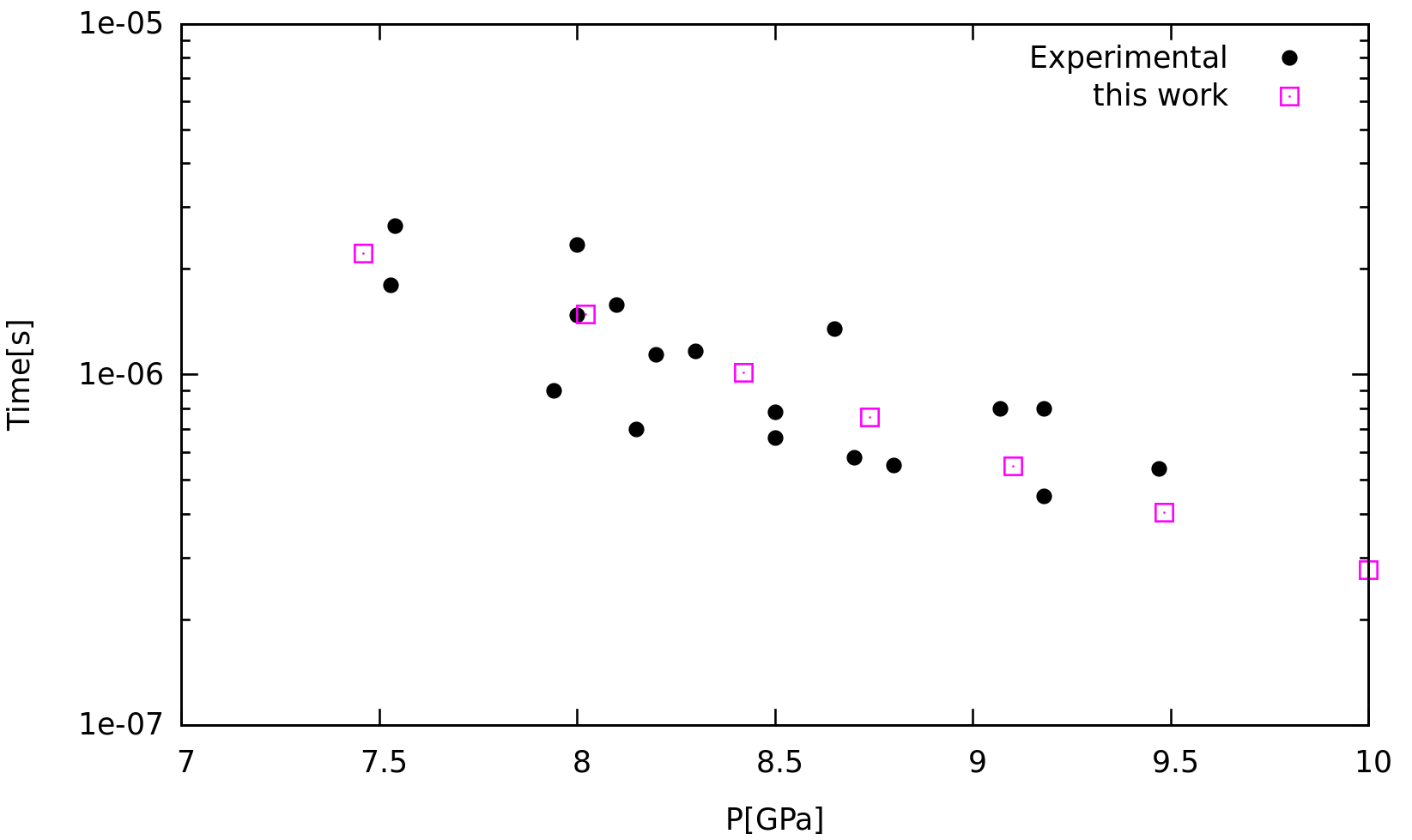}
\caption{ Ignition time \emph{vs} input pressure Pop-plot. The black filled circles represent experimental data and the 
open squares numerically calculated data. }\label{ignitionTimes}
\end{figure} 
The exact time when ignition occurs is problematic, since ideally, one would use the same definition of time of ignition for both 
numerical and experimental results. In numerical simulations, the time of ignition can be defined as the time when a specific 
fraction of the explosive material has reacted. The selection of the appropriate percentage should be based on experiments.
 However, experimentally it is rather difficult, where possible at all, to measure the chemical species, and ignition is usually 
measured using luminosity, something that is not trivial  to accurately and consistently retrieve from the simulations. 
Thus, we arbitrarily define ignition to be the time when a small percentage (namely $10\%$) of the explosive material has 
reacted and compare our numerical results against experimental data taken from Berke et al.\ \cite{berke1970shock},
 Hardesty \cite{hardesty1976investigation} and Chaiken \cite{chaiken1978correlation}. 
This comparison is presented in Fig.\  \ref{ignitionTimes} and shows that the numerical results fall well within the range 
of the experimental ignition data.

\section {Ignition of neat nitromethane}

\label{Sec:neat2000}

In condensed phase explosives, initiation can be achieved by the passage
of a shock wave through the quiescent explosive, raising the temperature
and pressure and triggering the start of reaction. 

To study the initiation process of nitromethane,
simulations of the shock-induced ignition of neat, liquid nitromethane are
performed.

Since the ignition process of this application is purely one-dimensional, we consider a domain of dimensions $[\SI{0}{\milli \meter},\SI{3.2}{\milli\meter}]$ 
and four levels of refinement, each with a refinement factor of $2$, resulting
in an effective resolution of $2560$ cells ($dx=\SI{0.85}{\micro 
\meter}$). This resolution resolves well the steady state reaction zone of
liquid nitromethane which is calculated from experiment \cite{sheffield2002particle} to have width of  $\sim$\SI{300}{\micro \meter}. The initial conditions for this test are given in Table \ref{ICneat}. 

\begin{table*}
\centering
\caption {Initial conditions for the initiation of neat nitromethane.}
\label{ICneat}
\begin{tabular}{lcccccc}
\hline\noalign{\smallskip}
Region & $\rho_1 [\SI{}{\kilogram \per \meter \tothe{3}}]$ & $\rho_{2} [\SI{}{\kilogram \per \meter \tothe{3}}]$ & $u [\SI{}{\meter \per \second}]$ & $p [\SI{}{\pascal}]$  & $\lambda$ &$z$  \\
\noalign{\smallskip}\hline\noalign{\smallskip}
 shocked nitromethane  & 1934 & 1934 & 2000  & $10.98\times10^9$ & 0 &$10^{-6}$ \\
ambient nitromethane  & 1134 & 1134 & 0 &   $10^{5}$ & 1 & $10^{-6}$ \\
\noalign{\smallskip}\hline
\end{tabular}
\end{table*}

When a shock wave is set up numerically as an initial condition, a start up
error is generated due to the symmetric Riemann problem \cite{leveque1998nonlinear}.
This error has the form of a small well in the density distribution behind
the shock wave, which translates into a small hill in the temperature field. 
Since the reaction rate we use to
model the reactions in liquid nitromethane depends exponentially on the temperature,
even a disturbance of a small magnitude in this field (here $\sim$\SI{20}{\kilogram
\per \meter \tothe{3}}) would rapidly grow, generating a spurious hot-spot.
The hot-spot would, in turn, lead to ignition earlier than it would be expected
if the density field was clean. In order to remove the disturbance by  extrapolating
from the non-disturbed shocked state, or by cutting the domain at a point
after the start up error, the disturbance has to be sufficiently formed. Thus,
before treating the error, the shock wave is  allowed to travel some small
but significant
distance from its initial position.
If reactions were turned on during this travel time, the numerical hot spot
would affect the state behind the shock wave. To overcome this, an  inert
shock wave is allowed to  travel the distance required
for the start up error to be adequately formed and then the part of the domain
that is more than 5 cells behind the shock wave is cut off.
After the cutoff, the simulation
is restarted with the reactions turned on.

 The cut of the domain is performed at time $0.02291\SI{}{\micro \second}$
and reaction is allowed to start at time $0.02291\SI{}{\micro \second}$.
All the times referred hereafter are relative to the reaction start
time. Also, all the positions are relative to the cut-off position ($x=\SI{0.123}{\milli
\meter}$), as the domain is considered to be repositioned at $x=0$ after the
cut-off.

\begin{figure*}[!t]
\centering
\begin{minipage}{2\columnwidth}
\centering
\begin{subfigure}[b]{0.45\textwidth}
\includegraphics[width=0.95\textwidth]{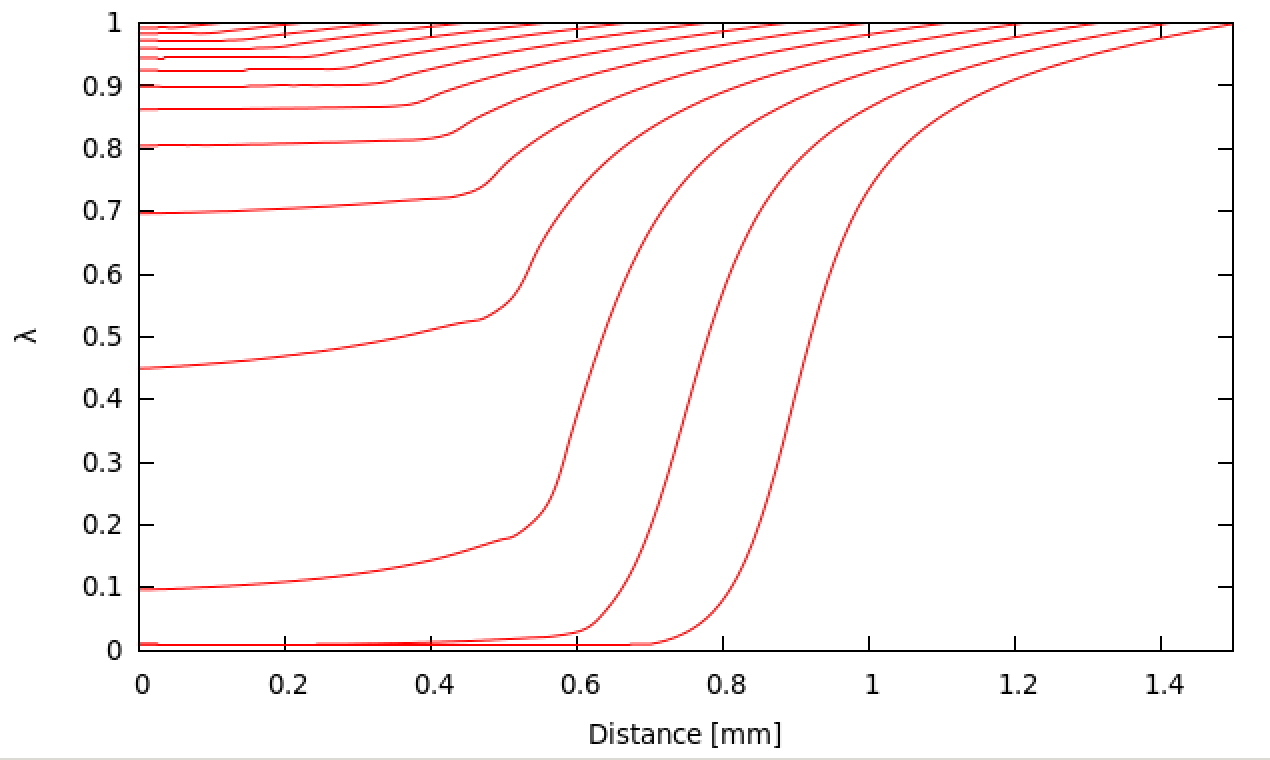}
 \caption{\label{a}}
\end{subfigure}
        \begin{subfigure}[b]{0.45\textwidth}
\includegraphics[width=0.95\textwidth]{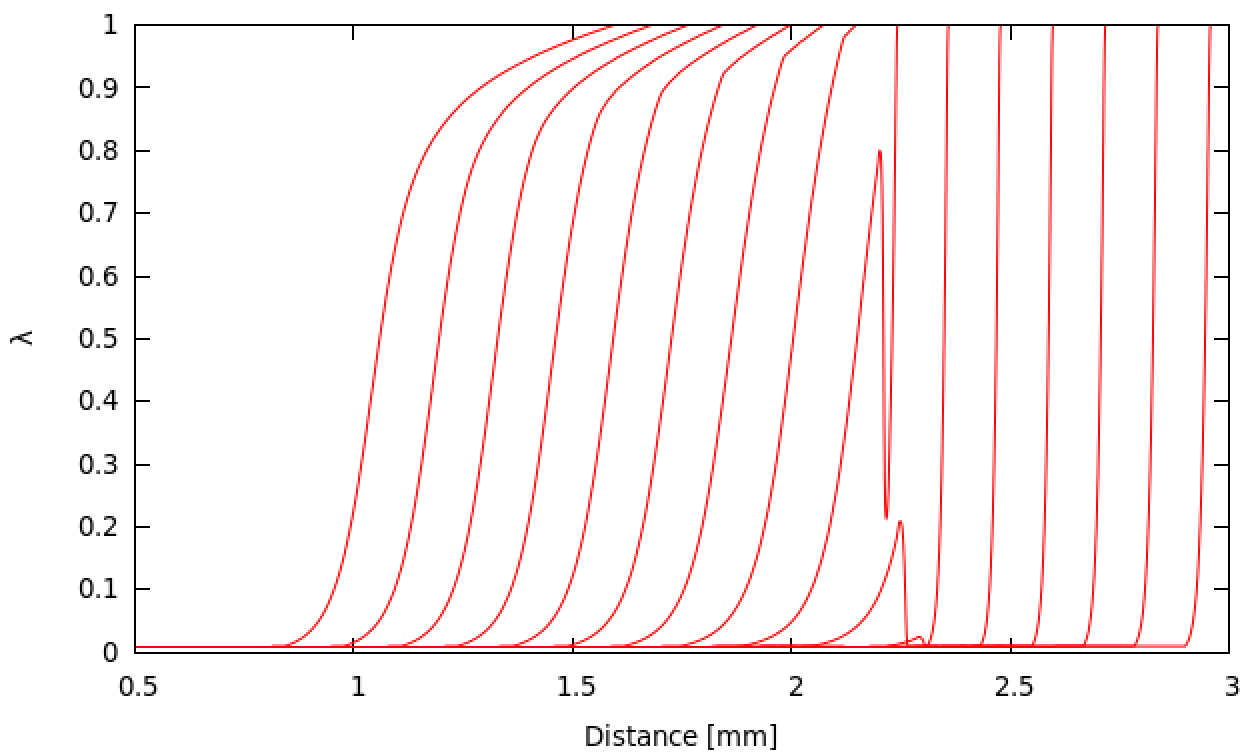}
 \caption{\label{b}}
        \end{subfigure}
\end{minipage}
\begin{minipage}{2\columnwidth}
\centering
        \begin{subfigure}[b]{0.45\textwidth}
\includegraphics[width=0.95\textwidth]{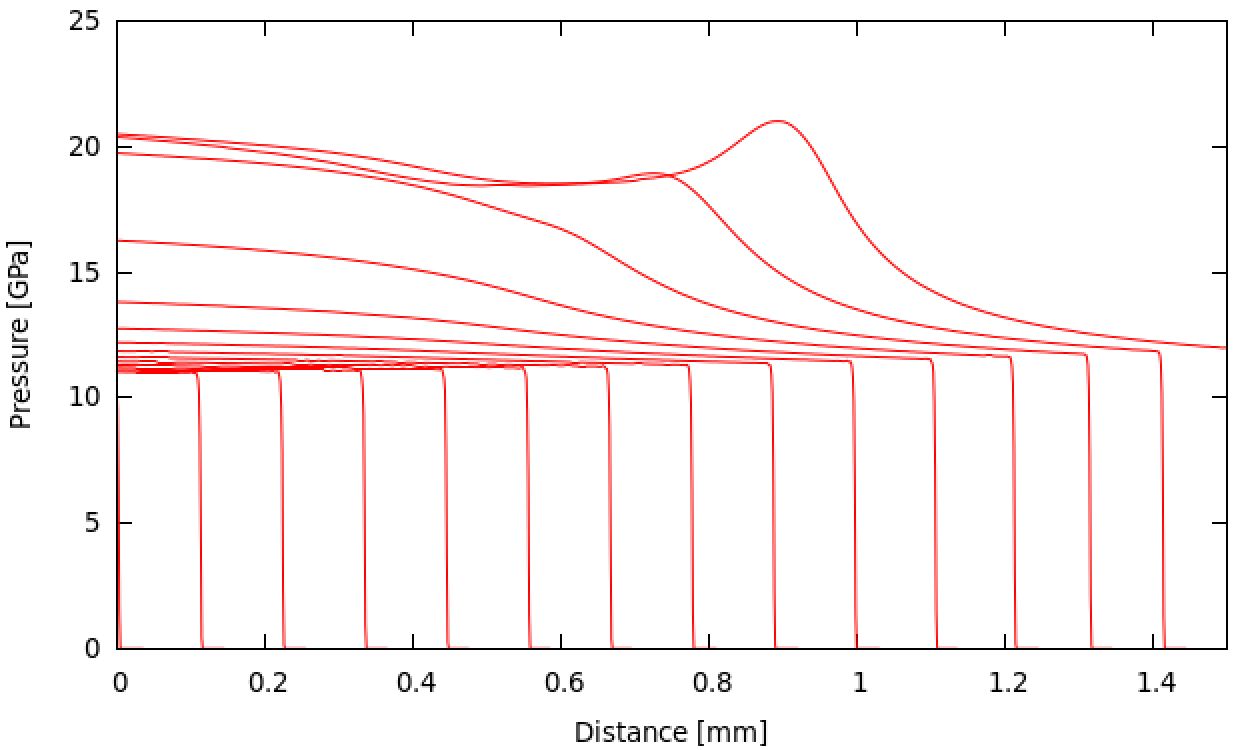}
 \caption{\label{c}}
\end{subfigure}
        \begin{subfigure}[b]{0.45\textwidth}
\includegraphics[width=0.95\textwidth]{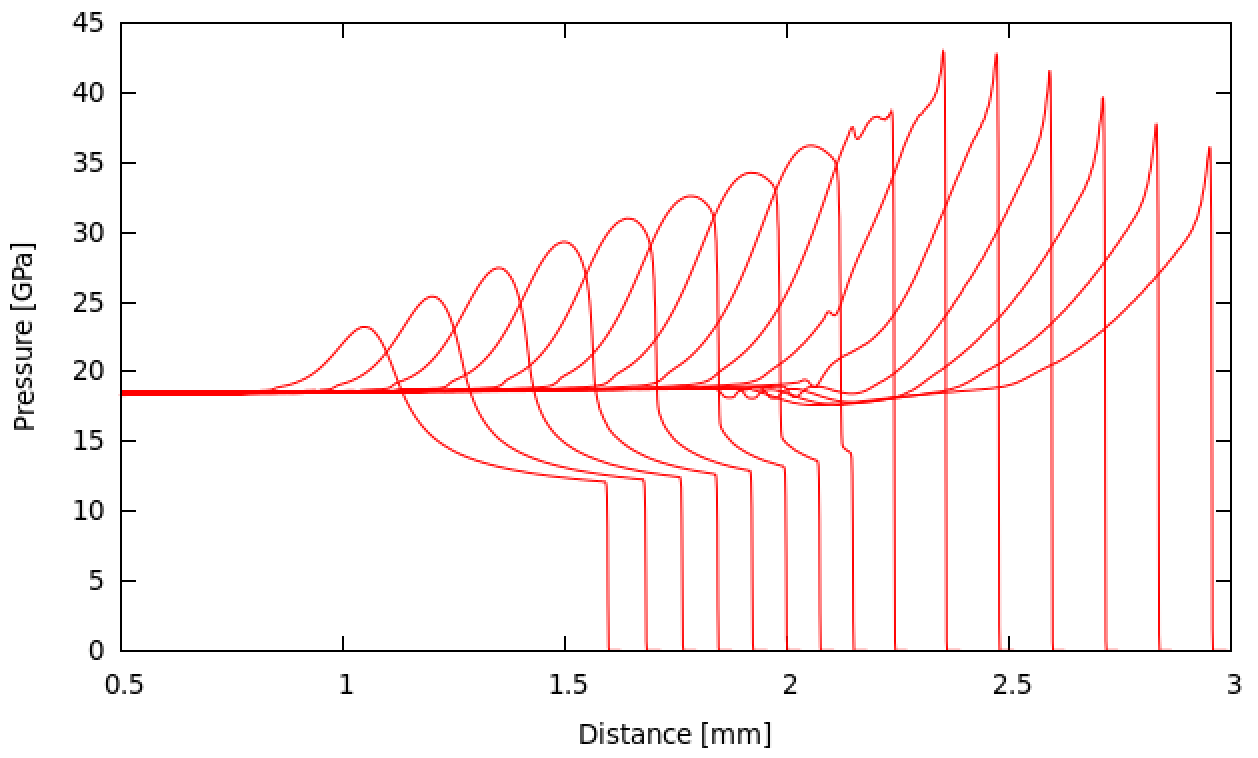}
 \caption{\label{d}}
        \end{subfigure}
\end{minipage}
\begin{minipage}{2\columnwidth}
\centering        
\begin{subfigure}[b]{0.45\textwidth}
\includegraphics[width=0.95\textwidth]{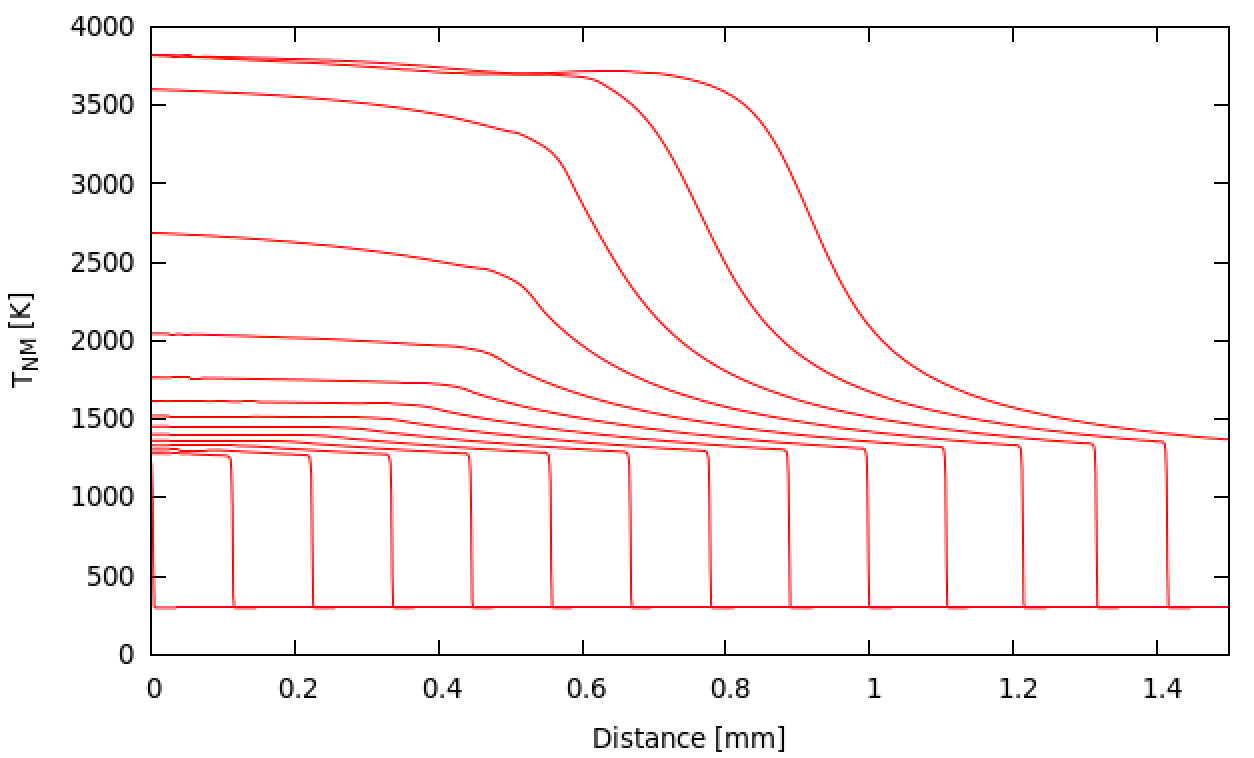}
 \caption{\label{e}}
\end{subfigure}
        \begin{subfigure}[b]{0.45\textwidth}
\includegraphics[width=0.95\textwidth]{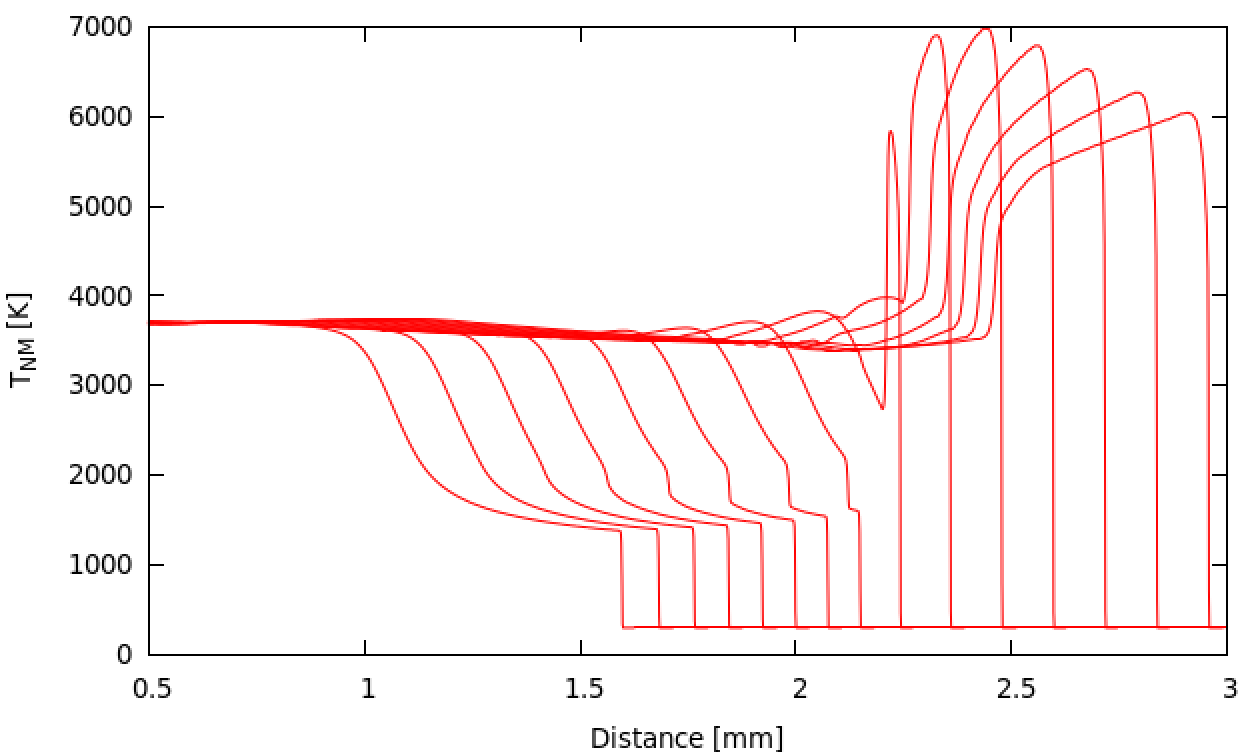}
 \caption{\label{f}}
        \end{subfigure}
\end{minipage}
\caption{One-dimensional evolution of the ignition and transition to detonation
of neat nitromethane under the influence of a\SI{10.98}{\giga \pascal} shock-wave
at selected times. On the left, stages $0-14$ are illustrated and on the
right stages $15-29$. The top row illustrates the reaction progress variable
($\lambda$), the middle row
 the pressure ($p$) and the bottom row the nitromethane
temperature
($T_{NM}$).  }
\label{neat1d}
\end{figure*}

The evolution of the reaction progress variable ($\lambda$),
pressure ($p$) and nitromethane temperature ($T_{NM}$) is illustrated in Fig.\ \ref{neat1d}.
As can be seen in the early stages of Figs.\  \ref{neat1d}(a), (c) and (e), the
fuel that is closer to the incident shock wave is shocked and heated first.
Hence, it has more time to react than the fuel that is further away from
the shock. As a result,   temperature and mass-fraction gradients are generated.
During this process, explosive material is burnt and the reaction progress
variable starts to deviate away from 1. By defining as the ignition
time the time when $\lambda =0.9$, we observe that here ignition occurs at stage 7 of {\color{black}Fig.\  \ref{neat1d}(a) [$\lambda$-plot]}, corresponding
to 
$t_{ign}=\SI{0.16}{\micro \second}$.  At the end of this slowly-evolving
induction phase, the fluid cannot sustain the high pressure and temperature
behind the shock wave, resulting to the generation of  a signal, as seen
during stages $13$--$14$ of {\color{black}Fig.\  \ref{neat1d}(c) [pressure plot]}. At this time ($t\sim\SI{0.32}{\micro
\second}$), thermal runaway is considered to occur.  A rapid
reaction stage follows the ignition stage, usually called the transition
to detonation phase,
 during which the generated pulse is growing (Fig.\  \ref{neat1d}(d)). At
stage 23, the reaction wave overtakes the leading shock wave depicted in
Fig.\  \ref{neat1d}(d). The overtake is accompanied by a rapid increase of
pressure, temperature and reaction (decrease of $\lambda$). Thereafter, the
detonation structure  settles down towards a steady-state solution.

\section {The collapse of a single cavity in reactive liquid nitromethane}

\label{Sec:singleReac2000}
In this section, we consider an isolated gas-filled cavity of radius \SI{0.08}{\milli \meter} collapsing in reacting liquid nitromethane in the domain spanning $[\SI{0}{\milli \meter},\SI{0.2}{\milli \meter}]\times[\SI{0.75}{\milli \meter},\SI{0.54}{\milli \meter}]$. The initial conditions in the shocked, pre-shocked and cavity regions are given in Table \ref{IC}. The domain is longer than in the inert simulations to allow for sufficient reaction to take place. It is also taller, to avoid the minor reflection of waves from the top boundary (even with transmissive boundaries) that could affect the ignition process. We adopt the same abbreviations as in Part I. 
\begin{table*}
\centering
\caption {Initial conditions for the initiation of nitromethane by cavity collapse.}
\label{IC}
\begin{tabular}{lccccccc}
\hline\noalign{\smallskip}
Region & $\rho_1 [\SI{}{\kilogram \per \meter \tothe{3}}]$ & $\rho_{2} [\SI{}{\kilogram \per \meter \tothe{3}}]$ & $u [\SI{}{\meter \per \second}]$ & $v [\SI{}{\meter \per \second}]$ & $p [\SI{}{\pascal}]$  & $\lambda$ &$z$  \\
\noalign{\smallskip}\hline\noalign{\smallskip}
 shocked nitromethane  & 2.388 & 1934 & 2000 & 0 & $10.98\times10^9$ & 0 &$10^{-6}$ \\
ambient nitromethane  & 1.2 & 1134 & 0 & 0 &  $10^{5}$ & 1 & $10^{-6}$ \\
air cavity  & 1.2 & 1134 & 0 & 0 &  $10^{5}$ & 1 & $1-10^{-6}$ \\
\noalign{\smallskip}\hline
\end{tabular}
\end{table*}

The evolution of the reaction progress variable $\lambda$ (left),  pressure, $p$ (middle) and  nitromethane temperature, $T_{NM}$ (right)
 is illustrated at selected times in Fig.\ \ref{single2000a}. 
{\color{black}
\begin{figure*}[!t]
\centering
\includegraphics[width=\textwidth]{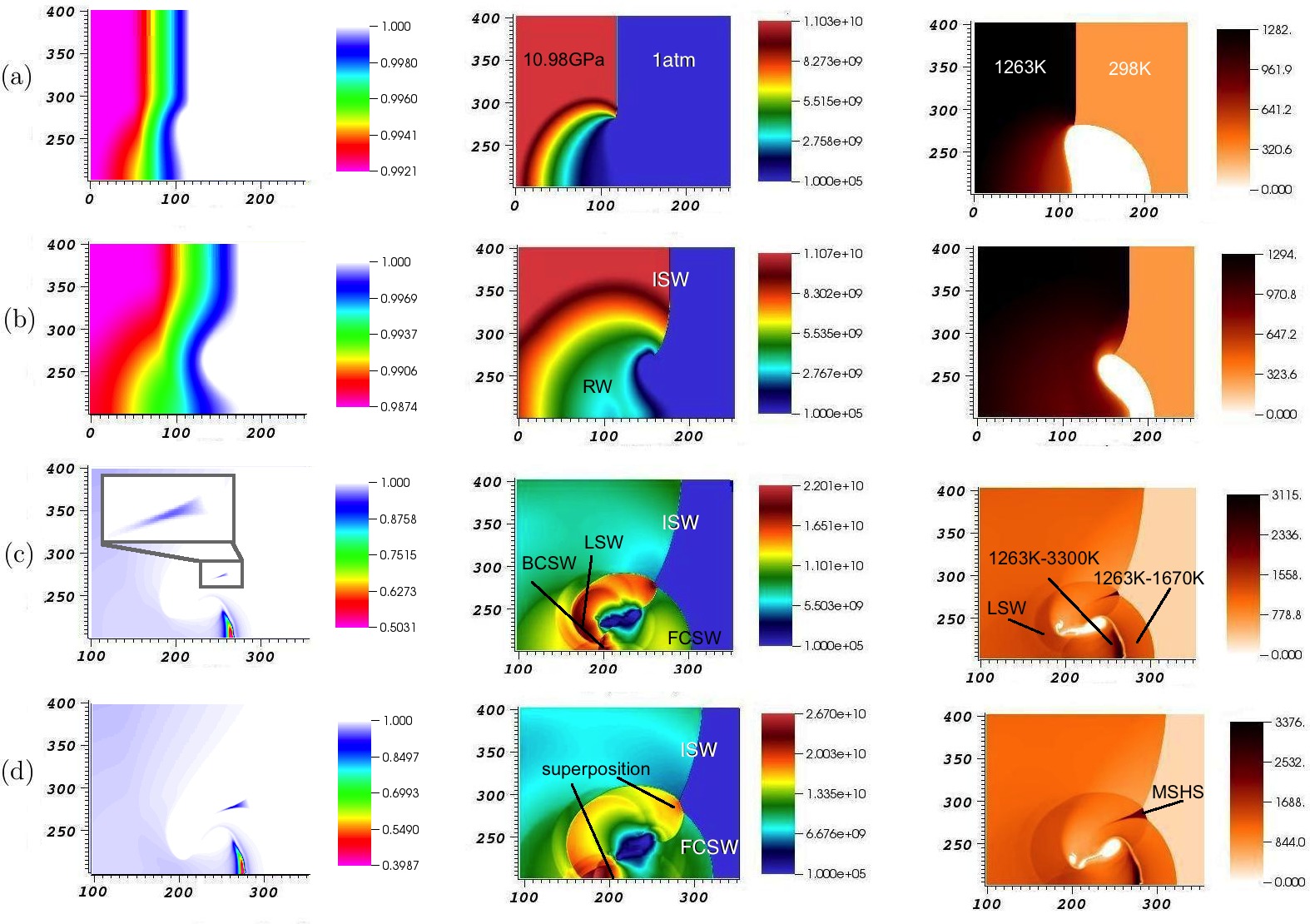}
\caption{The ignition process in the vicinity of a cavity embedded in liquid nitromethane following the passage 
of a $\SI{10.98}{\giga \pascal}$ shock wave. The evolution of mass fraction $\lambda$ (left), 
pressure $p$ (middle) and nitromethane temperature $T_{NM}$(right) at times $t=$
(a) \SI{0.0229}{\micro \second}, 
(b) \SI{0.0351}{\micro
\second}, (c) \SI{0.0591}{\micro
\second} and (d) \SI{0.0629}{\micro
\second} is presented. 
The horizontal and vertical axes are space ordinates in \SI{}{\micro
\meter}.}
\label{single2000a}
\end{figure*}

The incident shock wave (ISW)  travels within the nitromethane, compressing the material to $10.98$\SI{}{\giga \pascal} and 
raising the temperature to $1263$\SI{}{\kelvin}, as illustrated in Fig.\ \ref{single2000a}(a). In Part I of this work we followed the generated wave pattern in detail and studied the effect of each wave on the temperature field. The wave patterns in this scenario are the same as those in Part I and thus we will not repeat the details of their generation. We will only describe the additional effects observed due to the presence of reactions. 

At the initial stages of the collapse (see Fig.\  \ref{single2000a}(a,b)) and away from the cavity (where the rarefaction 
wave (RW) from the collapse  has not yet arrived), the reaction progress variable evolves as it would in a 1D neat nitromethane 
experiment. The expansion generated by the RW leads to a lowering of the temperature within the jet and 
this affects the shape of the reaction zone. As a result, before the cavity collapses, the highest 
temperature in the nitromethane is in the uniform, unperturbed region behind the ISW and only minimal reaction is observed. After the cavity collapse (at $t=\SI{0.040}{\micro \second}$) we observe for the first time in the collapse process, temperatures 
that are above the post-shock temperature. Specifically, the back collapse shock wave (BCSW) generates temperatures in the range  
$\SI{1263}{\kelvin}-\SI{3300}{\kelvin}$ (see Fig.\ \ref{single2000a}(c)) accompanied by a small amount of reaction 
of  $\lambda\approx0.97$. The front collapse shock wave (FCSW) generates temperatures within the range $\SI{1263}{\kelvin}-\SI{1670}{\kelvin}$ 
(see Fig.\ \ref{single2000a}(c)), in a region of almost no reaction at all ($\lambda\approx0.99$). 
The higher temperatures and the faster reaction at the rear of the cavity occur by the re-compression of the 
material that had already been shocked and was therefore pre-heated. 

Ignition (in at least one reaction site) 
is observed at time  $t_{ign}=\SI{0.0451}{\micro \second}$ in the back hot spot (BHS). 
The maximum temperature in the front hot spot (FHS) reaches at this point
$\SI{2068}{\kelvin}$ and $\SI{3089}{\kelvin}$ at the BHS. 

For this set up, the ISW proved slower than the nitromethane jet. Hence, at the time of collapse, the ISW is still 
traversing around the lobes. As a result of passing over the upper part of the upper lobe (and equivalently the lower 
part of the bottom lobe) many transmission/reflection processes take place.
These processes result in a coalescence of waves (LSW) that are transmitted from the lobes into the pre-shocked (by the ISW) 
nitromethane residing around the lobes (Fig.\ \ref{single2000a}(d)). This leads to the generation of temperatures higher 
than the post-shock temperature in the regions around the lobes. However, this temperature rise is not sufficient to 
generate a new ignition site in these timescales.

The superposition of the ISW and the FCSW generates the Mach stem hot spot (MSHS) as described in Part I (Fig.\ \ref{single2000a}(d)), which in the reactive scenario encloses temperatures of the range 1263K-1493K. At this point, the highest 
overall temperature is still observed in the BHS ($\SI{2773}{\kelvin}$). As the Mach stem region grows, the temperature
 it encloses increases rapidly, reaching values of the order of $\SI{2700}{\kelvin}$.Ignition is observed in the MSHS at $t=\SI{0.0630}{\micro\second}$. 

To the rear of the cavity, the waves emanating from the top and bottom lobe are superposed. Parts of them are 
superposed with the BCSW as well (see Fig.\ \ref{single2000a}(d)). This results in regions that have been shocked 
twice or even three times and hence leads to temperatures of the order of $\SI{1600}{\kelvin}$.

In the final stages of the collapse process, the remains of the cavity are advected downstream and more burning is observed 
in the reaction sites, with $\lambda$ reaching a value of $0.0745$ in the BHS and a value of $0.265$ in the MSHS  by the end 
of the simulation.
}

\begin{figure}[!ht]
\centering
\includegraphics[width=0.45\textwidth]{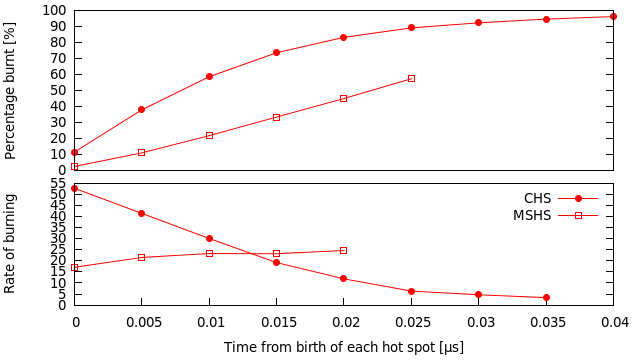}
\caption{Percentage of explosive burnt (top) and rate of burning (bottom) in the centreline hot spots and Mach stem hot spots from the time of their birth.}
\label{CHSvsMSHSignition}
\end{figure}

In Fig.\ \ref{CHSvsMSHSignition} (top) the percentage of burning given as the maximum value of $100\times(1-\lambda)$  in the centreline hot spots (BHS and FHS) and in the MSHS is shown over time. As time 0 we denote the time of birth of the hot spots which is the ignition time in each location identified earlier. It can be seen that the burning in the centreline hot spots (CHS) follows a $\sqrt{x}$ graph while the MSHS burning increases roughly linearly. In Fig.\ \ref{CHSvsMSHSignition} (bottom) the rate of burning in the two hot spots is presented. It can be seen that the reactions in the CHS grow faster initially than the reactions in the MSHS although this is reverted at later times and the reactions in the MSHS grow faster than the reaction in the CHS.

\subsection{Evolution along constant latitude lines}
\begin{figure*}[!th]
\centering
\begin{minipage}{2\columnwidth}
\centering
        \begin{subfigure}[b]{0.329\textwidth}
\includegraphics[width=\textwidth]{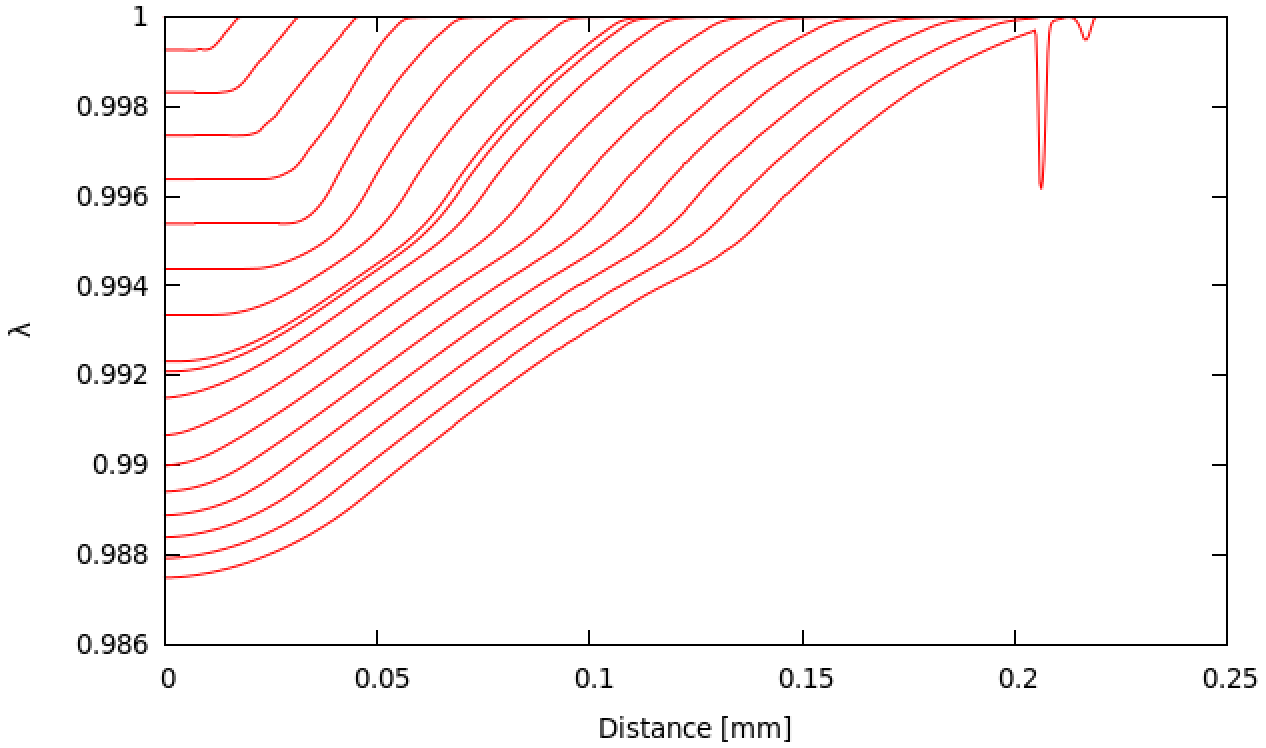}
\end{subfigure}
        \begin{subfigure}[b]{0.329\textwidth}
\includegraphics[width=\textwidth]{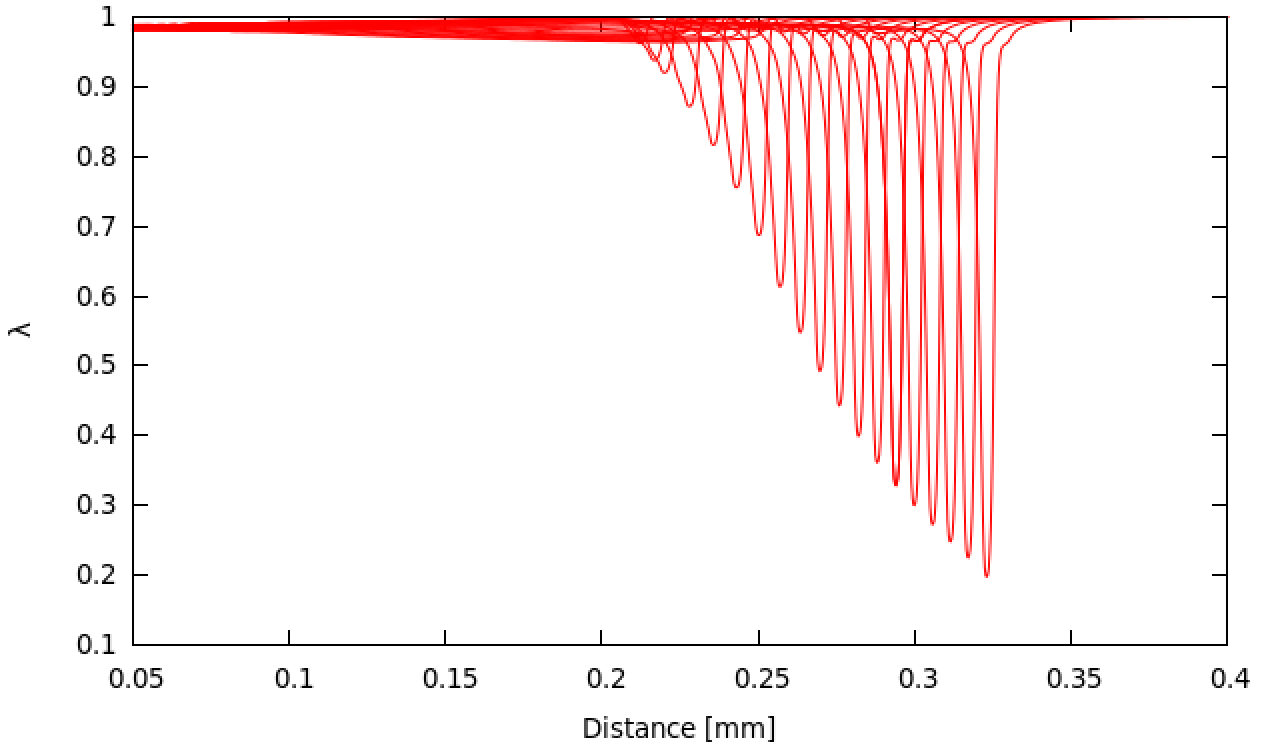}
        \end{subfigure}
\begin{subfigure}[b]{0.329\textwidth}
\includegraphics[width=\textwidth]{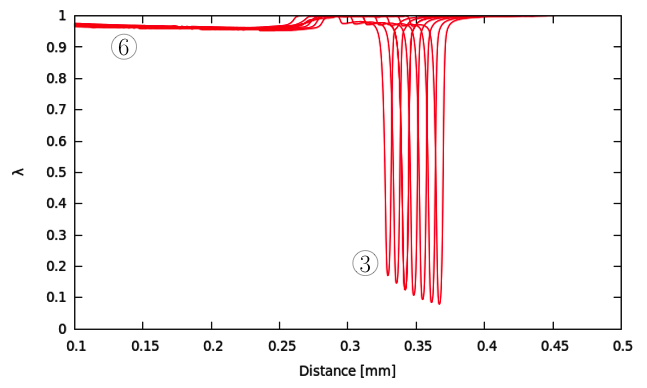}
\end{subfigure}
\end{minipage}
\begin{minipage}{2\columnwidth}
\centering       
 \begin{subfigure}[b]{0.329\textwidth}
\includegraphics[width=\textwidth]{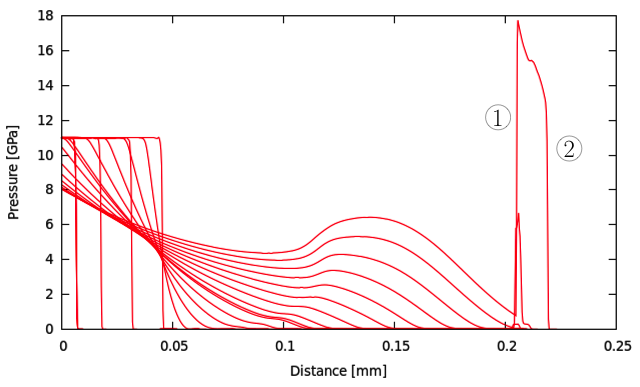}
\end{subfigure}
        \begin{subfigure}[b]{0.329\textwidth}
\includegraphics[width=\textwidth]{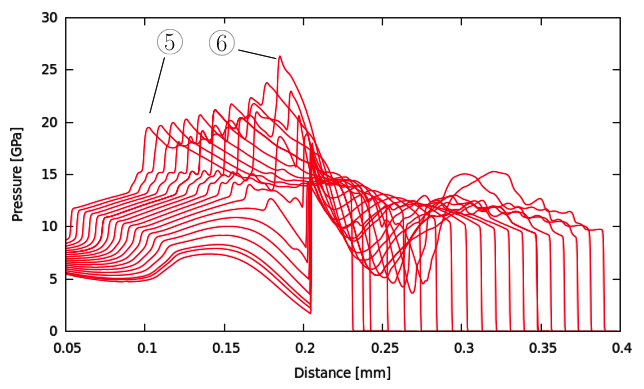}
        \end{subfigure}
\begin{subfigure}[b]{0.329\textwidth}
\includegraphics[width=\textwidth]{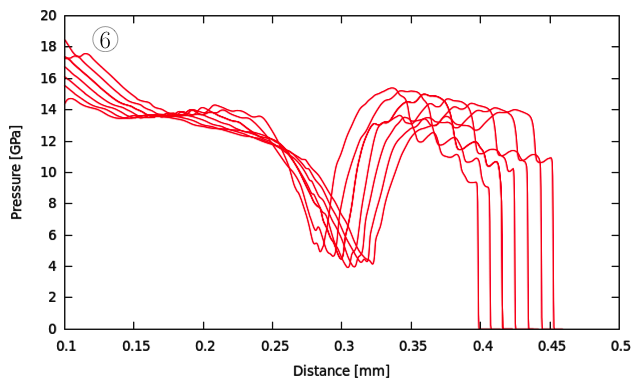}
\end{subfigure}
\end{minipage}
\begin{minipage}{2\columnwidth}
\centering        
\begin{subfigure}[b]{0.329\textwidth}
\includegraphics[width=\textwidth]{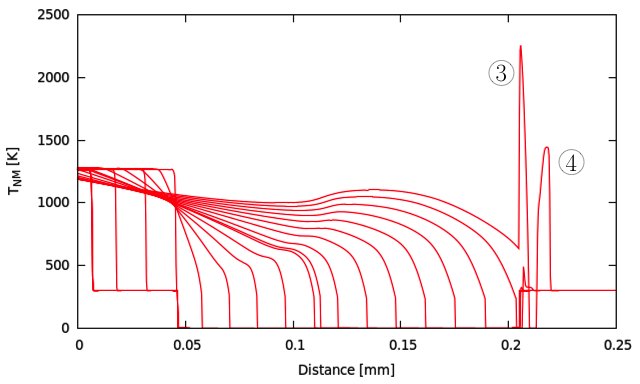}
\end{subfigure}
        \begin{subfigure}[b]{0.329\textwidth}
\includegraphics[width=\textwidth]{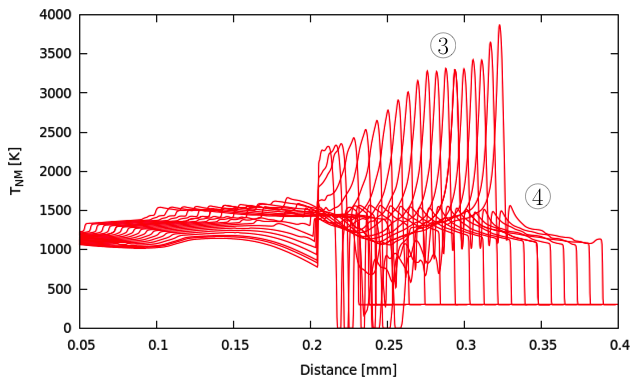}
        \end{subfigure}
\begin{subfigure}[b]{0.329\textwidth}
\includegraphics[width=\textwidth]{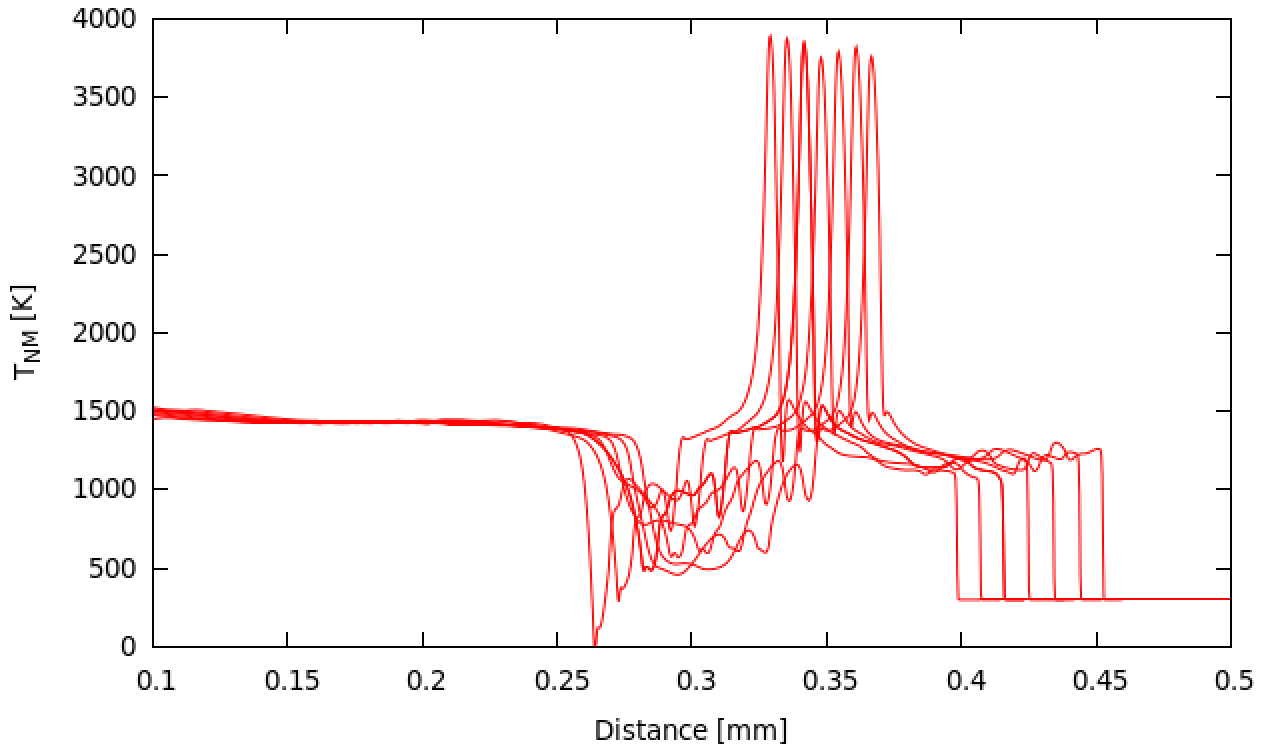}
\end{subfigure}
\end{minipage}
\caption{Lineouts along $y=\SI{0.21}{\milli \meter}$ from figure \ref{single2000a}.
The mass fraction (top row), the pressure (middle row) and the nitromethane
temperature (bottom row) are shown at stages (a) 1--18, (b) 19--37, (c)
38--44, corresponding to times (a) \SI{0.00228}{\micro\second}--\SI{0.0424}{\micro\second},
(b) \SI{0.0445}{\micro\second}--\SI{0.0780}{\micro\second} and (c)  \SI{0.0798}{\micro\second}--\SI{0.0915}{\micro\second}.}
\label{single20001d210}
\end{figure*}

It is informative to consider the evolution of the flow field and its effect on the temperature along lines of constant latitude. 
First, we discuss events along $y=\SI{0.21}{\milli\meter}$.  
This provides insight to the hot spot generation at the rear of the cavity (BHS), the minimal reaction at the front of the cavity (FHS) 
and the temperature distribution close to the centreline of the cavity. The evolution of the reaction rate variable $\lambda$, 
pressure and nitromethane  temperature along this line is shown in Fig.\  \ref{single20001d210}.

At the early stages, a slight increase in the temperature field by the passage of the ISW is seen. Upon the interaction of the ISW with the cavity, a downstream-travelling air shock is generated and an upstream travelling RW. The effect of the RW is a decrease in pressure and temperature. 
As it propagates, its effect becomes more profound, manifested as a further local decrease in the pressure and temperature fields 
and rate of reaction. The evolving crest-like feature (following the local decrease) seen in the pressure and temperature plots 
(Fig.\  \ref{single20001d210}(a)), is due to the  formation of the nitromethane jet. 

As the cavity collapses, the lineout crosses both shock waves (FCSW and BCSW), seen as two high-pressure and high-temperature 
fronts moving away from each other, labelled as $\circled{1},\circled{2}$ in Fig.\  \ref{single20001d210}(a). The BCSW compresses 
the material to a considerably higher pressure than the FCSW, generating a higher temperature at the rear of the cavity, compared to the 
front. Initially, the high-pressure fronts (i.e.\  BCSW and FCSW) coincide with the high-temperature fronts. However, as the collapse 
shock waves (CSWs) move away from the cavity, the pressure fronts ($\circled{1},\circled{2}$) propagate faster than the temperature fronts ($\circled{3},\circled{4}$).  The formation of the two 
CSWs leads to a considerable increase in the reaction rate, especially within  the BHS (as indicated by the $\lambda$-plot in 
Fig.\  \ref{single20001d210}(b)). The reaction in the FHS is observed to be comparatively slow. Within the BHS and along this lineout, 
 ignition is seen to occur at $t=\SI{0.0498}{\micro \second}$. 

Waves emanate from the lobes after the collapse of the cavity and are are labelled $\circled{5}$ as they intersect this line. By stage 26 they are superposed  with the BCSW 
along this lineout.  
Neither the downward waves nor their superposition with the BCSW result in a considerable increase in the temperature field. In fact, the highest temperature still occurs in the BHS. At stage 27, 
 the upward wave from the lower lobe (not modelled explicitly but with its behaviour implicitly taken into account by the reflective lower boundary) reaches the line $y=\SI{0.21}{\milli \meter}$. Part of this wave is superposed with both the BCSW and the wave entities emanating from the upper lobe 
and part of it is superposed with the waves emanating from  the upper lobe  
only.  
These superpositions increase the pressure locally  but do not have a significant effect on the temperature. As these shock waves (labelled $\circled{6}$) move away from the cavity, the temperature and pressure along the line $y=\SI{0.21}{\milli \meter}$ do not increase further, but the reaction in the BHS still increases. 

Focusing on the reaction progress variable field illustrated in the $\lambda-$plot of Fig.\  \ref{single20001d210}(c), it is observed that, at all late stages,
the largest amount of reaction  occurs in the BHS. Some reaction is also seen in the FHS and in the region that is traversed by the BCSW and
the waves emanating from the lobes. 
After the ignition at stages 21--22, the fuel continues burning until
it reaches the threshold of $\lambda=0.01$, where no more fuel is considered to be available
for burning. 

After stage 35, 
 the advection of the rear remaining parts of the cavity reach the lineout, leading to the low pressure and temperature parts of the lineouts and the $\lambda=1$ plateau seen in Fig.\ \ref{single20001d210}(c). 

\begin{figure*}
\centering
\begin{minipage}{2\columnwidth}
\centering
        \begin{subfigure}[b]{0.329\textwidth}
\includegraphics[width=\textwidth]{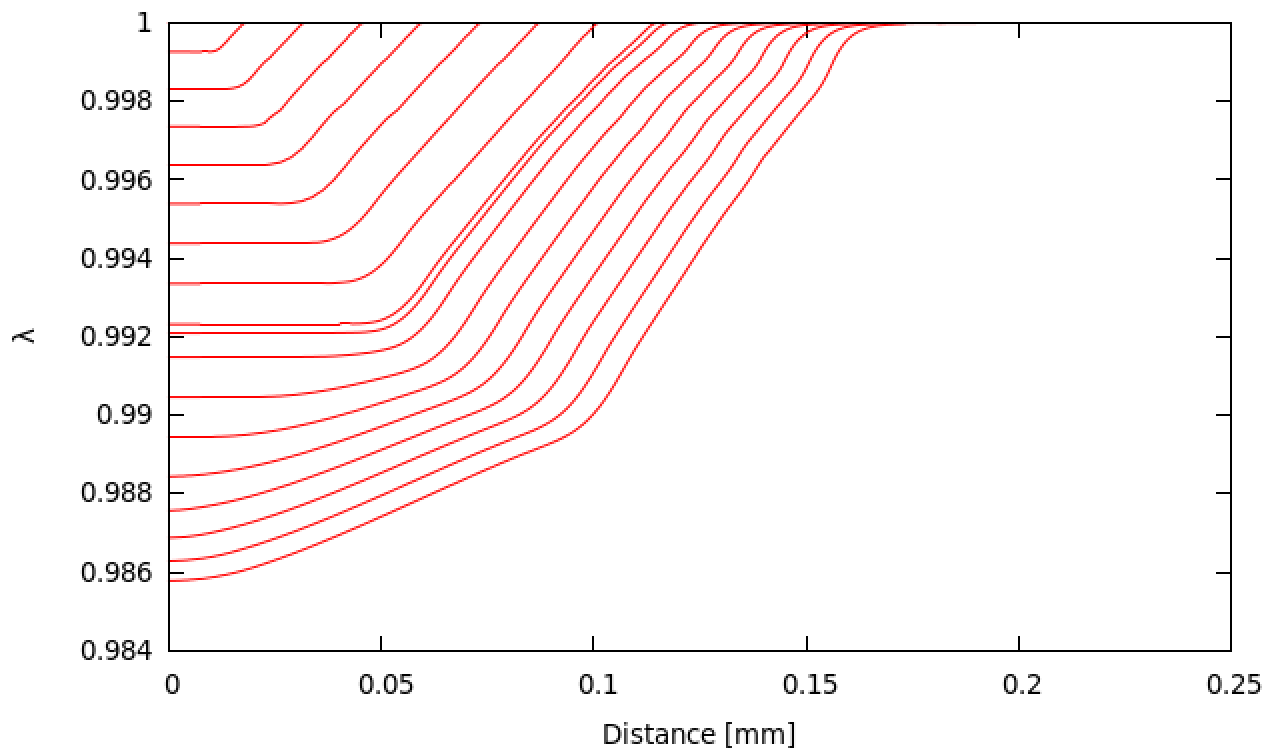}
        \end{subfigure}
        \begin{subfigure}[b]{0.329\textwidth}
\includegraphics[width=\textwidth]{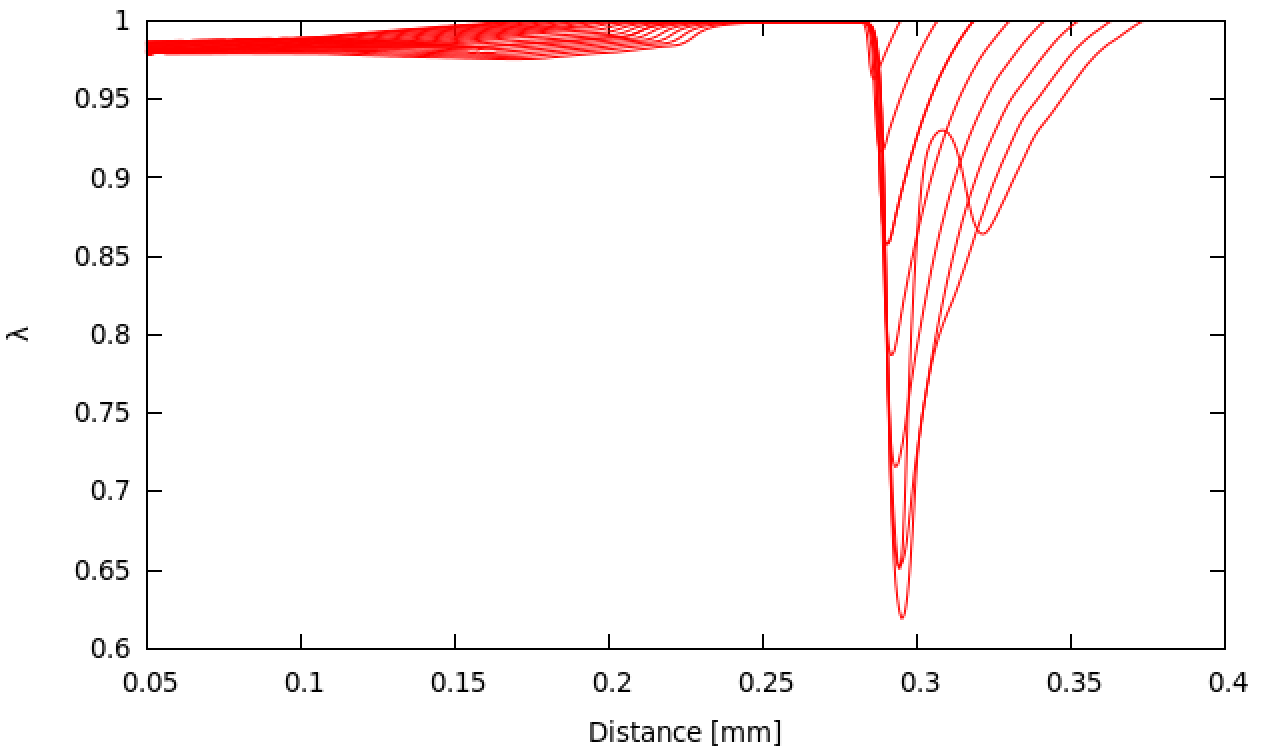}
        \end{subfigure}
\begin{subfigure}[b]{0.329\textwidth}
\includegraphics[width=\textwidth]{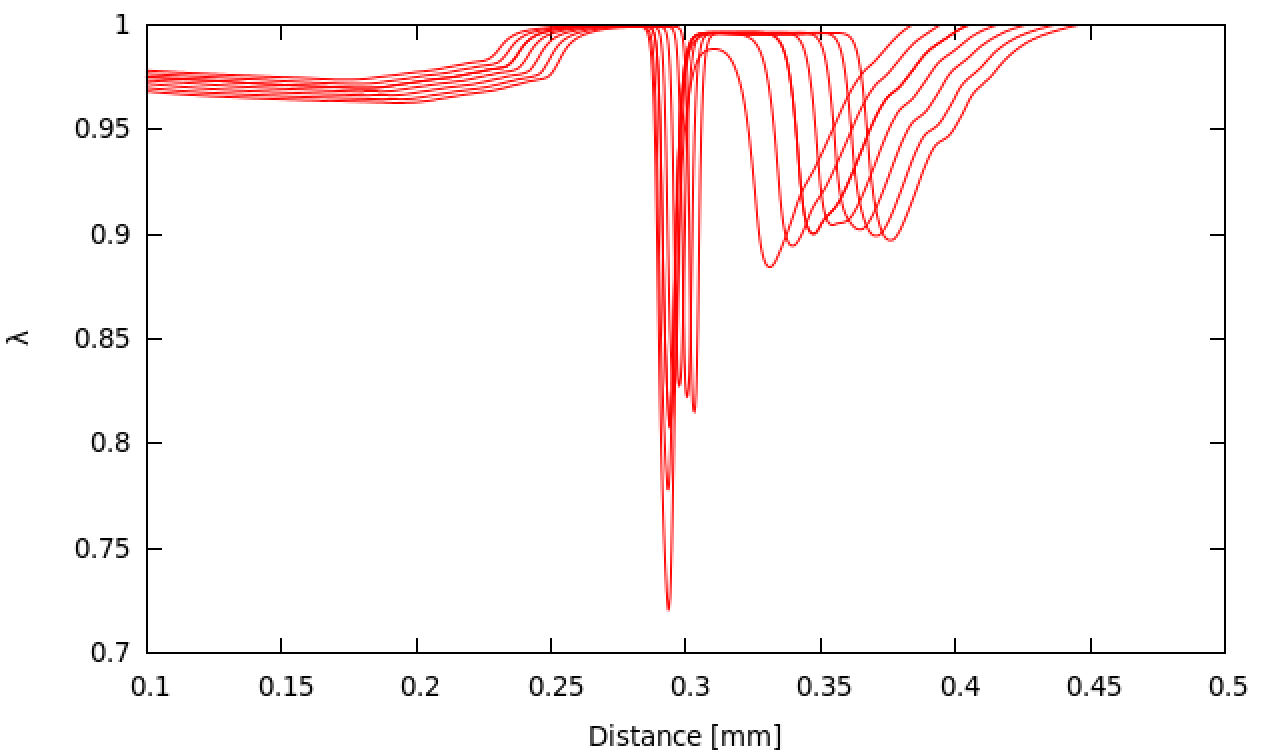}
        \end{subfigure}
\end{minipage}
\begin{minipage}{2\columnwidth}
\centering
        \begin{subfigure}[b]{0.329\textwidth}
\includegraphics[width=\textwidth]{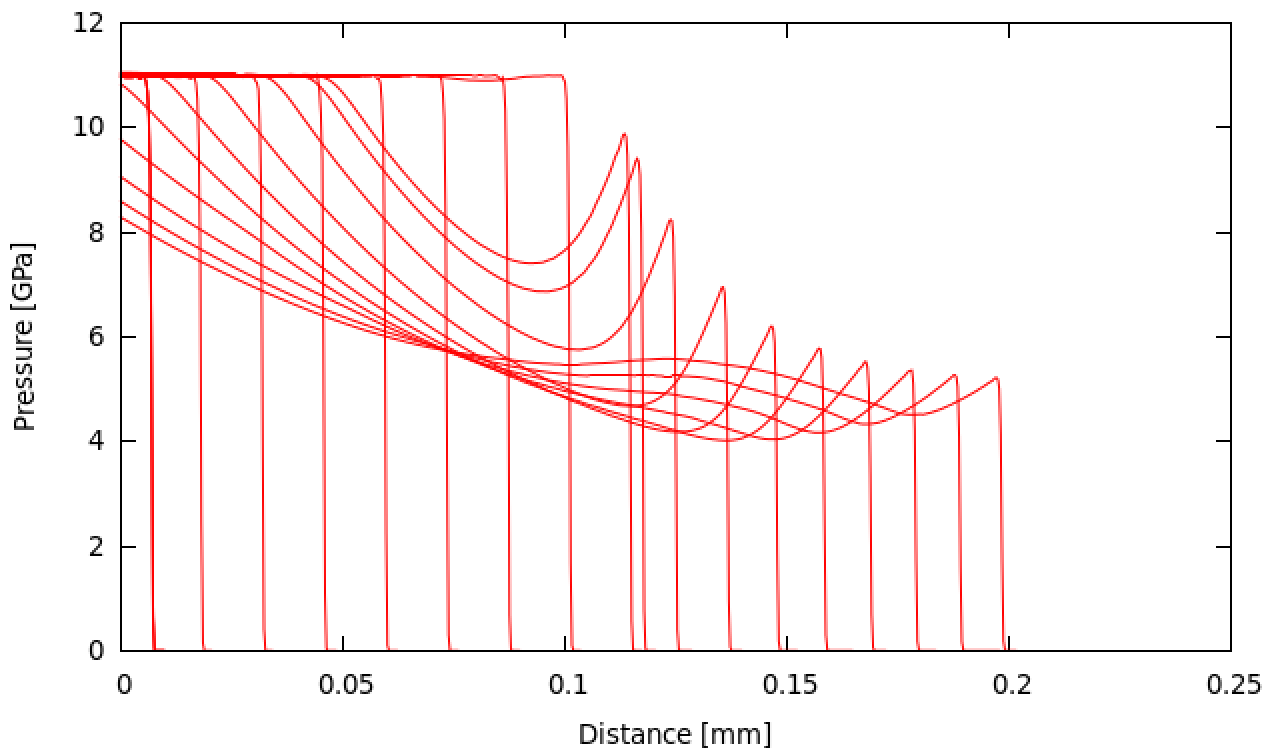}
        \end{subfigure}
        \begin{subfigure}[b]{0.329\textwidth}
\includegraphics[width=\textwidth]{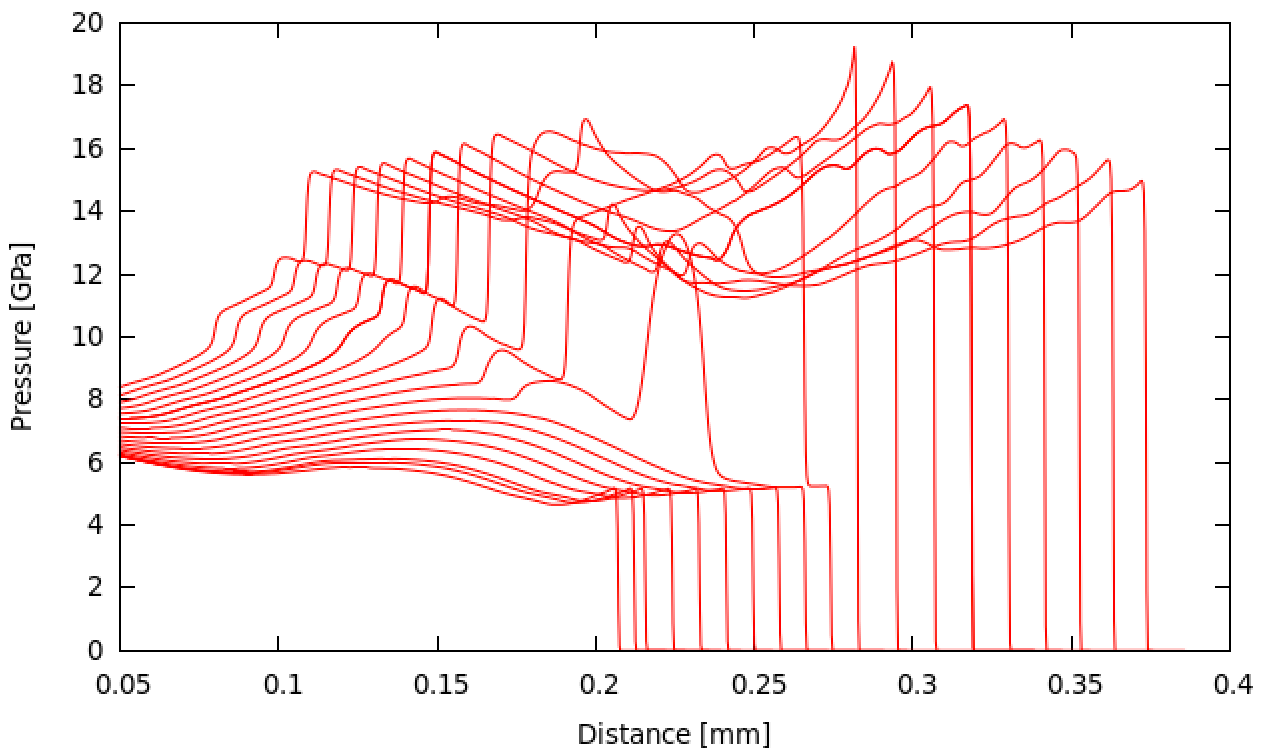}
        \end{subfigure}
\begin{subfigure}[b]{0.329\textwidth}
\includegraphics[width=\textwidth]{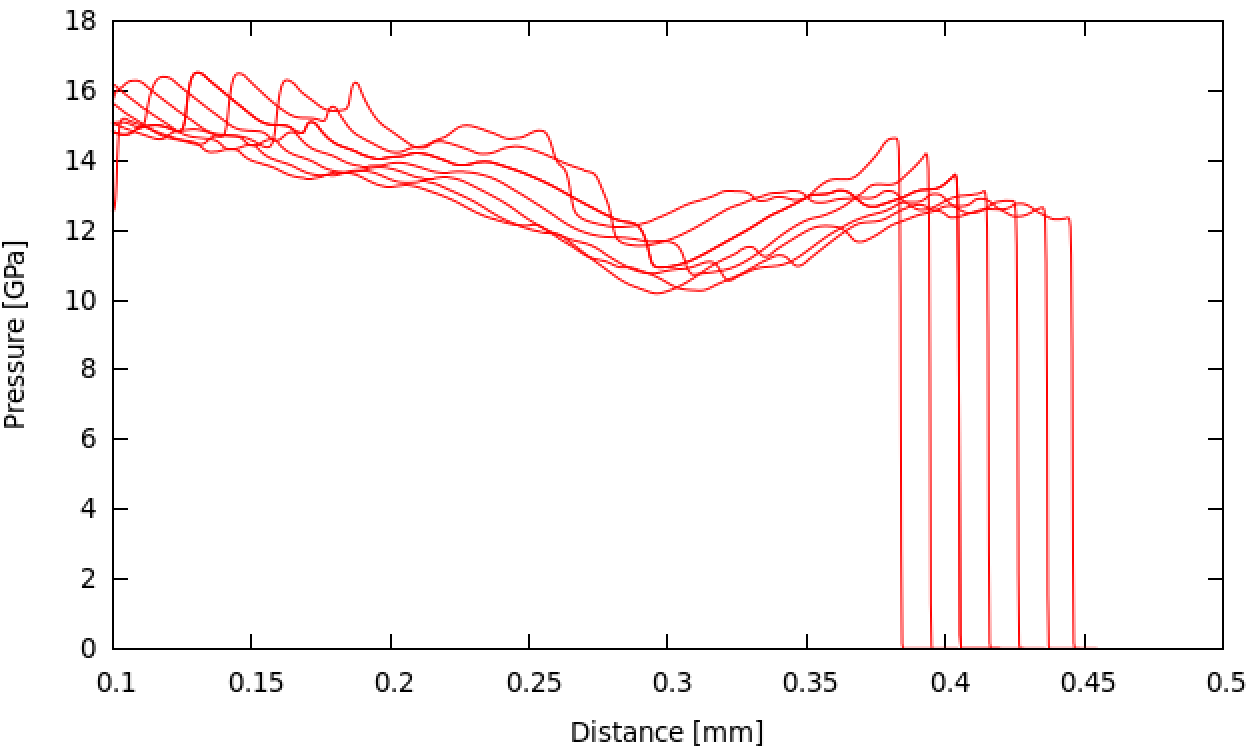}
\end{subfigure}
\end{minipage}
\begin{minipage}{2\columnwidth}
\centering
        \begin{subfigure}[b]{0.329\textwidth}
\includegraphics[width=\textwidth]{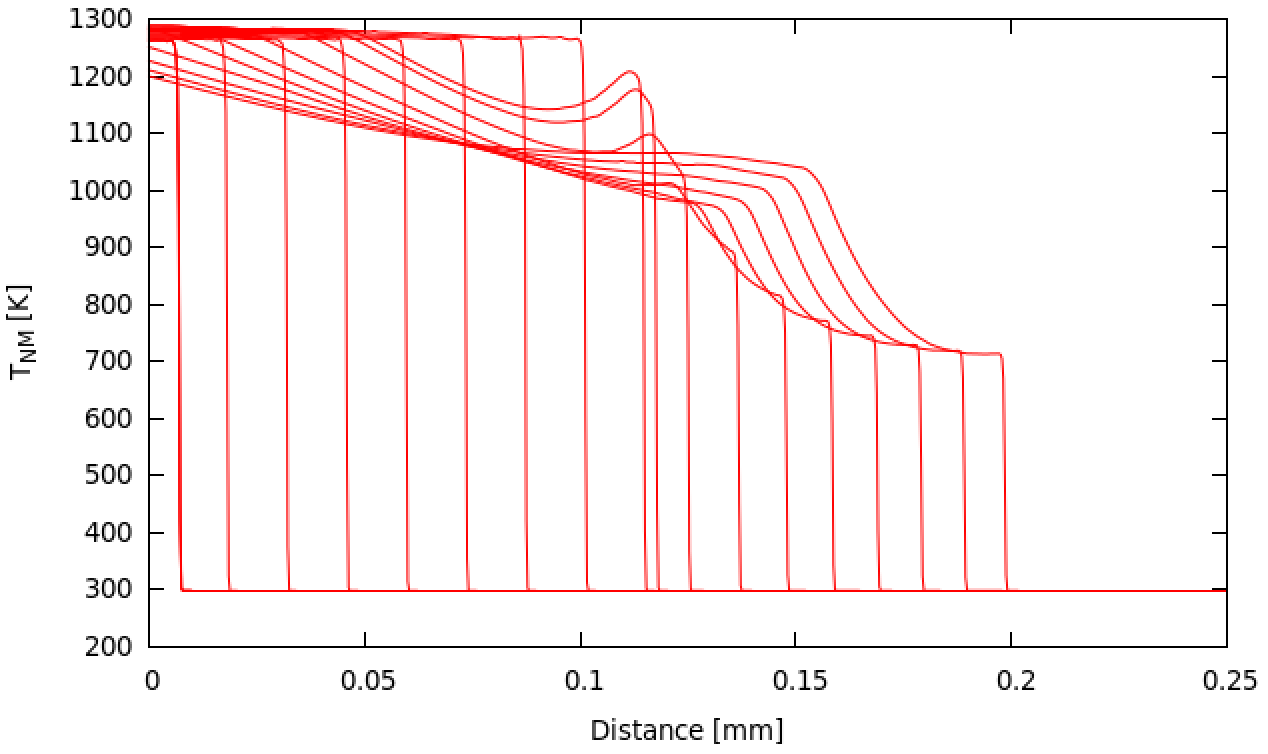}
        \end{subfigure}
        \begin{subfigure}[b]{0.329\textwidth}
\includegraphics[width=\textwidth]{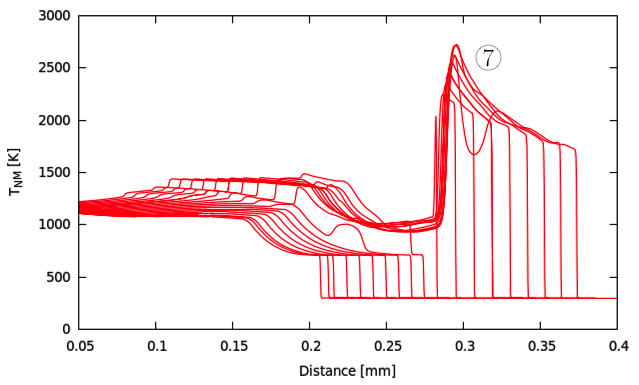}
        \end{subfigure}
\begin{subfigure}[b]{0.329\textwidth}
\includegraphics[width=\textwidth]{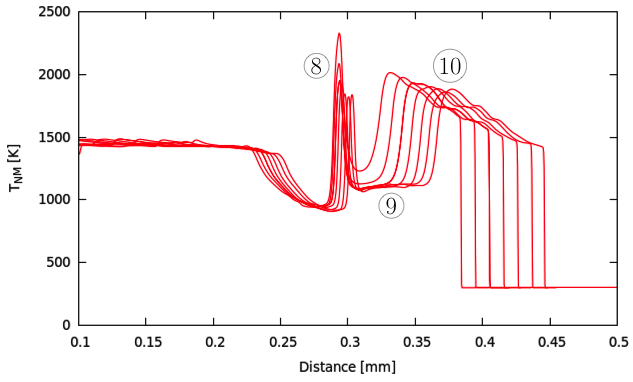}
\end{subfigure}
\end{minipage}
\caption{Lineouts along $y=\SI{0.29}{\milli \meter}$ from figure \ref{single2000a}.
The mass fraction (top row), the pressure (middle row) and the nitromethane
temperature (bottom row) are illustrated at stages (a) 1--18, (b) 19--37,
(c)
38--44, corresponding to times (a)  $0.00228$ \SI{}{\micro
\second}--\SI{0.0424}{\micro
\second},
(b) \SI{0.0445}{\micro
\second}--\SI{0.0780}{\micro
\second} and (c)  \SI{0.0798}{\micro
\second}--\SI{0.0915}{\micro
\second}.}
\label{single20001d290}
\end{figure*}

Lineouts at  $y=\SI{0.29}{\milli
\meter}$ are used to give insight about the MSHS and in general about the
temperature distribution above the cavity. The evolution of $\lambda$, pressure and nitromethane
temperature on this line is illustrated in Fig.\  \ref{single20001d290}.

 The effect of the rarefaction wave is seen in all three fields from
stages 8-9 
 in this case. 
At stage 26 
the effect of the $S_{12,14,16}$ wave entity as described in Part I of this work is seen in the plots
of Fig.\ \ref{single20001d290}(b).
The FCSW overtakes the ISW at stage 28 
This results in  the increase of the temperature
and a subsequent increase of the reaction. As the line  $y=\SI{0.29}{\milli
\meter}$ goes though a large part of the MSHS, the elevated temperature is
not found in an isolated peak, but is distributed within a region behind
the moving Mach stem front (Fig.\ \ref{single20001d290}(b), labelled as $\circled{7}$).
This elevated temperature region results in the increase of reaction, as
illustrated by the $\lambda$ plot in Fig.\  \ref{single20001d290}(b). From
stages 44 
 onward, the temperature is first seen to increase in a hill-like form (labelled
as $\circled{8}$), then decreases taking a well-like form ($\circled{9}$)
and then increases again taking the form of a hill, but with a shallower
gradient than before ($\circled{10}$). This is because at these late stages,
the Mach stem feature has attained a horn-like shape and
it is advected upwards. As a result, the line  $y=\SI{0.29}{\milli
\meter}$ intersects the stem, then goes through a region below the stem and
then intersects the stem again. At the location of the second intersection,
the temperature is elevated compared to the regions outside the stem, but
it is lower than at the location of the first intersection. This translates
as a well-hill-well feature in the $\lambda-$plot, indicating more reaction
along the part of the line  $y=\SI{0.29}{\milli
\meter}$ that intersects the stem the first time, far less reaction outside
the stem and a moderate amount of reaction at the second intersection.

\begin{figure*}[!t]
\centering
\begin{minipage}{2\columnwidth}
\centering        
\begin{subfigure}[b]{0.329\textwidth}
\includegraphics[width=\textwidth]{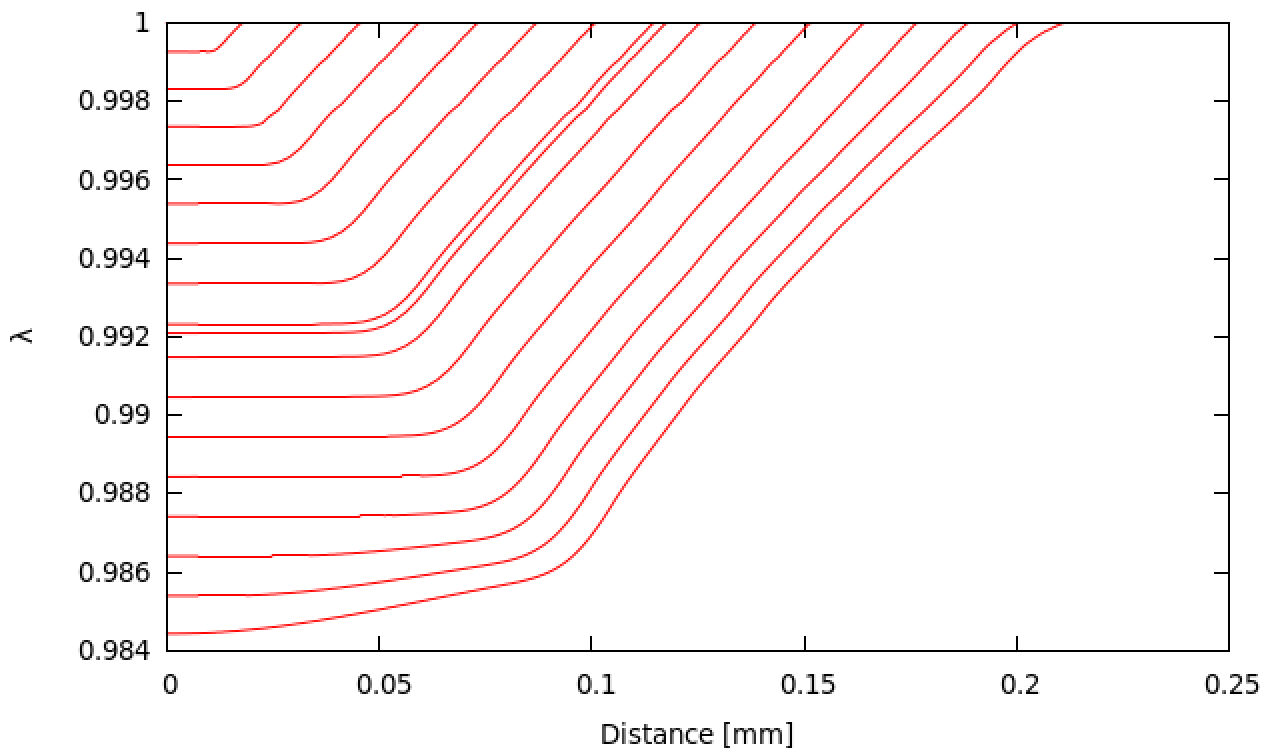}
        \end{subfigure}
        \begin{subfigure}[b]{0.329\textwidth}
\includegraphics[width=\textwidth]{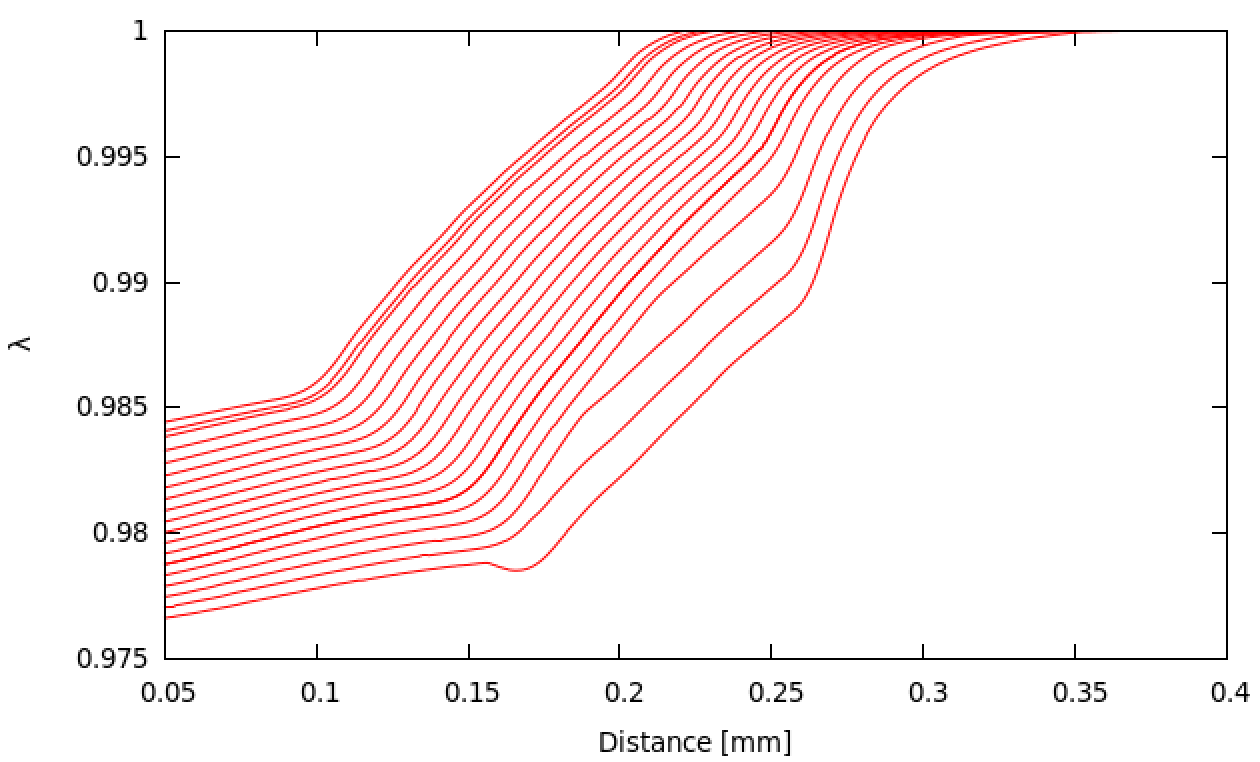}
        \end{subfigure}
\begin{subfigure}[b]{0.329\textwidth}
\includegraphics[width=\textwidth]{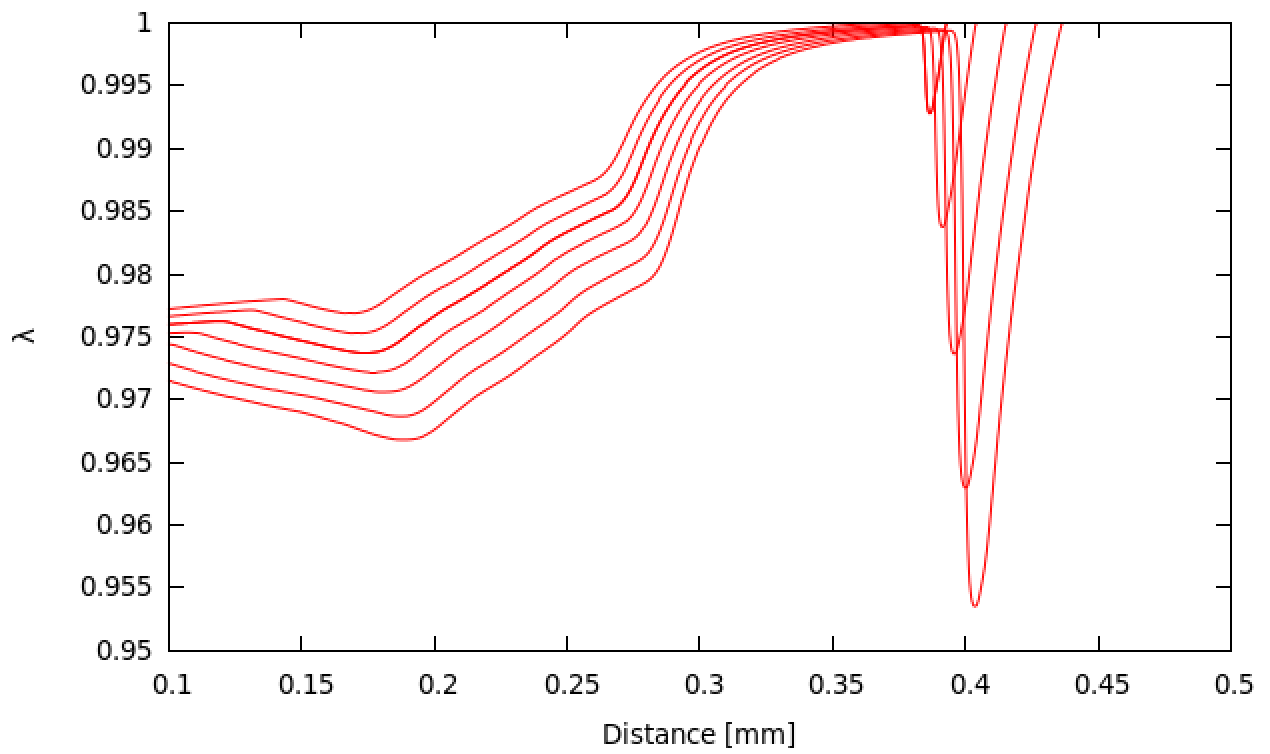}
        \end{subfigure}
\end{minipage}
\begin{minipage}{2\columnwidth}
\centering        
\begin{subfigure}[b]{0.329\textwidth}
\includegraphics[width=\textwidth]{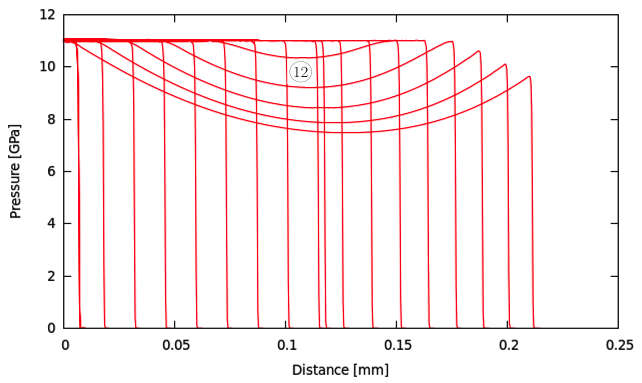}
        \end{subfigure}
        \begin{subfigure}[b]{0.329\textwidth}
\includegraphics[width=\textwidth]{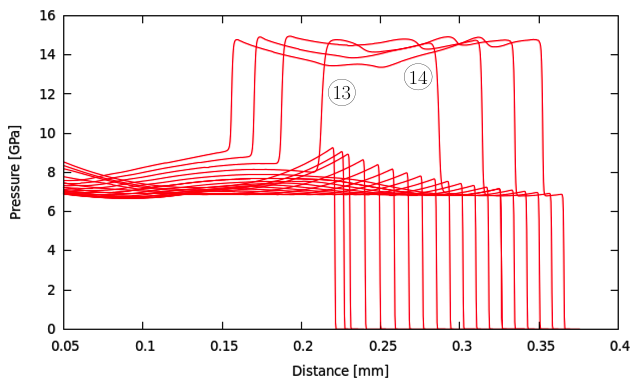}
        \end{subfigure}
\begin{subfigure}[b]{0.329\textwidth}
\includegraphics[width=\textwidth]{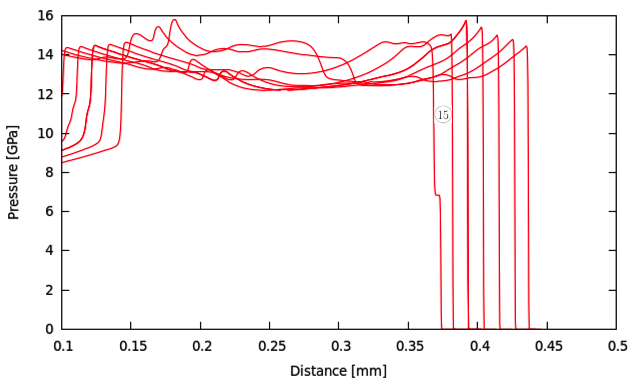}
        \end{subfigure}
\end{minipage}
\begin{minipage}{2\columnwidth}
\centering        
\begin{subfigure}[b]{0.329\textwidth}
\includegraphics[width=\textwidth]{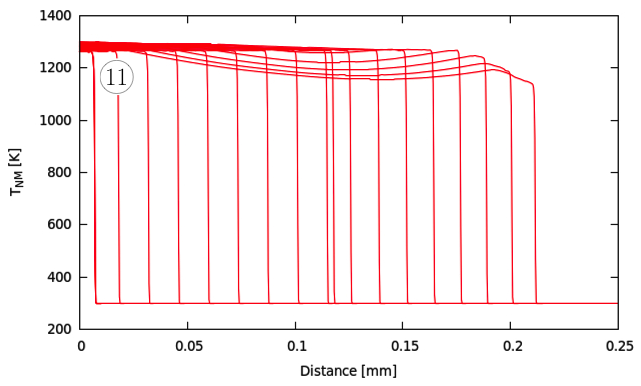}
        \end{subfigure}
        \begin{subfigure}[b]{0.329\textwidth}
\includegraphics[width=\textwidth]{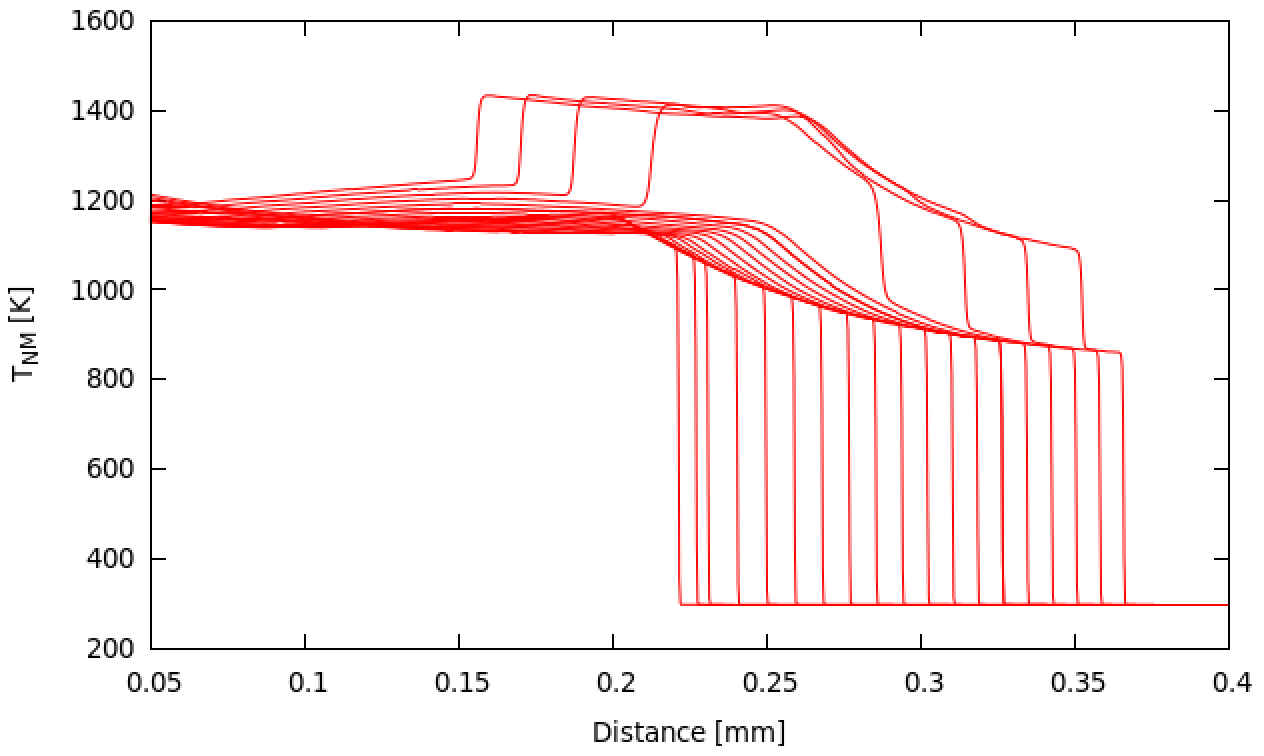}
        \end{subfigure}
\begin{subfigure}[b]{0.329\textwidth}
\includegraphics[width=\textwidth]{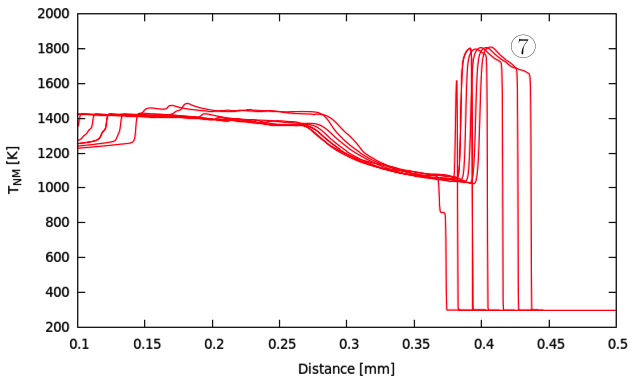}
        \end{subfigure}
\end{minipage}
\caption{Lineouts along $y=\SI{0.35}{\milli \meter}$ from figure \ref{single2000a}. The mass fraction (top row), the pressure
(middle row) and the nitromethane temperature (bottom row) are illustrated
at stages (a) 1--18, (b) 19--37, (c) 38--44, corresponding to times (a) 
\SI{0.00228}{\micro
\second}--\SI{0.0424}{\micro
\second}, (b)  \SI{0.0445}{\micro
\second}--\SI{0.0780}{\micro
\second} and (c)  \SI{0.0798}{\micro
\second}--\SI{0.0915}{\micro
\second}.  }
\label{single20001d350}
\end{figure*}

Another horizontal lineout, at $y=\SI{0.35}{\milli \meter}$, is used to give insight about the
generation of the MSHS and temperature distribution and initiation around the cavity. The evolution of the $\lambda$-field, the pressure and the nitromethane temperature
along the line are given in Fig.\  \ref{single20001d350}. On this line, the effect of all the waves emanating from the cavity collapse process are seen later than on the previous lineouts considered.

In the early stages 
 of Fig.\ \ref{single20001d350}(a), a slight increase in the pressure and
a more noticeable increase in $T_{NM}$ due to the passage of the shock wave
is seen (labelled as $\circled{11}$ in the temperature plot). This is accompanied
by the start of the reaction behind the shock wave, seen in the $\lambda-$plot.
The slight increase of the post-shock pressure, the formation of a temperature
gradient and the shape of the $\lambda-$plot are as observed in the shock-induced
ignition of neat nitromethane, because no effects from the collapse of the
cavity have reached the line  $y=\SI{0.35}{\milli
\meter}$ yet. 

By  stage 14, 
 the rarefaction wave (RW) has reached  $y=\SI{0.35}{\milli
\meter}$ (as opposed to stages 8-9 on $y=\SI{0.29}{\milli
\meter}$) and its effect is seen as a descent in the pressure and temperature
plot (labelled as $\circled{12}$ in the pressure plot).  The effect of the
RW is increased as the wave propagates upwards,  seen as growth of the dip
in pressure and temperature (Fig.\ \ref{single20001d350}(b)). In the $\lambda-$plot,
the shape of the lower part of a graph is affected, presenting a slight increase
of the gradient of the straight-line part of the graph. This indicates that
less reaction is taking place when the rarefaction wave is present compared
to the reaction that would have taken place in the absence of the rarefaction
wave. 

At stage 34, 
 the entity of waves $S_{12,14,16}$
is seen to have crossed the line  $y=\SI{0.35}{\milli
\meter}$ (as opposed to stage 26 for $y=\SI{0.29}{\milli
\meter}$). Upon reaching $y=\SI{0.35}{\milli
\meter}$, the wave increases the pressure and temperature along the part of
the line that intersects it. As this wave is circular (spherical in 3D) and
propagates outwards from the cavity, the area cut by the line  increases with
time. As a result, in the one-dimensional pressure plot of Fig.\  \ref{single20001d350}(b),
this entity of waves is seen to be composed of a front pressure wave that
propagates downstream (labelled as $\circled{13}$) and a back pressure wave,
that propagates upstream  (labelled as $\circled{14}$). The effect of the
wave entity $S_{12,14,16}$ on the one-dimensional temperature field is seen
in Fig.\  \ref{single20001d350}(b) as an increase in $T_{NM}$, in the form
of two temperature fronts. Its effect on the reaction progress variable is
seen in \ref{single20001d350}(b). It appears  as a dip in $\lambda$ attributed
to the  backward moving part of the wave,  as a sudden but not very distinct
drop in $\lambda$ attributed to the front part of the wave and as an overall
decrease of the gradient of the straight-line part of the graph. All these
features of the $\lambda-$plot indicate rapid increase of reaction as soon
as the new shock waves re-shock the area.

Stage 38 (Fig.\  \ref{single20001d350}(c)) corresponds to the time before
the FCSW overtakes the ISW on line  $y=\SI{0.35}{\milli
\meter}$ (as opposed to stage 28 for $y=\SI{0.29}{\milli
\meter}$) and stage 39 
to the time immediately after the overtake\footnote{Overtake in this context
refers to the overtake of the ISW by the CSW and not the overtake of the
ISW by a detonation wave.} (labelled as $\circled{15}$ in the pressure plot).
After the overtake, the lineout goes through the high-temperature Mach stem
region, leading to the generation of the temperature peaks in the temperature
plot (labelled as $\circled{7}$). This leads to an increase of the reaction,
as indicated by the sudden decrease of the reaction progress variable in
the $\lambda-$plot. The temperature of the hot spot in the Mach stem continues
to gradually increase. This  translates into the temperature peaks of Fig.\
 \ref{single20001d350}(c), resulting in the increase of the reaction and
the sudden drop in $\lambda$  in  Fig.\  \ref{single20001d350}(c).

\section{Comparing the three-dimensional and two-dimensional, cavity collapse in reacting nitromethane}

\begin{figure*}[!ht]
\centering
\begin{minipage}{2\columnwidth}
\centering
     \begin{subfigure}[b]{0.47\textwidth}
\includegraphics[width=\textwidth]{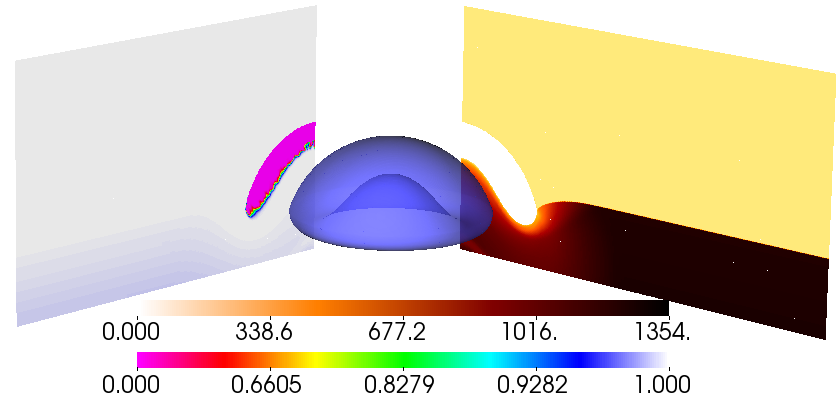}
\end{subfigure}                                                                                                                                                                                    \begin{subfigure}[b]{0.47\textwidth}
\includegraphics[width=\textwidth]{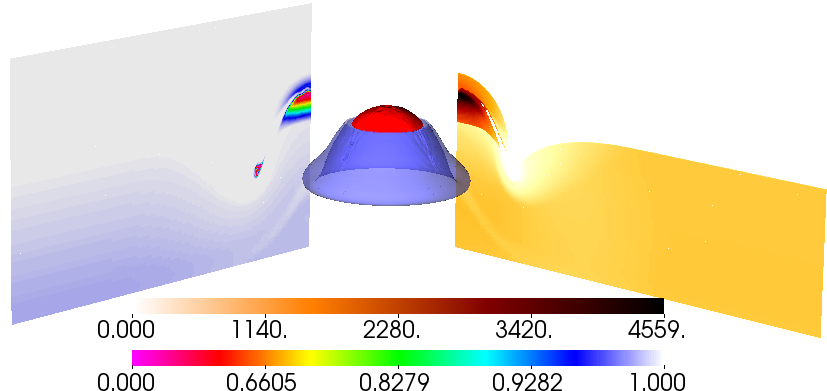}     
        \end{subfigure}
\end{minipage}
\begin{minipage}{2\columnwidth}
\centering   
\includegraphics[width=0.47\textwidth]{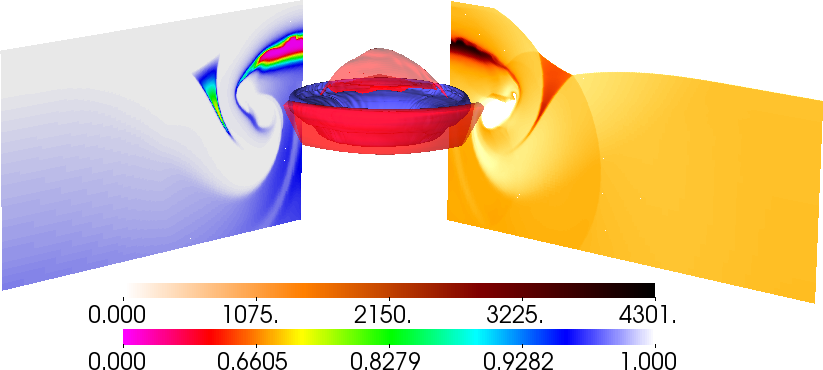}
\end{minipage}
\caption{Snapshots of the three-dimensional collapse of a cavity in reacting liquid nitromethane. The blue, three-dimensional $z=0.5$ contour represents the cavity boundary. The
red, $\lambda=0.9$ contour represents the generated hot-spot, i.e. the region where
reaction takes place. A two-dimensional slice  through
the centre of the cavity is taken for the $\lambda$ and nitromethane temperature fields. The slices are projected on the left half ($\lambda$ field) and right half (nitromethane temperature field) of each figure. Note that the collapse process is shown here to move from bottom to top rather than left to right. The images here can be considered as rotated by 90-degrees counter-clockwise compared to Fig. \ref{single2000a}. This is done solely for illustration purposes.}
\label{3Dpseudo}
\end{figure*}

\begin{figure*}[!ht]
\centering
\begin{minipage}[t]{2\columnwidth}
\centering
        \begin{subfigure}[t]{0.43\textwidth}
\includegraphics[width=\textwidth]{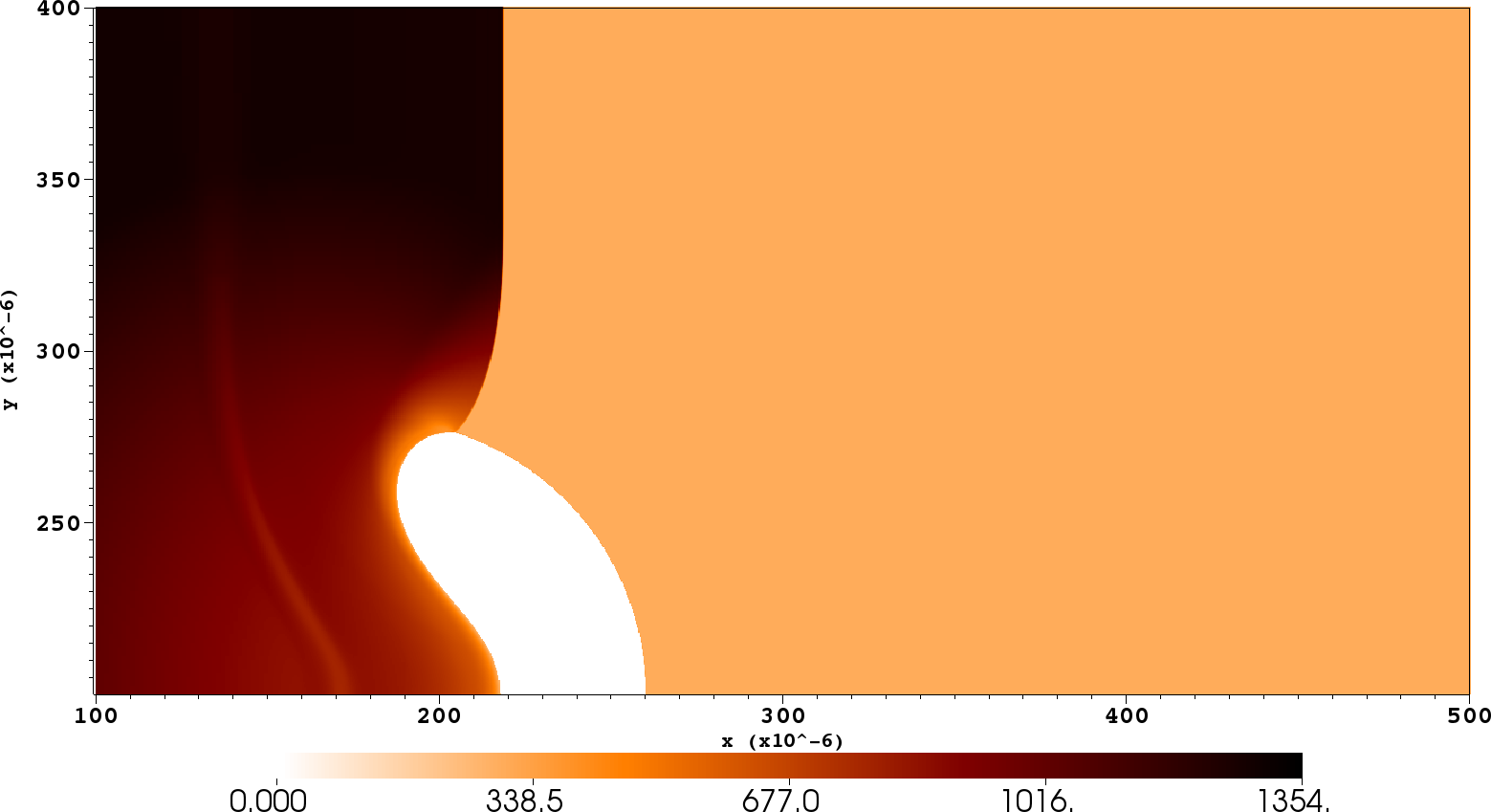}
\caption{2D $t=0.03\SI{}{\micro \second}$}        
\end{subfigure}
 \begin{subfigure}[t]{0.43\textwidth}
\includegraphics[width=\textwidth]{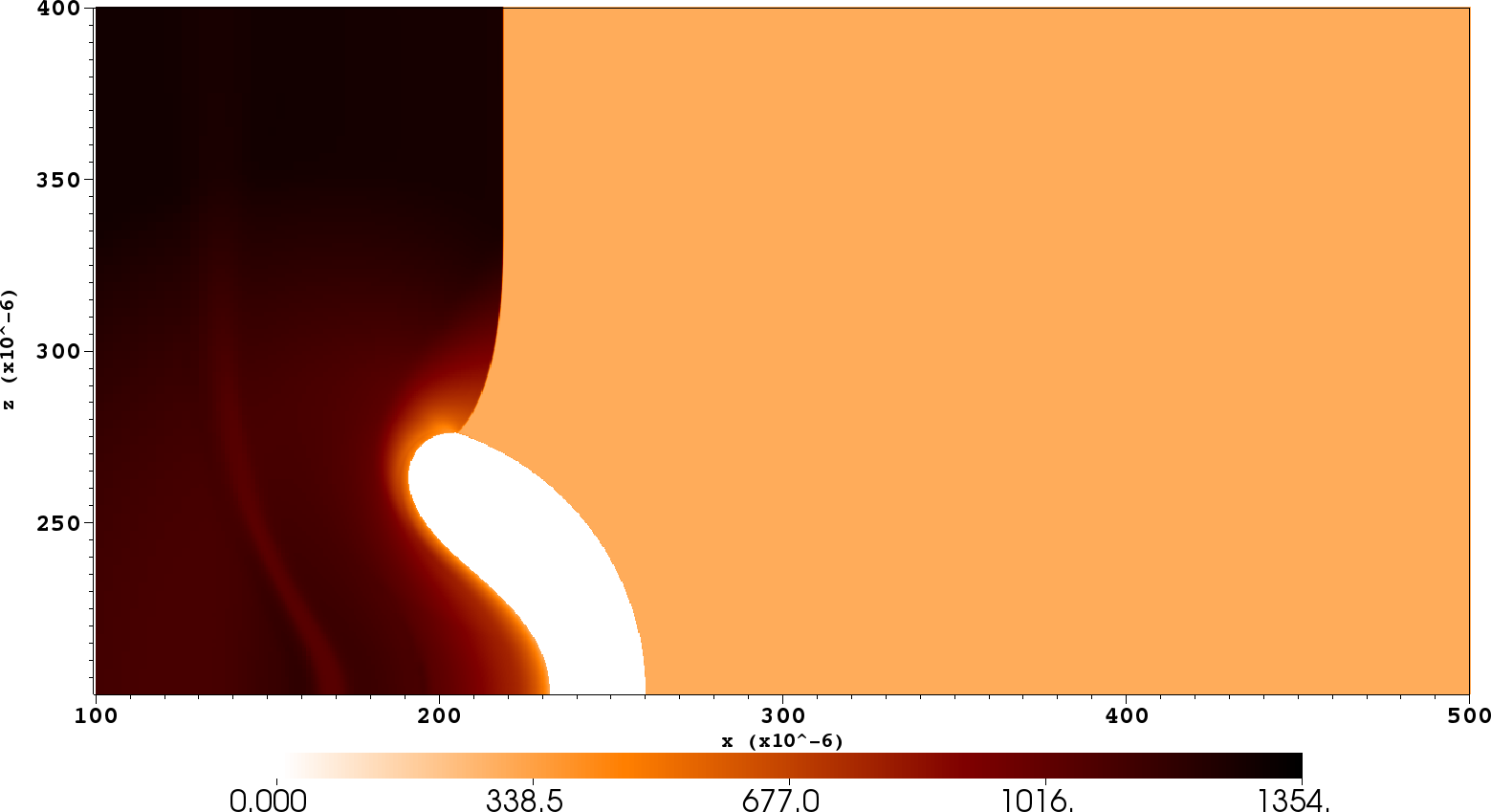}
\caption{3D $t=0.03\SI{}{\micro \second}$}
        \end{subfigure}
\end{minipage}
\begin{minipage}[t]{2\columnwidth}
\centering
        \begin{subfigure}[t]{0.43\textwidth}
\includegraphics[width=\textwidth]{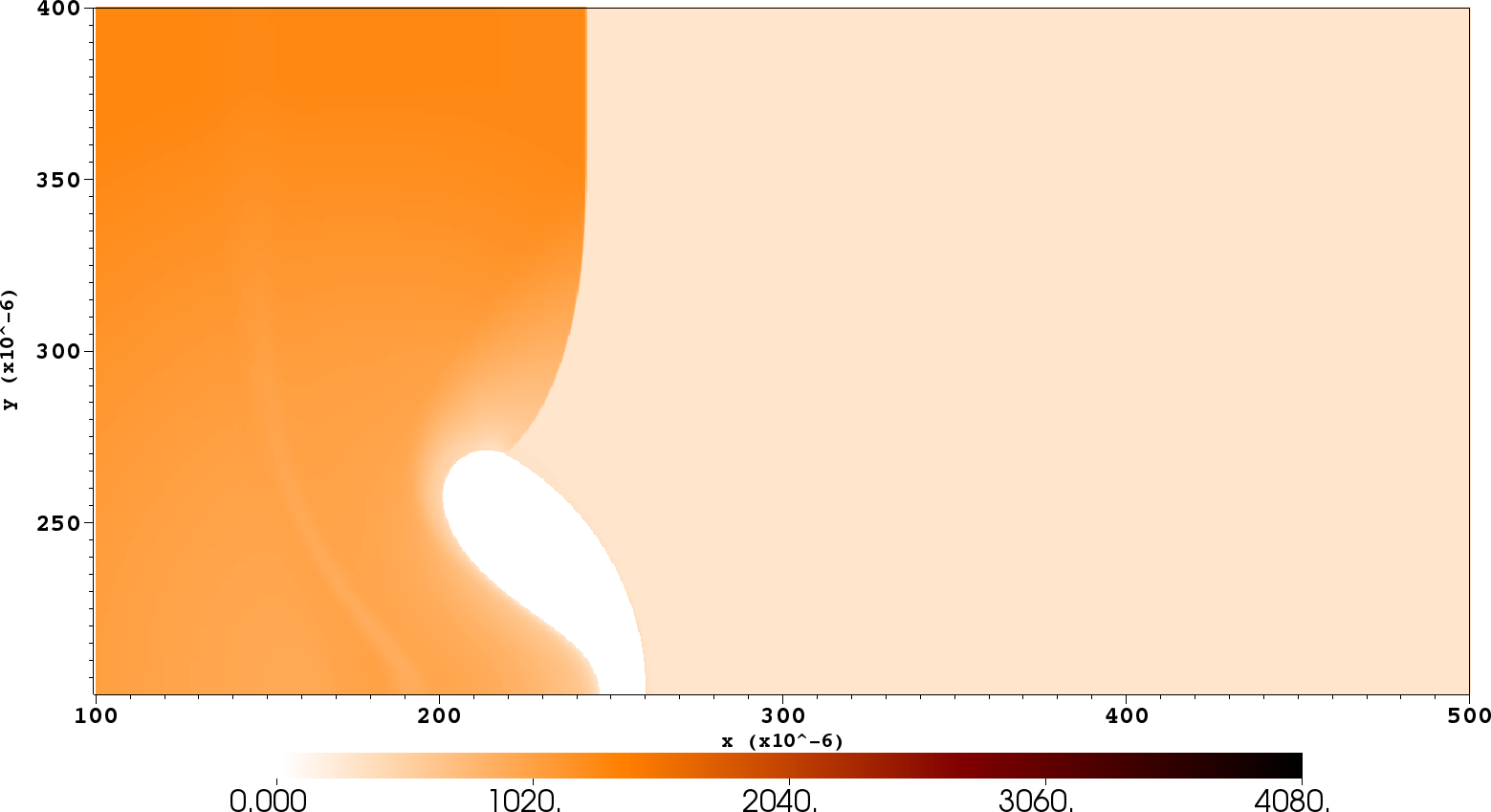}
 \caption{2D $t=0.035\SI{}{\micro \second}$}        
\end{subfigure}
 \begin{subfigure}[t]{0.43\textwidth}
\includegraphics[width=\textwidth]{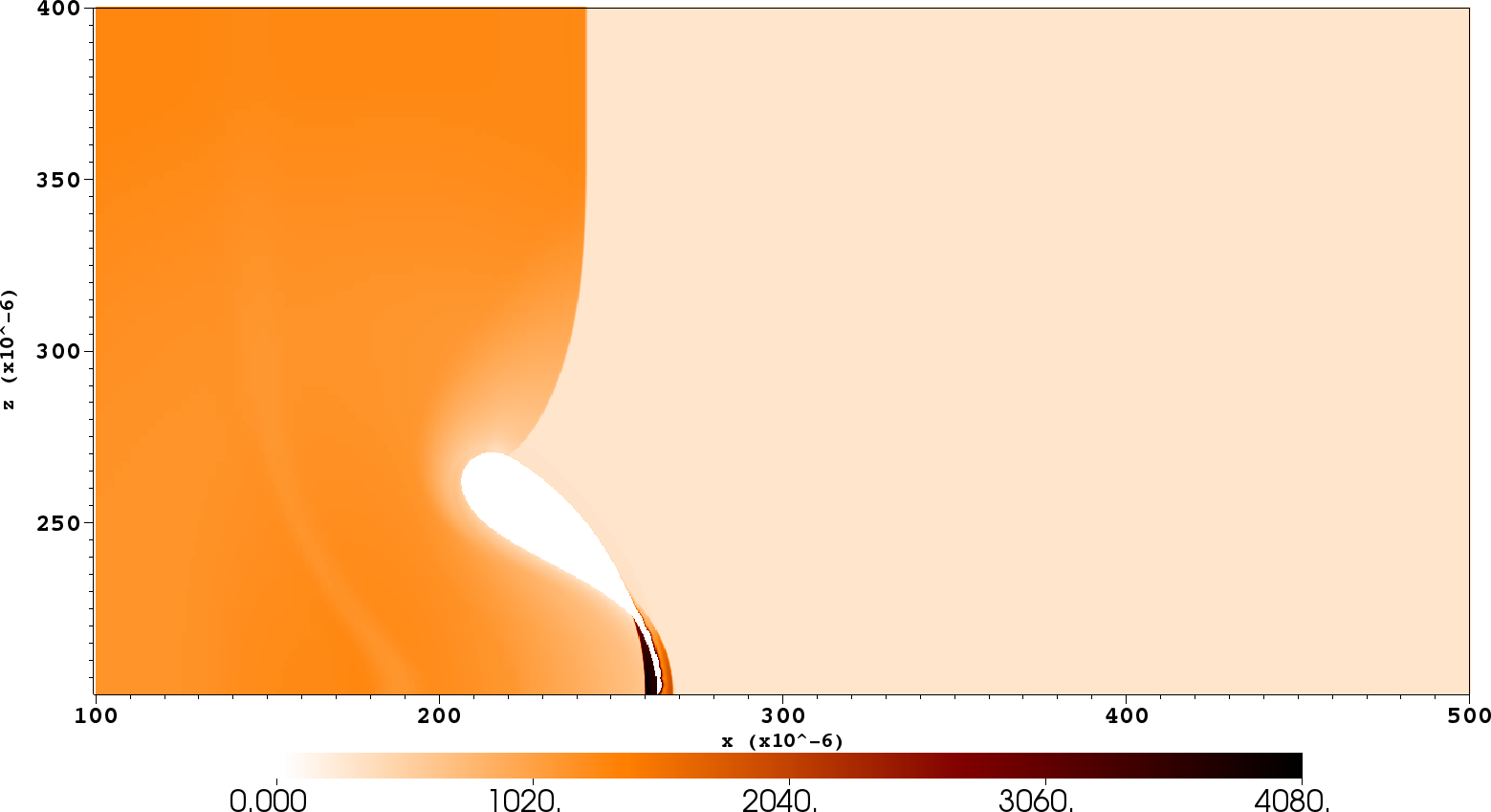}
 \caption{3D $t=0.035\SI{}{\micro \second}$}       
 \end{subfigure}
\end{minipage}
       \begin{minipage}[t]{2\columnwidth}
\centering
        \begin{subfigure}[t]{0.43\textwidth}
\includegraphics[width=\textwidth]{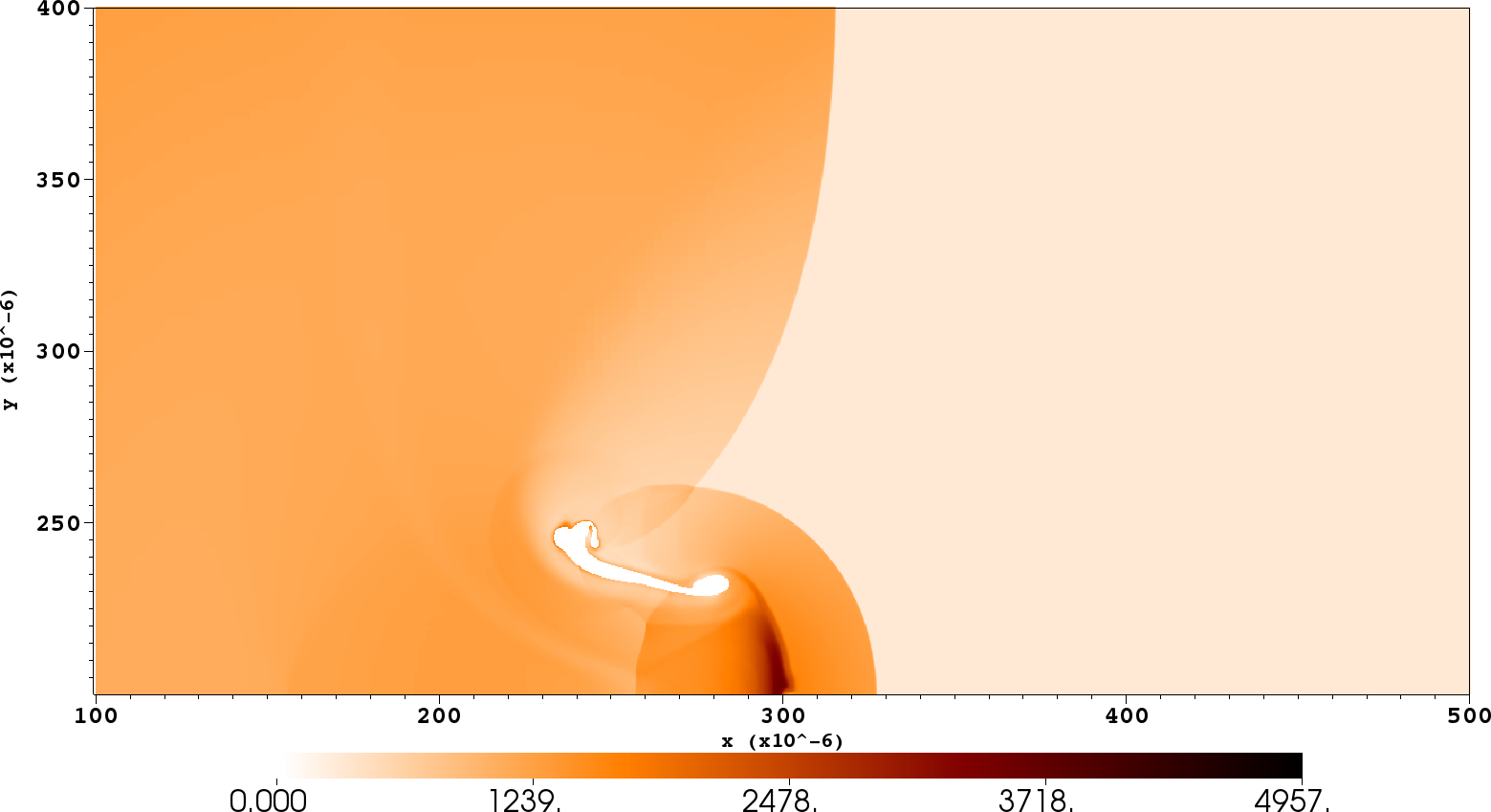}
        \caption{2D $t=0.05\SI{}{\micro \second}$}       
 \end{subfigure}
 \begin{subfigure}[t]{0.43\textwidth}
\includegraphics[width=\textwidth]{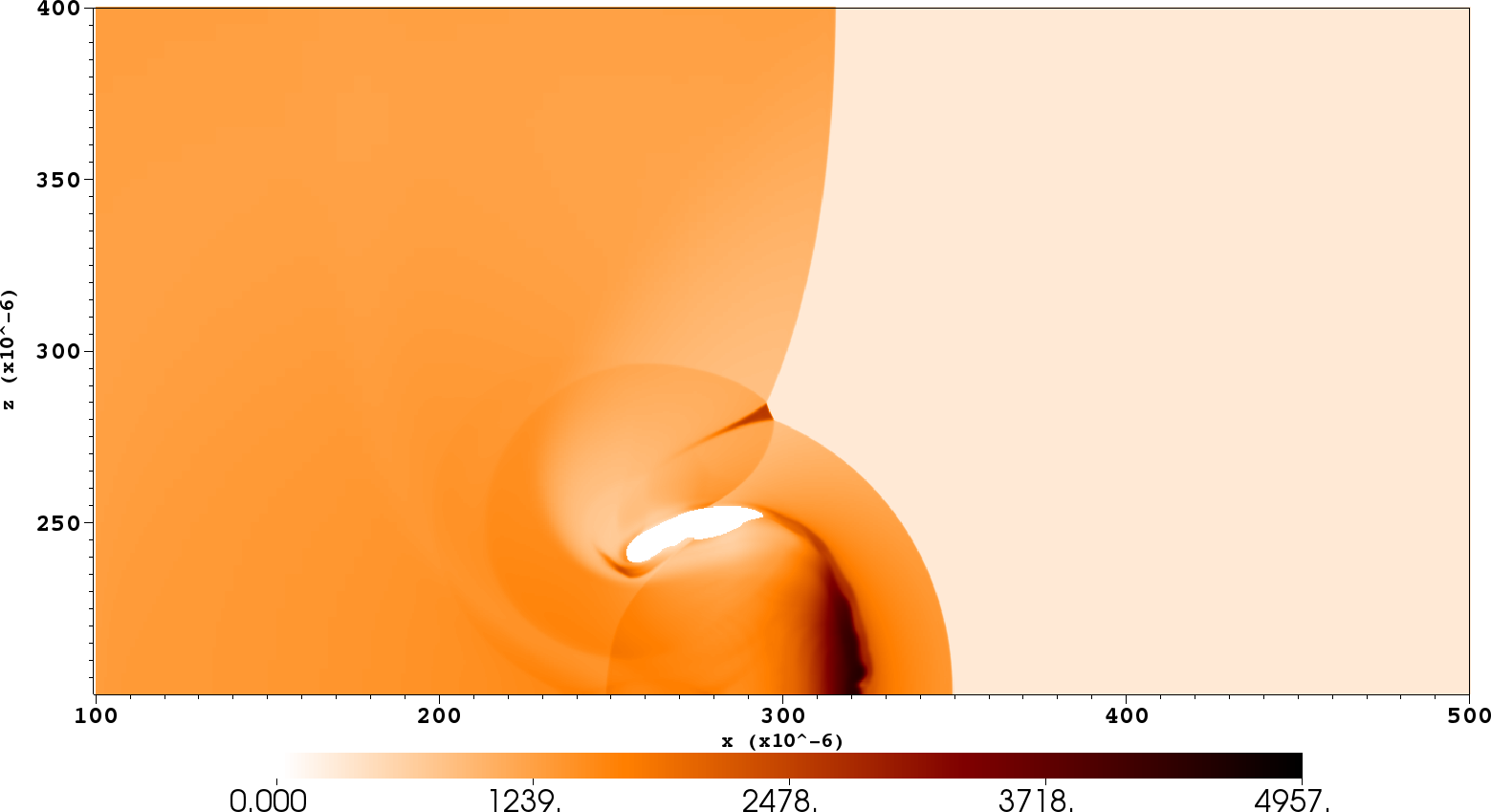}
        \caption{3D $t=0.05\SI{}{\micro \second}$}
\end{subfigure}
\end{minipage}
\begin{minipage}{2\columnwidth}
\centering
        \begin{subfigure}[b]{0.43\textwidth}
\includegraphics[width=\textwidth]{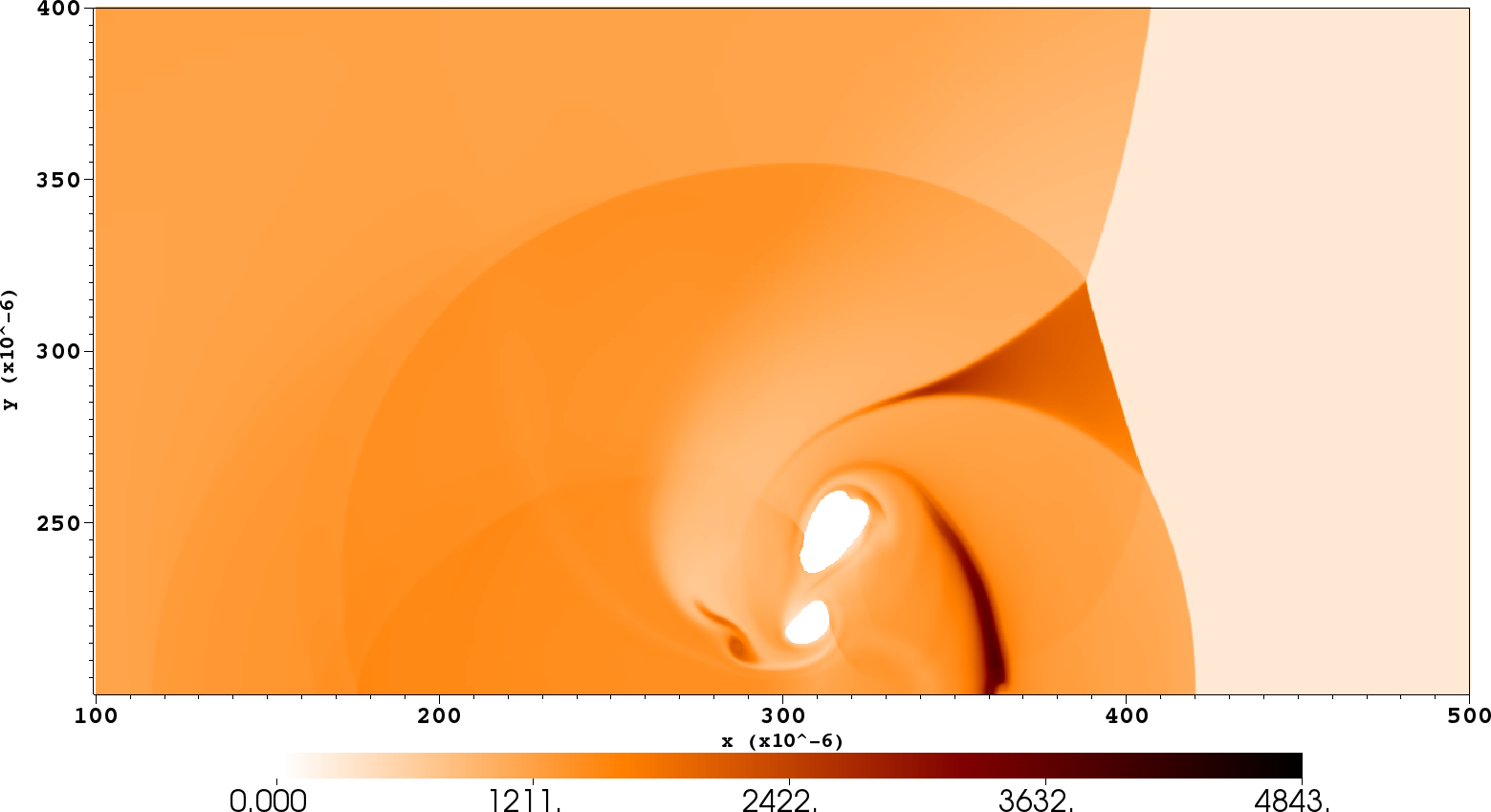}
        \caption{2D $t=0.07\SI{}{\micro \second}$}        
\end{subfigure}
 \begin{subfigure}[b]{0.43\textwidth}
\includegraphics[width=\textwidth]{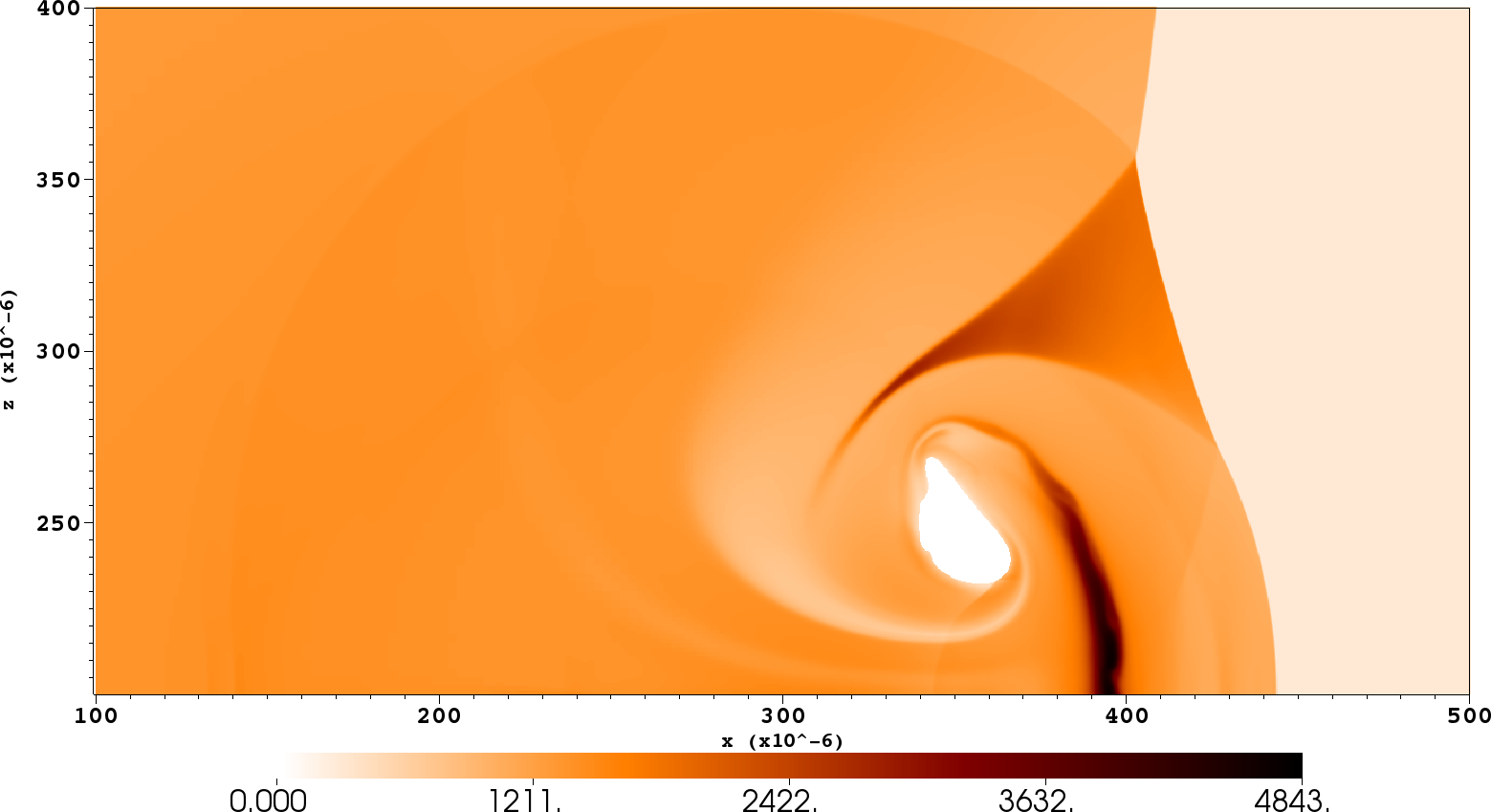}
        \caption{3D $t=0.07\SI{}{\micro \second}$}
        \end{subfigure}
\end{minipage}
\caption{Comparison of the temperature field and collapse times between a two-dimensional and a three-dimensional of the cavity collapse under a $10.98\SI{}{\giga \pascal}$ ISW at selected times.}
\label{2Dvs3Dpseudo}
\end{figure*}

\begin{figure*}[!ht]
\centering
\begin{minipage}{2\columnwidth}
\centering
        \begin{subfigure}[b]{0.45\textwidth}
\includegraphics[width=\textwidth]{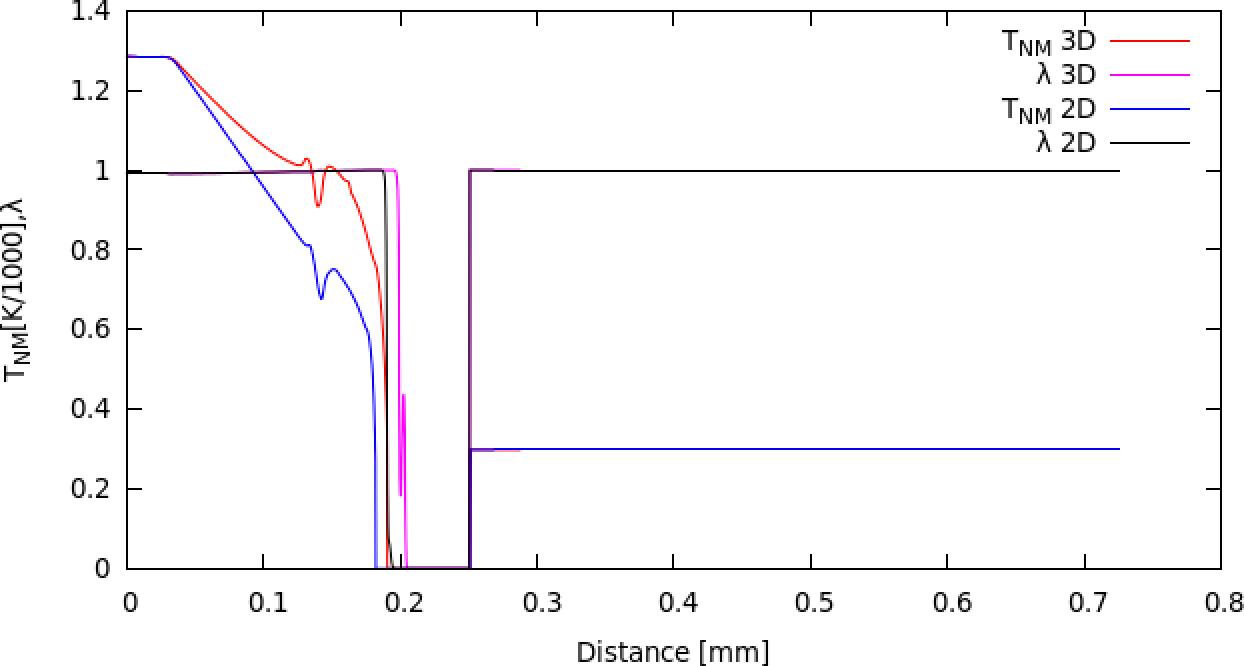}
    \caption{$t=0.025\SI{}{\micro \second}$}
        \end{subfigure}
 \begin{subfigure}[b]{0.45\textwidth}
\includegraphics[width=\textwidth]{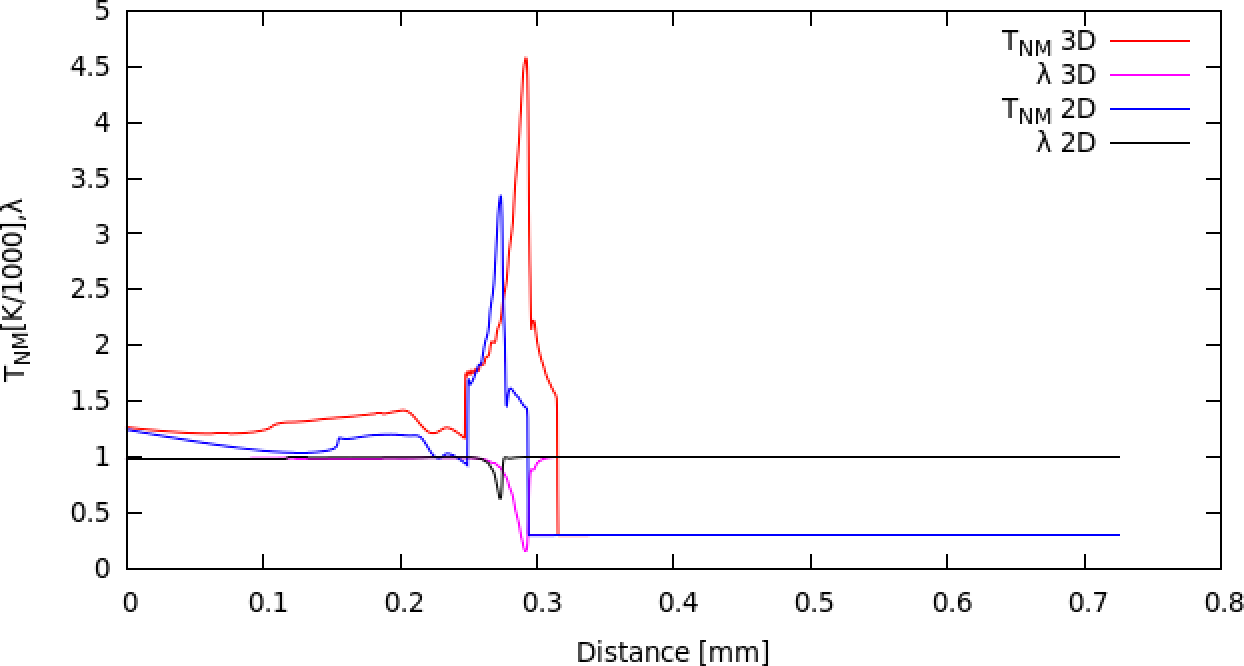}
    \caption{$t=0.045\SI{}{\micro \second}$}        
\end{subfigure}
\end{minipage}
\begin{minipage}{2\columnwidth}
\centering
        \begin{subfigure}[b]{0.45\textwidth}
\includegraphics[width=\textwidth]{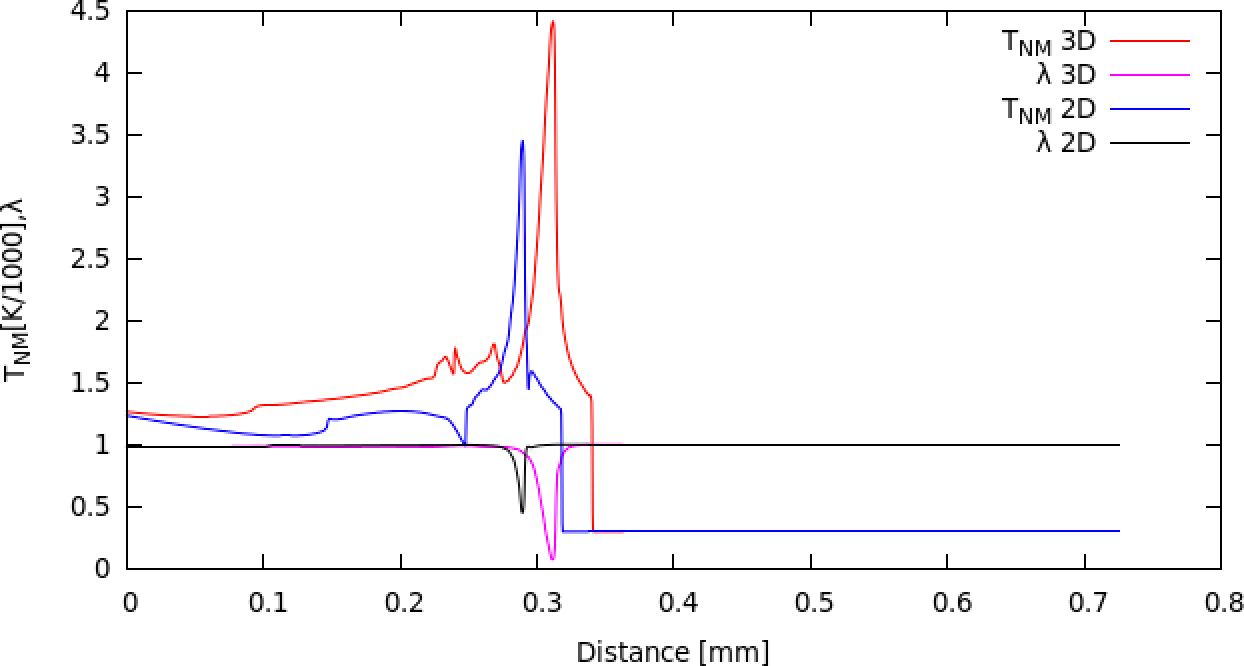}
    \caption{$t=0.05\SI{}{\micro \second}$}        
\end{subfigure}
 \begin{subfigure}[b]{0.45\textwidth}
\includegraphics[width=\textwidth]{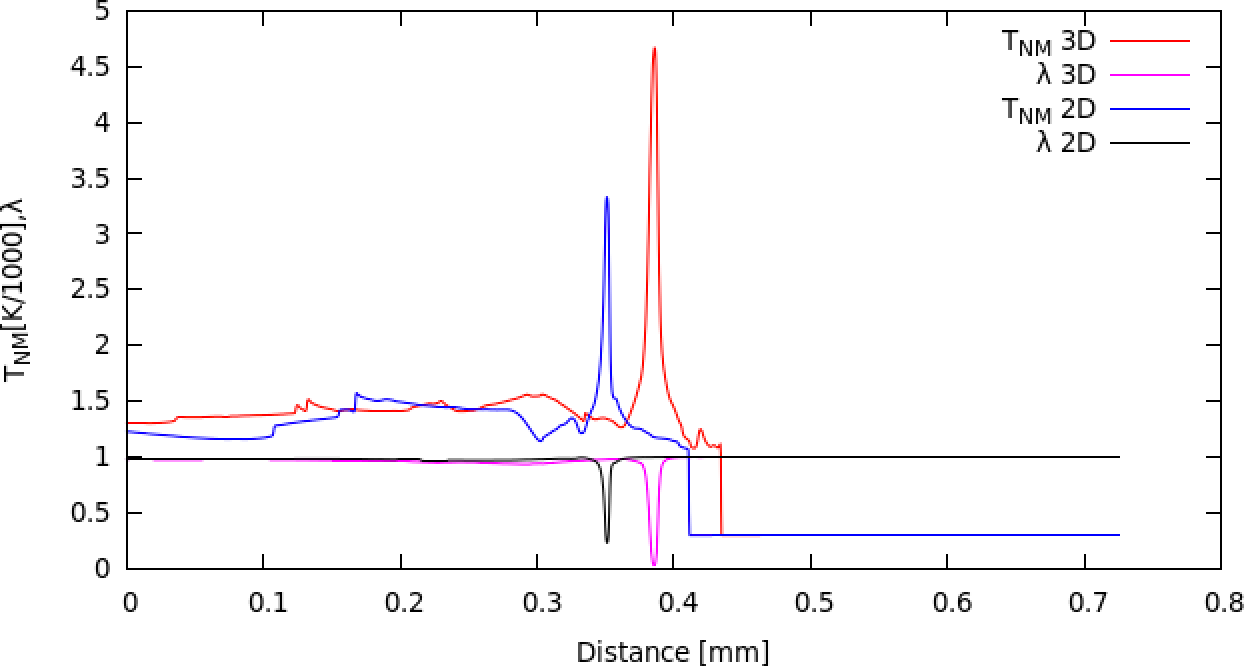}
    \caption{$t=0.07\SI{}{\micro \second}$}        
\end{subfigure}
\end{minipage}
\caption{Comparison of the temperature field and reaction along $y=\SI{0.21}{\milli \meter}$ in the 2D and 3D configurations.}
\label{2Dvs3D200}
\end{figure*}

\begin{figure*}[!ht]
\centering
\begin{minipage}{2\columnwidth}
\centering
        \begin{subfigure}[b]{0.45\textwidth}
\includegraphics[width=\textwidth]{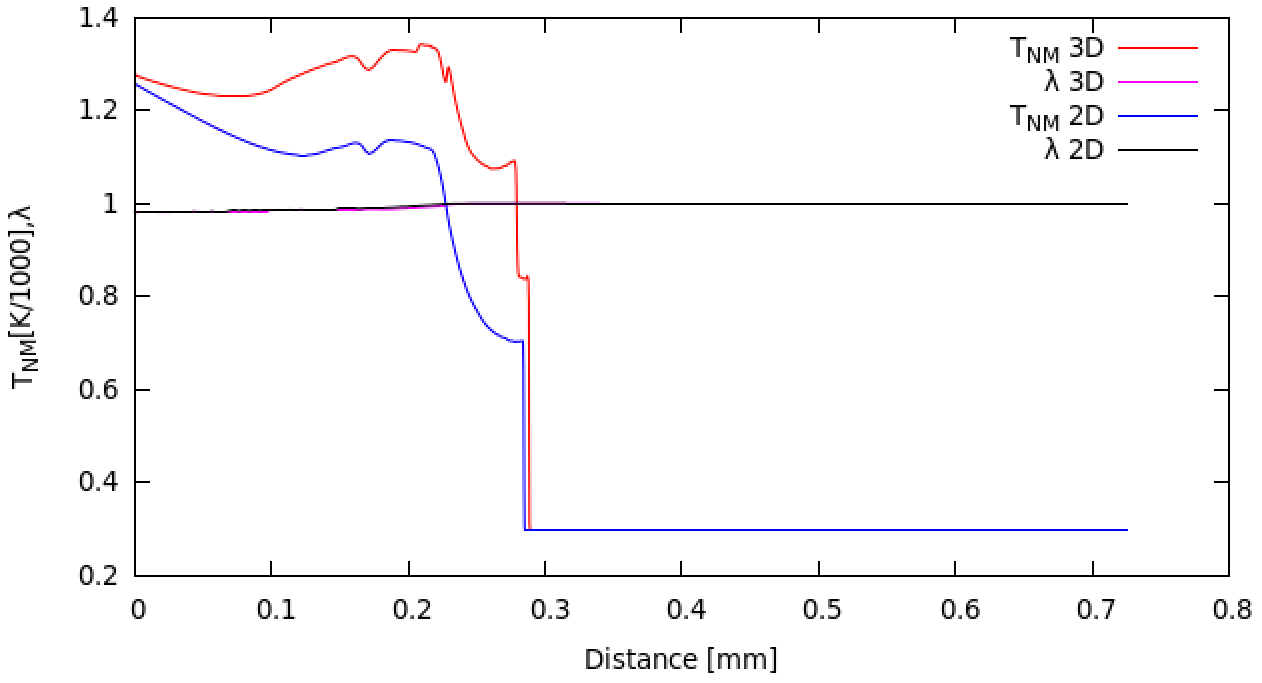}
    \caption{$t=0.05\SI{}{\micro \second}$}
        \end{subfigure}
 \begin{subfigure}[b]{0.45\textwidth}
\includegraphics[width=\textwidth]{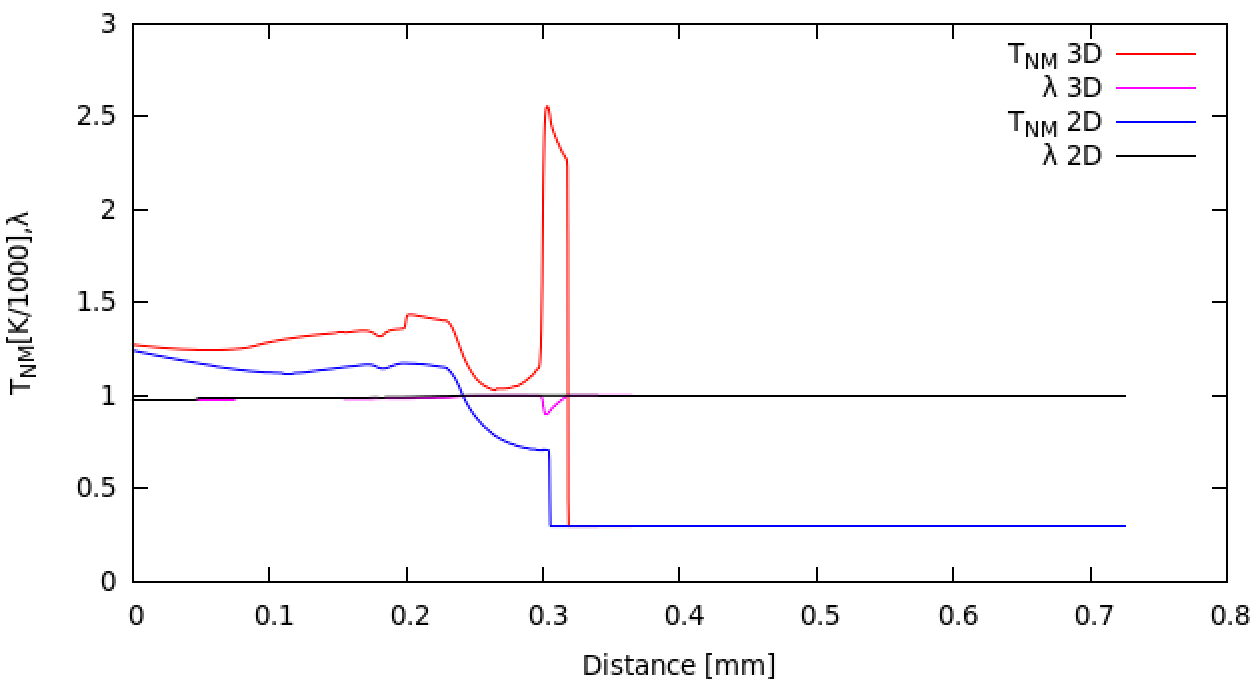}
    \caption{$t=0.055\SI{}{\micro \second}$}
        \end{subfigure}
\end{minipage}
\begin{minipage}{2\columnwidth}
\centering
        \begin{subfigure}[b]{0.45\textwidth}
\includegraphics[width=\textwidth]{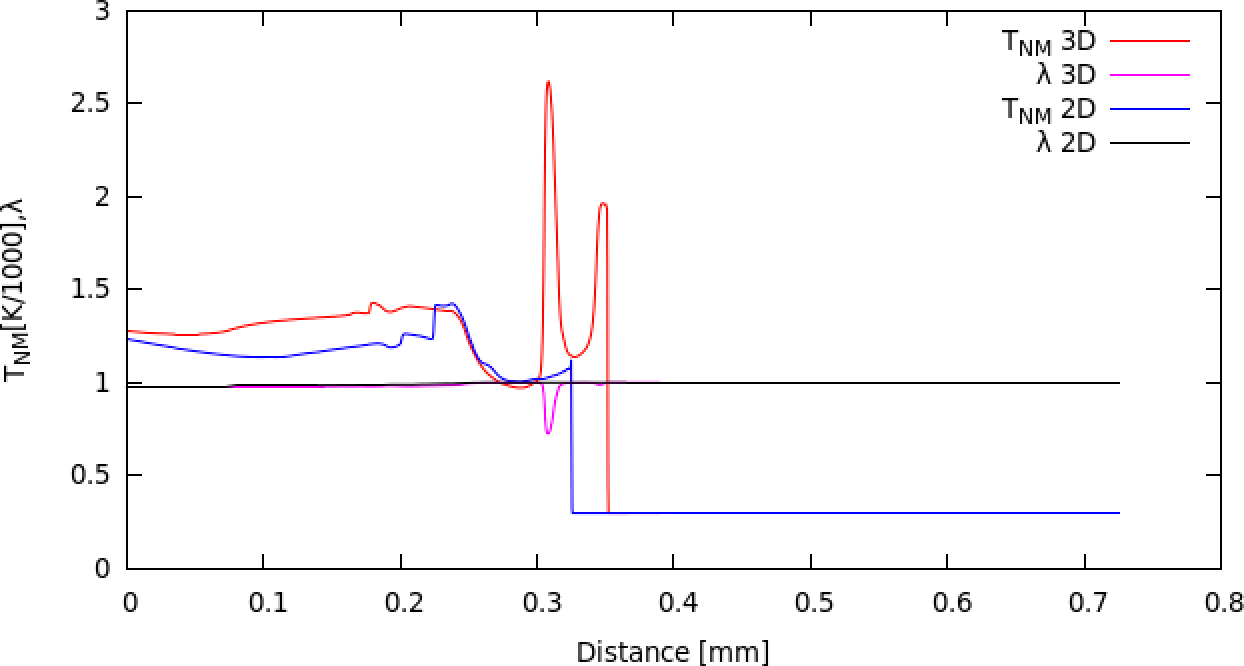}
    \caption{$t=0.06\SI{}{\micro \second}$}
        \end{subfigure}
 \begin{subfigure}[b]{0.45\textwidth}
\includegraphics[width=\textwidth]{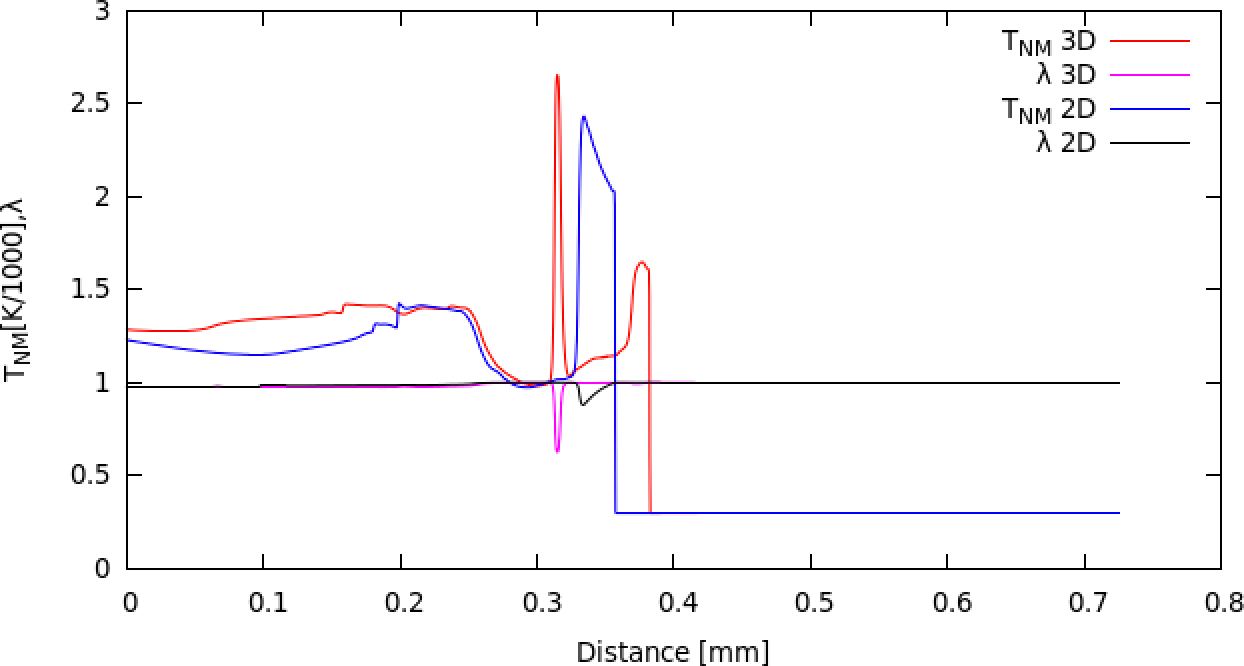}
    \caption{$t=0.065\SI{}{\micro \second}$}
        \end{subfigure}
\end{minipage}
\caption{Comparison of the temperature field and reaction along $y=\SI{0.29}{\milli \meter}$ in the 2D and 3D configurations.}
\label{2Dvs3D290}
\end{figure*}

In this section, the three-dimensional (3D) collapse of a cavity in reacting nitromethane is presented and compared to the two-dimensional (2D) equivalent simulation presented in the previous section. Selected stages of the collapse are shown in Fig.\ \ref{3Dpseudo}. The three-dimensional $z=0.5$ contour represents the cavity material boundary, while the $\lambda=0.9$ contour represents the generated hot-spot(s), i.e.\ the region where
reaction takes place. We take a two-dimensional slice of the $\lambda$ field through  the centre of the cavity and project it on the left half of each
figure. This again illustrates the reaction regions. Similarly, the projection of the nitromethane temperature field on the same plane is seen on the right
half of each figure. It is therefore presented how the generation of locally
high temperatures leads to the generation of hot-spots. The highest temperatures are located in front and behind the point of collapse of the cavity (FHS and BHS) as well as in the Mach stem generated at late stages of the collapse due to wave superposition. These lead to the generation of bell-shaped hot spots, mainly behind the point of collapse (BHS) and a torus-shaped hot spot corresponding to the three-dimensional Mach stem. 
The comparison between the 3D and 2D collapse process is performed by using planar pseudocolour plots of the temperature field (Fig.\ \ref{2Dvs3Dpseudo}) and the evolution of the temperature and $\lambda$ fields along lines of constant latitude, specifically along the centreline of the cavity and $y=29\SI{}{\micro \meter}$ (Figs.\ \ref{2Dvs3D200},\ref{2Dvs3D290}).

\begin{figure}[!ht]
\centering
\includegraphics[width=0.45\textwidth]{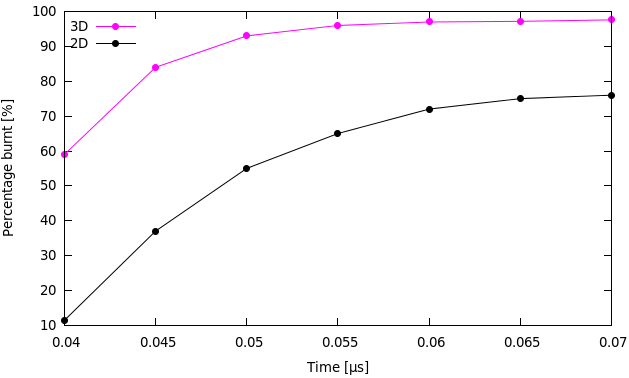}
\caption{Percentage of explosive burnt over time in the three-dimensional and two-dimensional scenarios.}
\label{2Dvs3Dignition}
\end{figure}

The three-dimensional nature of the flow field around the cavity results to a faster jet compared to the two-dimensional equivalent scenario. This is seen in Fig.\ \ref{2Dvs3Dpseudo}(a)-(b) and Fig.\ \ref{2Dvs3D200}(a) and effectively results in a faster nitromethane jet (1.4 times faster) in the 3D case than in the 2D case and the earlier observation of all subsequent features of the collapse phenomenon. The cavity collapses, as a result, faster by $\sim \SI{0.5}{\micro \second}$ in the 3D case (Fig.\ \ref{2Dvs3Dpseudo}(c)-(d)) and the generated collapse shock waves lead to a quicker ignition in the front and back hot spots (Fig.\ \ref{2Dvs3D200}(b)). The temperatures achieved upon collapse range between 400K and 1500K higher in the 3D scenario than the 2D. The percentage of explosive burnt in the centreline hot-spots in the centreline hot spot region is shown as a function of time in Fig. \ref{2Dvs3Dignition}. It can be seen that the higher temperature attained upon the 3D collapse generates a higher immediate burn of the material in 3D ($59\%$ burnt) compared to 2D ($11\%$ burnt) and a faster overall ignition and burning procedure.  The Mach stem and hence the MSHS are also generated quicker than in the 2D case (Fig.\ \ref{2Dvs3Dpseudo}(e)-(f) and Fig.\ \ref{2Dvs3D290}(b)). The Mach stem and the MSHS grow faster in the 3D case as well, as seen in Fig.\ \ref{2Dvs3Dpseudo}(g)-(h) and Fig.\ \ref{2Dvs3D290}(c)-(d). In general, in the 3D case, higher temperatures are also achieved compared to the 2D case, as it is evident in Fig.\ \ref{2Dvs3D200}(c)-(d) and  Fig.\ \ref{2Dvs3D290}(c)-(d) leading to faster immediate ignition.


\section{Comparing the cavity collapse in inert and reacting nitromethane}
\begin{figure}[!ht]
\centering
\includegraphics[width=0.45\textwidth]{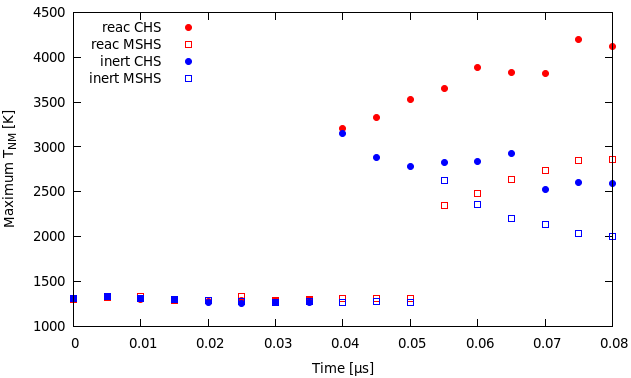}
\caption{Maximum nitromethane temperature observed
in the centreline hot spots (BHS and FHS, labelled collectively as CHS) plotted using circles and in the MSHS plotted using squares. The
red color denotes results from the reactive
setup and the blue color from the inert
set up.}
\label{Tmax}
\end{figure}

In Part I of this work the collapse of a cavity in non-reacting nitromethane was studied and this work extended that to the equivalent reacting medium. In this section the temperature field generated through the collapse process in the inert and reacting media is compared, to examine the effect of the reactions on the temperature field and the hot spot topology. The comparison uses the two-dimensional scenarios for simplicity. The number of hot spots generated in the two cases is identical and the location is roughly the same; any difference observed is of the order of $3-4\%$ of the bubble radius, and mainly in the BHS and FHS. The significant difference in the two collapse cases is seen in the temperature field. At the early stages of the simulation, before the collapse takes place, some initial reaction is observed behind the shock wave, which is translated to some increase in the temperature field in the reactive case, compared to the inert case. The difference in the temperature fields of the two cases is small, ranging from 0 to 50K. Once the cavity collapses, the BCSW and FCSW generate significant reaction in the BHS and FHS leading to an even larger increase of temperature in the reactive case and a difference of 65K from the inert case. This difference grows as the reaction progresses and the hot spots grow in size and strength. This can be seen in Fig.\ \ref{Tmax}, where the maximum temperature in the centreline hot spots (CHS) i.e.\ BHS and FHS and in the MSHS is shown. The general trend is that after the collapse, the maximum temperature observed in the simulation in the reactive case increases, whereas it decreases in the inert scenario. This is expected as the reaction supports the shock waves and vice versa. In the reactive case, the maximum temperature observed can be higher than in the inert scenario by as much as 1500K. For this configuration and the timescales studied the temperatures in the MSHS are always lower than the CHS. Moreover, the difference between the CHS and MSHS is larger in the reactive case than in the inert case.

\begin{figure}[!t]
\centering
        \begin{subfigure}[b]{0.45\textwidth}
\includegraphics[width=0.99\textwidth]{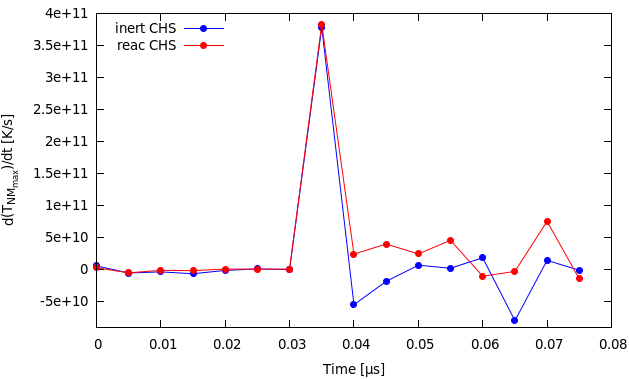}
\label{rateCHS}
\subcaption{CHS}
        \end{subfigure}
        \begin{subfigure}[b]{0.45\textwidth}
\includegraphics[width=0.99\textwidth]{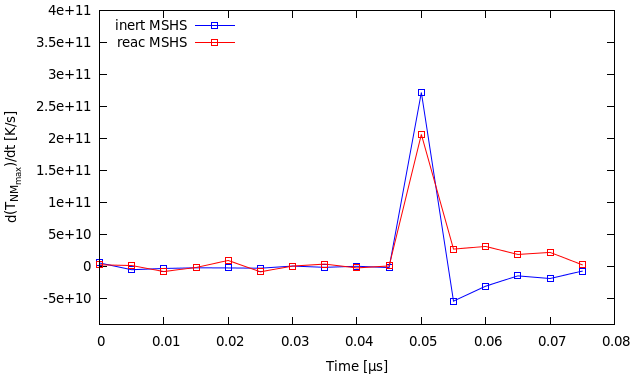}
\subcaption{MSHS}
        \end{subfigure}
\label{rateCHS}
\caption{Comparison of the rate of increase of maximum temperature (a) in the centreline hot spots (CHS) and (b) in the Mach stem hot spot (MSHS) in the inert and reactive scenarios.}
\label{rate}
\end{figure}

\begin{figure}[!t]
\centering
        \begin{subfigure}[b]{0.45\textwidth}
\includegraphics[width=0.99\textwidth]{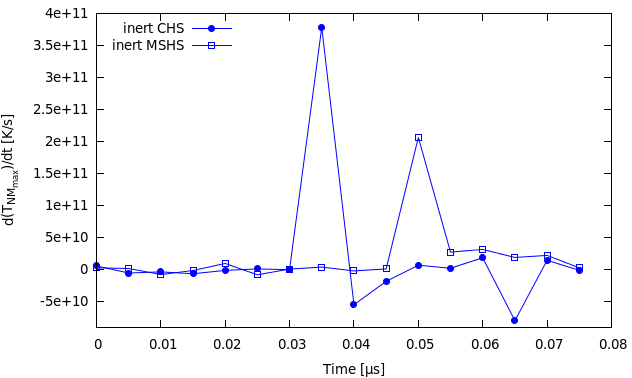}
\subcaption{inert}
        \end{subfigure}
        \begin{subfigure}[b]{0.45\textwidth}
\includegraphics[width=0.99\textwidth]{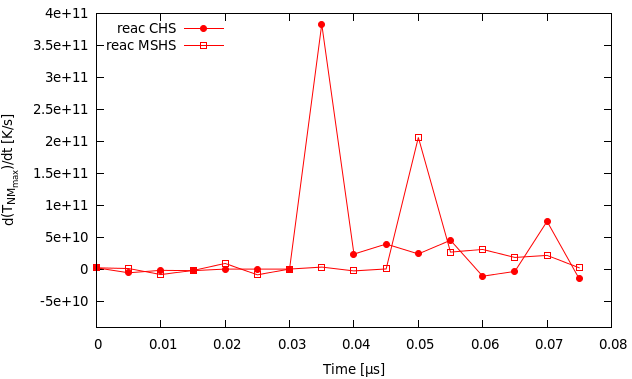}
\subcaption{reactive}
        \end{subfigure}
\caption{Comparison of the rate of increase of maximum temperature in the two hot spots (a) in the inert and (b) the reactive scenarios.}
\label{rate2}
\end{figure}

The comparison of the rate of increase of the temperature in the CHS and MSHS between the inert and reactive scenarios is presented in Fig.\ \ref{rate}. It is clear that the rate of increase of the maximum temperature in both hot spots is positive in the reactive case, i.e.\ the maximum temperature is always increasing, whereas this is not true for the inert scenario. The first peak in both loci corresponds to the birth of the hot spots. In the CHS a second, late peak is observed which corresponds to the superposition of the waves emanating from the two lobes along the centreline of the cavity. In the MSHS the increase of the maximum temperature relies purely on the waves generating the Mach stem. In Fig.\ \ref{rate2} we observe that the rate of increase in the MSHS in the inert case is always higher than the CHS (except at the birth of the hot spot). This is, however not true for the reactive case.  Comparing information from Figs.\ \ref{Tmax}-\ref{rate2}, we can conclude that in maximum temperature terms and in this configuration, the CHS are stronger in absolute value than the MSHS, although the growth rate of reaction in the MSHS was seen to be higher than the CHS at late stages of the collapse in Fig.\ \ref{CHSvsMSHSignition}. In the inert case, however, the rate of temperature increase in the MSHS is higher than in the CHS (except at the bubble collapse time) so the MSHS could, in longer timescales lead to higher temperatures than the CHS. This could also be true in the reactive case, though more detailed, late time simulations would be needed to determine that.  

\begin{figure*}[!ht]
\centering
        \begin{subfigure}[b]{0.45\textwidth}
\includegraphics[width=\textwidth]{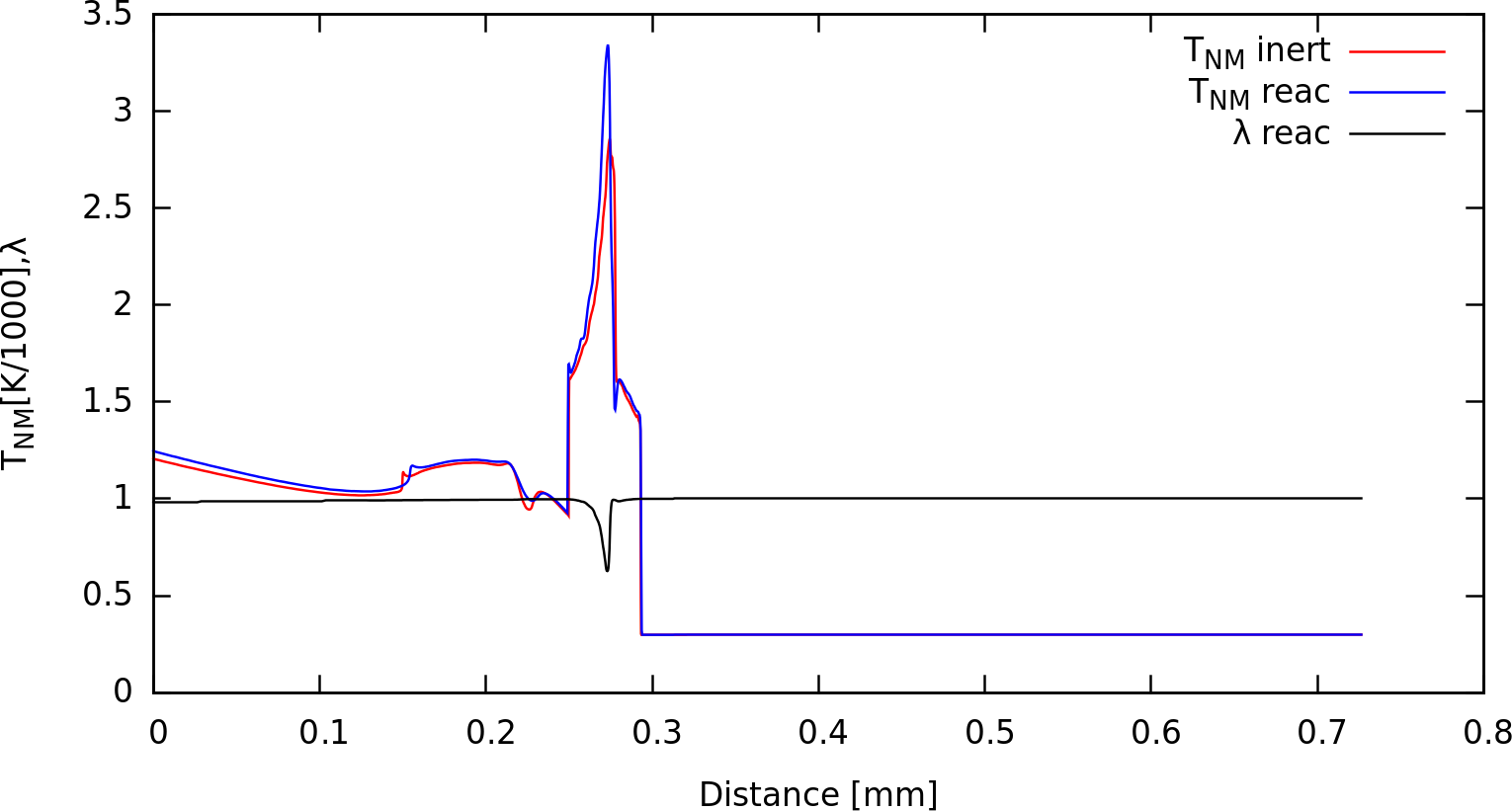}
    \caption{$t=0.045\SI{}{\micro \second}$}
        \end{subfigure}
 \begin{subfigure}[b]{0.45\textwidth}
\includegraphics[width=\textwidth]{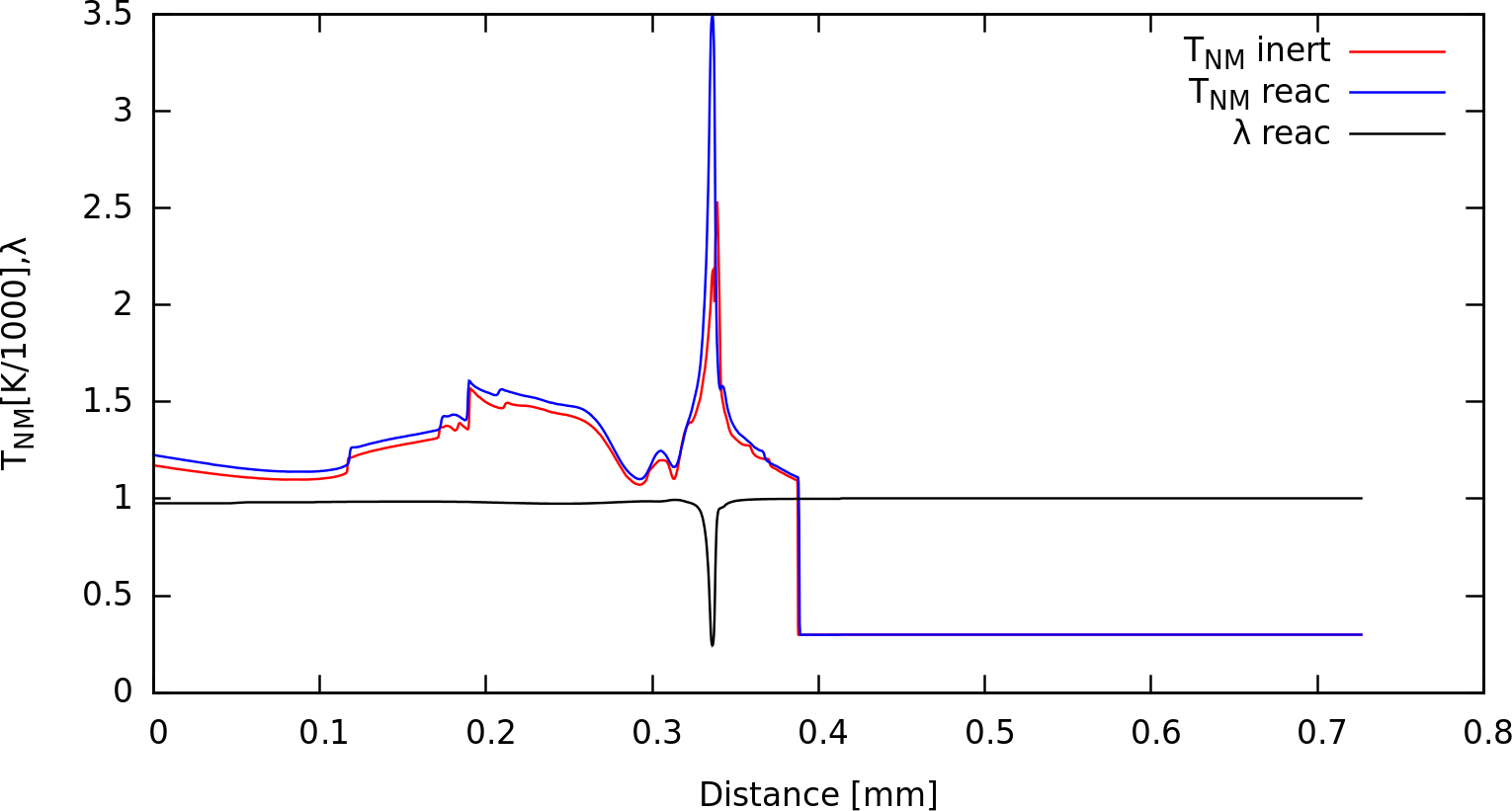}
    \caption{$t=0.065\SI{}{\micro \second}$}
        \end{subfigure}
\caption{Comparison of the temperature field and reaction along $y=\SI{0.21}{\milli \meter}$ in the inert and reactive configurations}
\label{inertVSreac200}
\end{figure*}

\begin{figure*}[!ht]
\centering
        \begin{subfigure}[b]{0.45\textwidth}
\includegraphics[width=\textwidth]{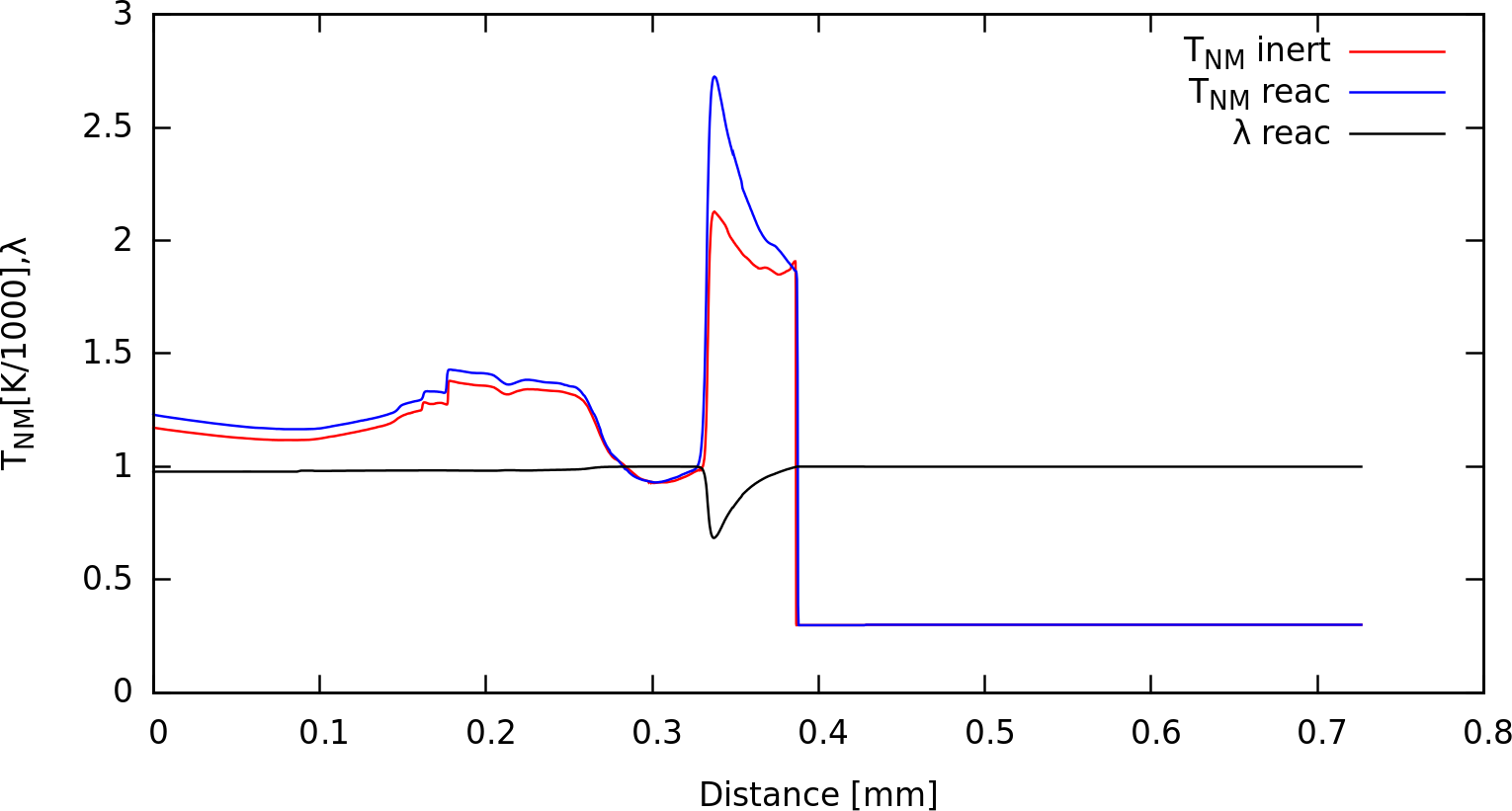}
    \caption{$y=\SI{0.29}{\milli \meter}$}
        \end{subfigure}
 \begin{subfigure}[b]{0.45\textwidth}
\includegraphics[width=\textwidth]{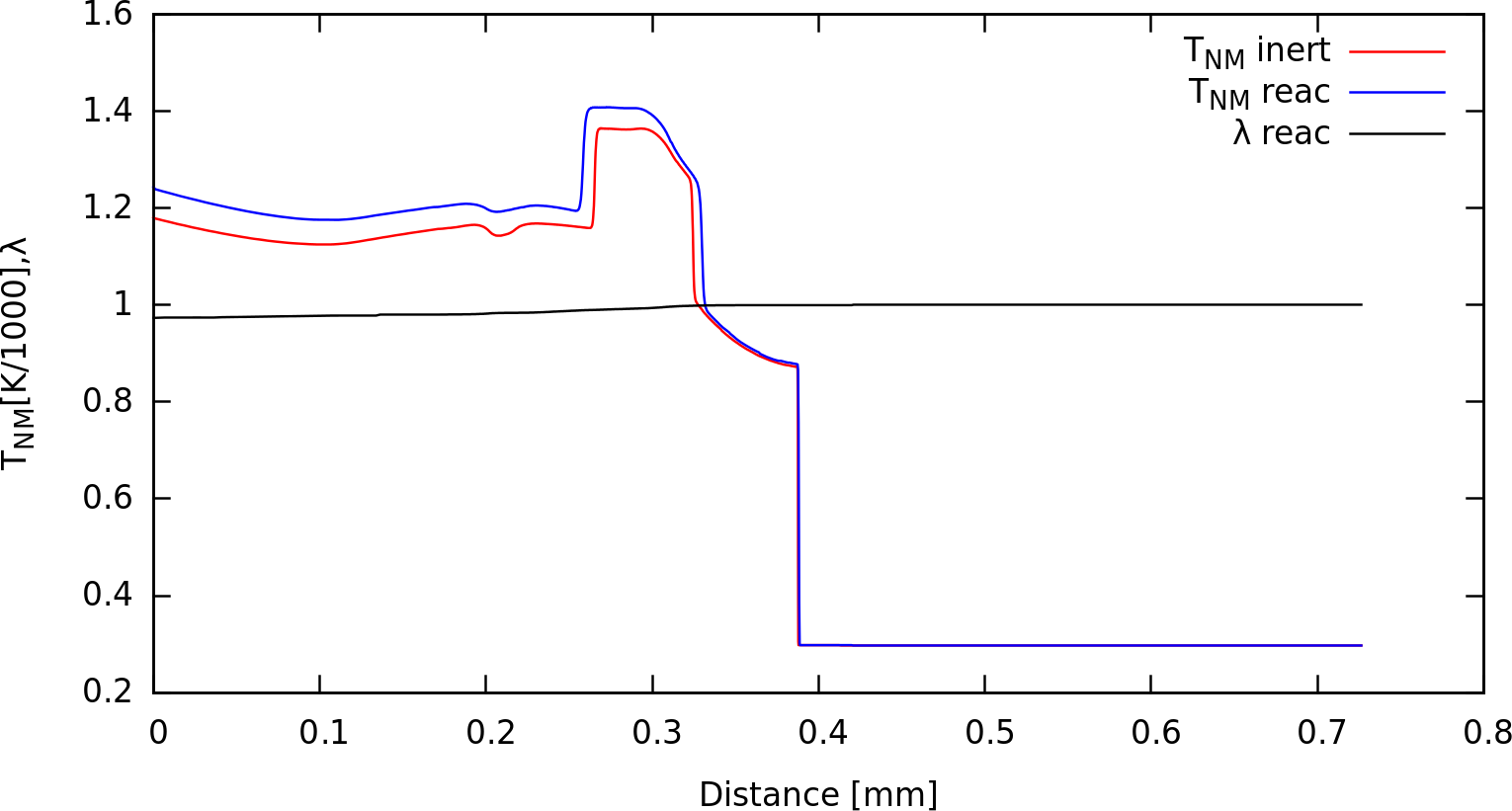}
    \caption{$y=\SI{0.35}{\milli \meter}$}
        \end{subfigure}
\caption{Comparison of the temperature field and reaction along (a) $y=\SI{0.29}{\milli \meter}$ and (b) $y=\SI{0.35}{\milli \meter}$ at $t=0.07\SI{}{\micro \second}$ between the inert and reactive configurations.}
\label{inertVSreac290350}
\end{figure*}

Using lineouts, we look at the temperatures in the different hot spots. The local temperature difference is higher in the late stages after the collapse so we omit illustrations of the early times. On $y=\SI{0.210}{\milli \meter}$ at $t=\SI{0.045}{\micro \second}$  (Fig.\ \ref{inertVSreac200}(a)), the temperature difference in the BHS is more significant than the FHS. This continues to be true at later times, until the two hot spots merge ($t=\SI{0.065}{\micro \second}$) in Fig.\ \ref{inertVSreac200}(b). The temperature difference in the MSHS is also significant, as seen on $y=\SI{0.29}{\milli \meter}$ in Fig.\ \ref{inertVSreac290350}(a) on $y=\SI{0.35}{\milli \meter}$ in Fig.\ \ref{inertVSreac290350}(b) for $t=\SI{0.07}{\micro \second}$). The width of the MSHS is also larger in the reactive case.


\section{Conclusions}

In this work we perform resolved numerical simulations of cavity collapse in liquid nitromethane
using a multi-phase formulation, which can recover reliable temperatures in the vicinity of the cavity. Considerable care is taken regarding the form of equations and numerical algorithm to eliminate spurious numerical oscillations in the temperature field. The model is validated against experimental data. Specifically, we demonstrate that the deduced CJ and von Neumann values match the values
found in the literature and also that the experimentally determined ignition Pop-plot data are matched.
Following the validation, we study the shock-induced cavity collapse in reacting liquid nitromethane, in two and three dimensions and follow the events leading to the generation of local temperatures and initiation of the explosive. 

Working towards elucidating the relative contribution of fluid dynamics and chemical reaction, we examined in Part I of this work the details of the hydrodynamic effects that lead to local temperature elevations and identified a more complex hot spot topology than previously described in the literature. In this second part of the work  we demonstrate the effect of chemical reactions that acts additively yo the fluid dynamics. We identify which high temperature regions lead to reactive hot spots and observe much higher temperatures (40K-1500K) than in Part I.

We examine additionally the ignition of nitromethane in the absence of cavities and compare the ignition times for the neat and single-cavity material. We observe that the initiation of nitromethane in the presence of an isolated collapsing cavity is reduced to less than one third (in 2D simulations or less than a quarter in 3D simulations) of the required time for igniting the neat material. This quantifies the sensitisation character of the cavities in this configuration.

It is observed that the highest nitromethane temperatures still occurs in the back hot spot (BHS) as also observed in Part I and this leads to the first ignition site that resides along the centreline of the cavity. Moreover, the temperatures in the Mach stem hot spot (MSHS) are proven high enough that ignition occurs in this hot spot as well, at a time less than half (both for 2D and 3D simulations) the time required for the ignition of the neat material. By studying the maximum percentage burnt in the two centreline hot spots (CHS) and Mach stem hot spot (MSHS) we observed that the burning in the CHS is, at these timescales, always higher than in the MSHS. However, the maximum burning of the CHS grows as $\sqrt{x}$ while the maximum burning in the MSHS grows linearly. As a result, the growth rate of the maximum burning in the CHS is higher than in the MSHS at first but this is reversed at late times.

By comparing two- and three-dimensional simulations we identify the change in topology of the hot spots due to the third dimension. The faster jet (1.4 times faster) in the 3D case results in an earlier collapse of the cavity ($\sim \SI{0.5}{\micro \second}$) and subsequent hot spot generation compared to 2D. The ignition in the BHS in the 3D case occurs between  $\SI{3.5}{\micro \second}$ and $\SI{4}{\micro \second}$ and in the MSHS between $\SI{5.5}{\micro \second}$ and $\SI{6}{\micro \second}$. Effectively the ignition is observed earlier in the 3D case by $\SI{0.5}{\micro \second}$, which is the amount of time by which the jet impact occurs earlier in 3D compared to 2D.  In the 3D scenario the temperatures can reach values of more than three times higher than the post-shock temperatures and in the 2D scenario more than twice the post-shock temperature. This leads to a higher percentage of maximum immediate burn of the material upon collapse in 3D (59\%) in 3D compared to 2D (11\%). The growth of the burning follows a similar trend, however, in the two cases. 

By comparing inert and reacting simulations we conclude that the effect of the reaction on the topology of the hot spots is negligible, whereas a large, additive effect on the temperature field is observed. We examine the maximum temperature in the centreline hot spots (CHS) and the MSHS in both inert and reactive scenarios. We demonstrate that after the birth of the hot spot the maximum temperature in the reactive case is increasing, as expected since the shock waves support the reactions and vice versa. In contrary, the maximum temperature in the inert case is decreasing ad the shock waves are not supported. An interesting feature observed is the superposition of waves emanating from the two lobes along the cavity centreline, leading to an additional short-lived maximum temperature peak in both cases.

The maximum temperatures describing the relative `strength' of the CHS and MSHS is studied as well. In the timescales considered, both in the inert and reactive scenarios, the CHS always exhibits higher temperatures than the MSHS. It is interesting to note that the temperatures of the MSHS in the reactive scenario are comparable to the temperatures in the CHS in the inert simulations.

The maximum temperature growth of the hot spots is also studied. Comparing the inert and reactive simulations, we observe that the rate of increase of maximum nitromethane temperature is positive for the reactive configuration but not in the inert case, for both the CHS and the MSHS. Comparing the rate of increase of maximum nitromethane temperature between the two hot spots it is observed that the rate of increase of the MSHS in the inert case is always higher than the CHS (except at the birth moment of the CHS i.e.\ upon the collapse of the cavity). In the reactive case this is however, not true. This is likely to have implications in configurations with multiple cavities collapsing in reactive media.

\section*{Acknowledgements}
This project was kindly funded by ORICA. The authors also benefited from conversations with A.
Minchinton, S.K. Chan and I.J. Kirby.

\bibliographystyle{spphys}       
\bibliography{shockwaves}   

\end{document}